\documentclass[aps,pra,floatfix,eqsecnum]{revtex4}
\usepackage{graphicx}% Include figure files
\usepackage{dcolumn}% Align table columns on decimal point
\usepackage{bm}% bold math
\usepackage{amsfonts}
\usepackage{latexsym}
\usepackage{amssymb}
\usepackage{verbatim}
\usepackage{amsthm}
\usepackage{amsmath}

\def\fun#1#2{\lower3.6pt\vbox{\baselineskip0pt\lineskip.9pt
  \ialign{$\mathsurround=0pt#1\hfil##\hfil$\crcr#2\crcr\sim\crcr}}}
\begin{document}

\title{
  Double balanced homodyne detection
}

\author{
  Kouji Nakamura
  ${}^{1}$\footnote{E-mail address: kouji.nakamura@nao.ac.jp},
  and
  Masa-Katsu Fujimoto
  ${}^{1}$\footnote{E-mail address: fujimoto.masa-katsu@nao.ac.jp}
}
%\email{kouji.nakamura@nao.ac.jp}
%\affil{%
\affiliation{%
  Gravitational-Wave Project Office,
  Optical and Infrared Astronomy Division,
  National Astronomical Observatory of Japan,
  Mitaka, Tokyo 181-8588, Japan
% \email{kouji.nakamura@nao.ac.jp}
}%

%%%% To generate auto affiliation numbers please use \author{}\affil{} command

%\author{Insert third author name here}
%\author[3]{Insert fourth author name here} %%% Use optional bracket [3] to change the respective address
%\affil{Insert third author address here}

%\author{Insert last author name here\thanks{These authors contributed equally to this work}}
%\affil{Insert last author address here}

%%% To include the collaborator name... Please use the command "\collaborator"
%%% For example: \collaborator{ATLAS Collaboration}

\date{\today}% It is always \today, today,

\begin{abstract}%
  In the context of the readout scheme for gravitational-wave
  detectors, the ``double balanced homodyne detection'' proposed in
  [K.~Nakamura and M.-K.~Fujimoto, arXiv:1709.01697.] is discussed in
  detail.
  This double balanced homodyne detection enables us to measure the
  expectation values of the photon creation and annihilation operators.
  Although it has been said that the operator
  $\hat{b}_{\theta}:=\cos\theta\hat{b}_{1}+\sin\theta\hat{b}_{2}$ can
  be measured through the homodyne detection in literature, we first show
  that the expectation value of the operator $\hat{b}_{\theta}$ cannot
  be measured as the linear combination of the upper- and
  lower-sidebands from the output of the balanced homodyne detection.
  Here, the operators $\hat{b}_{1}$ and $\hat{b}_{2}$ are the
  amplitude and phase quadrature in the two-photon formulation,
  respectively.
  On the other hand, it is shown that the above double balanced
  homodyne detection enables us to measure the expectation value of
  the operator $\hat{b}_{\theta}$ if we can appropriately prepare  the
  complex amplitude of the coherent state from the local oscillator.
  It is also shown that the interferometer set up of the eight-port
  homodyne detection realizes our idea of the double balanced homodyne
  detection.
  We also evaluate the noise-spectral density of the gravitational-wave
  detectors when our double balanced homodyne detection is applied as
  their readout scheme.
  Some requirements for the coherent state from the local oscillator
  to realize the double balanced homodyne detection are also
  discussed.
\end{abstract}

%\subjectindex{A60, E02}

%\pacs{
%  03.65.Ta, 03.65.Ca, 03.67.-a, 04.80.Nn, 42.50.-p, 42.50.Lc
%}% PACS, the Physics and Astronomy

\maketitle

%********************************************************************

%%%%%%%%%%%%%%%%%%%%%%%%%%%%%%%%%%%%%%%%%%%%%%%%%%%%%%%
%%%%%%%%%%%%%%%%%%%%%%%%%%%%%%%%%%%%%%%%%%%%%%%%%%%%%%%
%%%%%%%%%%%%%%%%%%%%%%%%%%%%%%%%%%%%%%%%%%%%%%%%%%%%%%%
\section{Introduction}
\label{sec:introduction}
%%%%%%%%%%%%%%%%%%%%%%%%%%%%%%%%%%%%%%%%%%%%%%%%%%%%%%%
%%%%%%%%%%%%%%%%%%%%%%%%%%%%%%%%%%%%%%%%%%%%%%%%%%%%%%%
%%%%%%%%%%%%%%%%%%%%%%%%%%%%%%%%%%%%%%%%%%%%%%%%%%%%%%%

%********************************************************************

In 2015, gravitational waves were directly observed by the Laser
Interferometer Gravitational-wave Observatory
(LIGO)~\cite{J.Aasi-et-al-2015}.
The first and second observing runs of the Advanced LIGO identified
binary black hole coalescence
signals~\cite{LIGO-GW150914-2016-GW151226-2016-GW170104-2017},
as well as a less significant
candidate~\cite{LIGO-PRD93-122003-2016-PRX6-041015-2016}.
During the second observing run of Advanced LIGO, European
gravitational-wave telescope Advanced Virgo~\cite{Advanced-Virgo-2015}
joined and reported a black hole coalescence
signal~\cite{LIGO-GW170814-2017} with more precise source location.
Furthermore, the first observation of gravitational wave from a
neutron-star binary coalescence was
identified~\cite{LIGO-GW170817-2017} and the electromagnetic counter
parts are observed~\cite{LIGO-GW170817-Multi-messenger-2017}.
Thus, the gravitational-wave astronomy and the multi-messenger
astronomy including gravitational-wave observation have finally
begun.
In addition, Japanese gravitational-wave telescope
KAGRA~\cite{KAGRA-2015} is also now constructing to create the global
network of the gravitational-wave detectors.
To develop this gravitational-wave astronomy as a more precise
science, it is important to continue the research and development of
the detector science together with source sciences.

%********************************************************************

The most difficult challenge in building a laser interferometric
gravitational-wave detector is isolating the test masses from the rest
of the world and keeping the device locked around the correct point of
the operation.
After these issues have been solved, we reach to the fundamental noise
that arises from quantum fluctuations in the system.
Actually, in Advanced LIGO, the sensitivity of detectors is limited by
the photon counting noise due to the quantum fluctuations of the
laser~\cite{J.Aasi-et-al-2015} and future gravitational-wave detectors
are also expected to be operated in a regime where the quantum physics
of both light and mirror motion couple to each other.
To achieve sensitivity improvements beyond the gravitational-wave
detectors of the next generation, a rigorous quantum-mechanical
description is required~\cite{S.L.Danilishin-F.Y.Khalili-2012}.

%********************************************************************

In 2001, Kimble et
al.~\cite{H.J.Kimble-Y.Levin-A.B.Matsko-K.S.Thorne-S.P.Vyatchanin-2001}
discussed quantum mechanical description of gravitational-wave
detectors in detail and their paper has been regarded as a milestone
paper of this topic.
In their paper, the idea of the frequency dependent homodyne detection
is proposed as one of techniques to reduce quantum noise and they
claimed that ``{\it we can measure the output quadrature
  $\hat{b}_{\theta}$ defined by
  \begin{eqnarray}
    \label{eq:hatbthehta-def}
    \hat{b}_{\theta}
    :=
    \cos\theta \hat{b}_{1} + \sin\theta \hat{b}_{2}
  \end{eqnarray}
  by the homodyne detection}''
through the citation of the works by Vyatchanin, Matsko and
Zubova~\cite{S.P.Vyatchanin-A.B.Matsko-1993}.
Here, $\hat{b}_{1}$ and $\hat{b}_{2}$ are the amplitude and phase
quadrature in the two-photon
formulation~\cite{C.M.Caves-B.L.Schumaker-1985} and $\theta$
is called the homodyne angle.
Furthermore, as a result of their analysis of the interferometric
gravitational-wave detectors, the output quadrature $\hat{b}_{\theta}$
includes gravitational-wave signal.
Apart from the leakage of the classical carrier field, we can formally
write their output quadrature as
\begin{eqnarray}
  \label{eq:hatbthehta-response}
  \hat{b}_{\theta}(\Omega)
  =
  R(\Omega,\theta) \left(
  \hat{h}_{n}(\Omega,\theta) + h(\Omega)
  \right)
  ,
\end{eqnarray}
where $h(\Omega)$ is a gravitational-wave signal in frequency domain,
$\hat{h}_{n}(\Omega)$ is the noise operator which is given by the
linear combination of the annihilation and creation operators of
photons which are injected to the main interferometer.
The idea of the frequency dependent homodyne
detection~\cite{H.J.Kimble-Y.Levin-A.B.Matsko-K.S.Thorne-S.P.Vyatchanin-2001}
is to choose the frequency dependent homodyne angle
$\theta=\theta(\Omega)$ so that the output noise is minimized at the
broad signal frequency range and their technique enables us to beat
the sensitivity limit called ``{\it standard quantum limit}'' of
gravitational-wave detectors which is also derived from Heisenberg's
uncertainty principle in quantum
mechanics~\cite{V.B.Braginsky-1968-1992}.

%********************************************************************

In gravitational-wave detectors, a homodyne detection is one of the
``readout schemes'' which enable us to measure the photon field which
includes gravitational-wave signals.
There are three typical readout schemes in gravitational-wave
detectors, which are called the heterodyne
detection~\cite{K.Somiya-and-A.Buonanno-Y.Chen-N.Mavalvala-2003}, the
DC readout scheme~\cite{LIGO-DC-readout-2006-2009}, and the homodyne
detection.
Current generation of detectors uses a DC readout scheme which
directly measures the power of the output field.
On the other hand, homodyne detections are not used in current
gravitational-wave detectors, yet, but are regarded as one of
candidates of readout schemes for future gravitational-wave detectors,
because these are expected to offers potential benefits over DC
readout~\cite{C.Bond-D.Brown-A.Freise-K.A.Strain-2016} as mentioned
above.
Thus, homodyne detections are one of targets of the research and
development activities for future gravitational-wave detectors.
Actually, there are some experimental papers on homodyne
detections~\cite{K.McKenzie-et-al-2007-M.S.Stefszky-et-al.-2012},
in spite that the measurement process of the operator $\hat{b}_{\theta}$
is still unclear.

%********************************************************************

In quantum measurement theory, a homodyne detection is known as the
measurement scheme of a linear combination of photon annihilation and
creation operators~\cite{H.M.Wiseman-G.J.Milburn-book-2009}.
In 1993, Wiseman and Milburn examined this homodyne detection in
detail~\cite{H.M.Wiseman-G.J.Milburn-1993} motivated by the
clarification of the property of the homodyne detection as a
non-projection measurement in quantum theory.
Based on the understanding of the homodyne detection in
Refs.~\cite{H.M.Wiseman-G.J.Milburn-book-2009,H.M.Wiseman-G.J.Milburn-1993},
we examine whether or not  ``the expectation value of
the operator $\hat{b}_{\theta}$ can be measured through the
balanced homodyne detection,'' in this paper.
As will be reviewed in
Sec.~\ref{sec:conventional-Homodyne-detections}, there are typically
two types of the homodyne detections, which are called
the ``simple homodyne detection'' and the ``balanced
homodyne detection,'' respectively.
Since the balanced homodyne detection gives us more precise results
than the simple homodyne detection, we regard that the homodyne
detection in
Ref.~\cite{H.J.Kimble-Y.Levin-A.B.Matsko-K.S.Thorne-S.P.Vyatchanin-2001}
is the balanced homodyne detection.
Furthermore, the operator $\hat{b}_{\theta}$ in the above statement is
not a self-adjoint operator, while ``observables'' are described by
self-adjoint operators in quantum theory.
The measurement of the expectation value of the non-self-adjoint
operator $\hat{b}_{\theta}$ is accomplished through the calculation of
a linear combination the expectation values of the self-adjoint
operators with complex coefficients.
This calculation corresponds to the simultaneous measurement of
the real- and imaginary-parts of the operator $\hat{b}_{\theta}$.
However, these real- and imaginary-parts do not commute with each
other and are regarded as ``non-simultaneously measureable
observables'' in quantum theory.
This non-commutable properties appear as the noise in the
measurement, in general, and this noise crucially depends on the
measurement process.
Therefore, the evaluation of the noise-spectral density is also
important to the measurements of the expectation values of
non-self-adjoint operators.

%********************************************************************

In the current quantum theoretical description of gravitational-wave
detectors, the two-photon
formulation~\cite{C.M.Caves-B.L.Schumaker-1985} is always used.
In this formulation, we consider the quantum fluctuations at the
frequency of the upper-sidebands $\omega_{0}+\Omega$ and the
lower-sidebands $\omega_{0}-\Omega$ around the classical carrier field
with the frequency $\omega_{0}$ at linear level.
The amplitude- and phase-quadratures in the two-photon formulation are
given by the linear combination of these sideband-quadratures.
When we consider the homodyne detection in two-photon formulation, we
have the information of the results of the balanced homodyne
detection with the upper- and lower-sidebands.
For these reasons, we regard that the general results from the balanced
homodyne detection is given by a linear combination of these upper-
and lower-sidebands.
Then, we examine the statement whether {\it `` the
  expectation value of the operator $\hat{b}_{\theta}$ can be measured
  as the linear combination of the upper- and lower-sidebands from the
  output of the balanced homodyne detection.''}
However, we reach to the negative conclusion against this statement,
which is our first assertion in this paper.

%********************************************************************

Based on the above examination, we also consider the problem, ``How to
realize the measurement of the expectation value of the operator
$\hat{b}_{\theta}$ in some way?''
As the result, we reach to the concept of the ``{\it double balanced
  homodyne detection}''.
As a realization of the ``double balanced homodyne detection'', we
rediscovered the eight-port homodyne detection in
literature~\cite{N.G.Walker-etal-1986-J.W.Noh-etal-1991-1993-M.G.Raymer-etal-1993}.
The eight-port homodyne detection was originally proposed as a device
to measure quasiprobability functions in quantum theory, and was used
in the discussion on problem what is the self-adjoint operator which
corresponds to the phase difference in the optical field.
Our double balanced homodyne detection was proposed in
Ref.~\cite{K.Nakamura-M.-K.Fujimoto-2017b} as a measurement scheme to
measure the expectation values of the photon annihilation and creation
operators and we show that {\it ``the expectation value of the
  operator $\hat{b}_{\theta}$ can be measured by the double balanced
  homodyne detection.''}
This is our second assertion in this paper.
We also evaluate the noise-spectral density of this measurement, which
is an important issue as mentioned above.

%********************************************************************

To realize the measurement through homodyne detections, we have to
prepare the optical field from the ``local oscillator'', appropriately.
In homodyne detections, an additional photon coherent state is mixed
to the output electric field from the main interferometer.
This additional light source is the ``local oscillator''.
In this paper, we assume that this additional field is in the coherent
state with an appropriate complex amplitude $\gamma(\omega)$.
In the above our examination, we also assume that we already knew the
complex amplitude $\gamma(\omega)$ and can control it.
Due to this assumption, we can look for the requirements for the
complex amplitude $\gamma(\omega)$ from the local oscillator to
realize the measurement of the expectation value of the operator
$\hat{b}_{\theta}$.

%********************************************************************

As an instructive example, we also consider the application of our
double balanced  homodyne detection to the Fabry-P\'erot
interferometer as the main interferometer of the gravitational-wave
detectors.
This example does show that the optimization of the signal-to-noise
ratio of gravitational-wave detectors must includes its readout scheme.
This is due to the fact that the total quantum measurement process
even for classical gravitational waves is completed through the
inclusion of its readout scheme.

%********************************************************************

The organization of this paper is as follows.
To avoid unnecessary confusion, in
Sec.~\ref{sec:conventional-Homodyne-detections}, we first review the
conventional homodyne detections from the view point of the Heisenberg
picture, while the arguments in
Refs.~\cite{H.M.Wiseman-G.J.Milburn-book-2009,H.M.Wiseman-G.J.Milburn-1993}
were based on the Schr\"odinger picture.
The descriptions in the Heisenberg picture is useful for the
arguments in this paper.
In Sec.~\ref{sec:BHD-two-photon}, we consider the balanced homodyne
detection in the two-photon formulation.
We show the statement, ``the expectation value of the operator
$\hat{b}_{\theta}$ cannot be measured as the linear combination of
the upper- and lower-sidebands from the output of the balanced
homodyne detection,'' in detail.
In Sec.~\ref{sec:Double_balanced_homodyne_detection}, we give the
detail explanation of the ``double balanced homodyne detection''
proposed~\cite{K.Nakamura-M.-K.Fujimoto-2017b} and show that the
statement, ``the expectation value of the operator $\hat{b}_{\theta}$
can be measured by the double balanced homodyne detection.''
In Sec.~\ref{sec:Noise_spectrum_GW_detector}, we show an example of an
application of our ``double balanced homodyne detection'' to
gravitational-wave detectors as their readout scheme.
The final section (Sec.~\ref{sec:Summary}) is devoted
to summary which includes the requirements for the coherent state from
the local oscillator to realize our double balanced homodyne detection
and the insight from the example of an application in
Sec.~\ref{sec:Noise_spectrum_GW_detector}.

%********************************************************************

Throughout this paper, we assume that the observed variable of optical
fields by the photodetector is ``photon number'' in the frequency
domain, though there is a long history of the controversy which
variable is the directly detected variable by the photodetectors in
the case of the detection of multi-frequency optical
fields~\cite{multi-frequency-filed-detection}.
In addition, we also assume that the optical field form the local
oscillator is the coherent state in which the each frequency modes do
not have any correlation.
Within this paper, we do not discuss about the experimental generation
of such states of the optical field, which is beyond current scope of
this paper.

%********************************************************************

This paper is also regarded as the full paper version of
Ref.~\cite{K.Nakamura-M.-K.Fujimoto-2017b}.

%********************************************************************

%%%%%%%%%%%%%%%%%%%%%%%%%%%%%%%%%%%%%%%%%%%%%%%%%%%%%%%
%%%%%%%%%%%%%%%%%%%%%%%%%%%%%%%%%%%%%%%%%%%%%%%%%%%%%%%
%%%%%%%%%%%%%%%%%%%%%%%%%%%%%%%%%%%%%%%%%%%%%%%%%%%%%%%
\section{Conventional homodyne detections in the Heisenberg picture}
\label{sec:conventional-Homodyne-detections}
%%%%%%%%%%%%%%%%%%%%%%%%%%%%%%%%%%%%%%%%%%%%%%%%%%%%%%%
%%%%%%%%%%%%%%%%%%%%%%%%%%%%%%%%%%%%%%%%%%%%%%%%%%%%%%%
%%%%%%%%%%%%%%%%%%%%%%%%%%%%%%%%%%%%%%%%%%%%%%%%%%%%%%%

%*******************************************************************

In this section, we review quantum theoretical descriptions of the
simple homodyne detection and the balanced homodyne
detection~\cite{H.M.Wiseman-G.J.Milburn-1993,H.M.Wiseman-G.J.Milburn-book-2009}
in the Heisenberg picture.
In Sec.~\ref{sec:Electric_field_notation}, we describe the notation of
the electric field in this paper.
In Sec.~\ref{sec:simple_homodyne_detection}, we describe the
explanation of the simple homodyne detector.
Then, in Sec.~\ref{sec:balanced_homodyne_detection}, we explain the
balanced homodyne detection.

%*******************************************************************

%%%%%%%%%%%%%%%%%%%%%%%%%%%%%%%%%%%%%%%%%%%%%%%%%%%%%%%
%%%%%%%%%%%%%%%%%%%%%%%%%%%%%%%%%%%%%%%%%%%%%%%%%%%%%%%
\subsection{Electric field notation}
\label{sec:Electric_field_notation}
%%%%%%%%%%%%%%%%%%%%%%%%%%%%%%%%%%%%%%%%%%%%%%%%%%%%%%%
%%%%%%%%%%%%%%%%%%%%%%%%%%%%%%%%%%%%%%%%%%%%%%%%%%%%%%%

%*******************************************************************

As well known, the electric field operator at time $t$ and the length
of the propagation direction $z$ in interferometers is described by
\begin{eqnarray}
  \label{eq:electric_field_original-1}
  \hat{E}_{a}(t-z)
  &=&
  \hat{E}^{(+)}_{a}(t-z) + \hat{E}^{(-)}_{a}(t-z)
  ,\\
  \label{eq:electric_field_original-2}
  \hat{E}_{a}^{(-)}(t-z)
  &=&
  \left[\hat{E}_{a}^{(+)}(t-z)\right]^{\dagger},
  \\
  \label{eq:electric_field_original-3}
  \hat{E}^{(+)}_{a}(t-z)
  &=&
  \int_{0}^{\infty} \sqrt{\frac{2\pi\hbar\omega}{{\cal A}c}}
  \hat{a}(\omega) e^{-i\omega(t-z)} \frac{d\omega}{2\pi},
\end{eqnarray}
where $\hat{a}(\omega)$ is the photon annihilation operator associated
with the electric field $\hat{E}_{a}(t-z)$ and ${\cal A}$ is the
cross-sectional area of the optical beam.
The annihilation operator $\hat{a}(\omega)$ satisfies the usual
commutation relation
\begin{eqnarray}
  \label{eq:usual-commutation-relation}
  \left[
  \hat{a}(\omega),
  \hat{a}^{\dagger}(\omega')
  \right]
  =
  2 \pi \delta(\omega-\omega')
  ,
  \quad
  \left[
  \hat{a}(\omega),
  \hat{a}(\omega')
  \right]
  =
  0
  .
\end{eqnarray}

%*******************************************************************

Throughout this paper, we denote the quadrature $\hat{a}$ as that for
the input field to the main interferometer.
The state associated with this quadrature $\hat{a}$ is usually the
vacuum state.
On the other hand, the output quadrature from the main interferometer
is denoted by $\hat{b}$.
Furthermore, in the homodyne detections, we inject the electric field
whose state is a coherent state as a reference field from the local
oscillator.
The quadrature associated with the electric field from the local
oscillator is denoted by $\hat{l}_{i}$.
As noted above, the state associated with the quadrature $\hat{l}_{i}$
is the coherent state $|\gamma\rangle_{l_{i}}$ which satisfies
\begin{eqnarray}
  \label{eq:hatli-coherent-state-def}
  \hat{l}_{i}(\omega) |\gamma\rangle_{l_{i}} = \gamma(\omega) |\gamma\rangle_{l_{i}}.
\end{eqnarray}
Here, $\gamma(\omega)$ is the complex eigenvalue for the coherent
state.
As well-known, the coherent state $|\gamma\rangle_{l_{i}}$ is given by
the operation of the operator $D[\gamma]$ to the vacuum state
$|0\rangle_{l_{i}}$ as
\begin{eqnarray}
  \label{eq:hatli-coherent-state-vacuum-state-relation}
  |\gamma\rangle_{l_{i}}
  =
  D[\gamma]|0\rangle_{l_{i}}
  :=
  \exp\left[
  \int \frac{d\omega}{2\pi} \left(
  \gamma(\omega) \hat{l}_{i}^{\dagger}(\omega)
  -
  \gamma(\omega)^{*} \hat{l}_{i}(\omega)
  \right)
  \right]|0\rangle_{l_{i}}
  .
\end{eqnarray}

%*******************************************************************

%%%%%%%%%%%%%%%%%%%%%%%%%%%%%%%%%%%%%%%%%%%%%%%%%%%%%%%
%%%%%%%%%%%%%%%%%%%%%%%%%%%%%%%%%%%%%%%%%%%%%%%%%%%%%%%
\subsection{Simple homodyne detection}
\label{sec:simple_homodyne_detection}
%%%%%%%%%%%%%%%%%%%%%%%%%%%%%%%%%%%%%%%%%%%%%%%%%%%%%%%
%%%%%%%%%%%%%%%%%%%%%%%%%%%%%%%%%%%%%%%%%%%%%%%%%%%%%%%

%*******************************************************************

\begin{figure}[ht]
  \centering
  \includegraphics[width=0.5\textwidth]{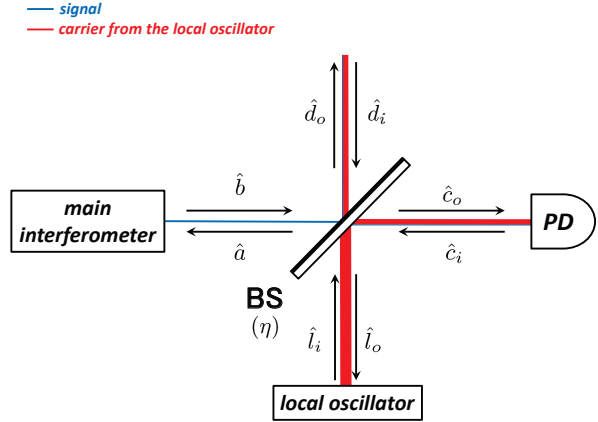}
  \caption{
    Configuration of the interferometer for the simple homodyne
    detection.
    The notations of the quadratures $\hat{a}$, $\hat{b}$,
    $\hat{c}_{o}$, $\hat{c}_{i}$, $\hat{d}_{o}$, $\hat{d}_{i}$,
    $\hat{l}_{o}$, and $\hat{l}_{i}$ are also given in this figure.
  }
  \label{fig:Simple_homodyne_detection}
\end{figure}

Here, we review the simple homodyne detection depicted in
Fig.~\ref{fig:Simple_homodyne_detection}.
In this paper, we want to evaluate the signal in the electric field
associated with the annihilation operator $\hat{b}(\omega)$.
The electric field from the local oscillator is the coherent state
(\ref{eq:hatli-coherent-state-vacuum-state-relation}) with the complex
amplitude $\gamma(\omega)$.
The output signal field associated with the quadrature $\hat{b}$ and
the additional optical field from the local oscillator is mixed
through the beam splitter with the transmissivity $\eta$.
In the ideal case, this transmissivity is $\eta=1/2$.
However, in this paper, we dare to denote this transmissivity of the
beam splitter for the homodyne detection by $\eta$ as a simple model
of the imperfection of the interferometer.
In the simple homodyne detection, the photon number is detected by the
photodetector (PD in Fig.~\ref{fig:Simple_homodyne_detection}) at
one of the port from the beam splitter as in
Fig.~\ref{fig:Simple_homodyne_detection}.

%*******************************************************************

%*******************************************************************

The output electric field $\hat{E}_{b}$ from the main interferometer
is given by
\begin{eqnarray}
  \hat{E}_{b}(t)
  =
  \int_{0}^{+\infty}
  \sqrt{\frac{2\pi\hbar|\omega|}{{\cal A}c}}
  \left(
  \hat{b}(\omega)
  e^{-i\omega t}
  +
  \hat{b}^{\dagger}(\omega)
  e^{+i\omega t}
  \right)
  \frac{d\omega}{2\pi}
  \label{eq:output_electric_field_form_main}
\end{eqnarray}
and the electric field from the local oscillator is
given by
\begin{eqnarray}
  \hat{E}_{l_{i}}(t)
  =
  \int_{0}^{+\infty}
  \sqrt{\frac{2\pi\hbar|\omega|}{{\cal A}c}}
  \left(
  \hat{l}_{i}(\omega)
  e^{-i\omega t}
  +
  \hat{l}_{i}^{\dagger}(\omega)
  e^{+i\omega t}
  \right)
  \frac{d\omega}{2\pi}
  .
  \label{eq:electric_field_form_local_oscillator}
\end{eqnarray}
Furthermore, the electric field to be detected through the
photodetector is given by
\begin{eqnarray}
  \hat{E}_{c_{o}}(t)
  =
  \int_{0}^{+\infty}
  \sqrt{\frac{2\pi\hbar|\omega|}{{\cal A}c}}
  \left(
  \hat{c}_{o}(\omega)
  e^{-i\omega t}
  +
  \hat{c}_{o}^{\dagger}(\omega)
  e^{+i\omega t}
  \right)
  \frac{d\omega}{2\pi}
  .
  \label{eq:detected_electric_field}
\end{eqnarray}

%*******************************************************************

At the beam splitter, the signal electric field $\hat{E}_{b}(t)$ and
the electric field $\hat{E}_{l_{i}}(t)$ are mixed and the beam
splitter output the electric field $\hat{E}_{c_{o}}(t)$ to the one of
the ports.
These fields are related by the relation
\begin{eqnarray}
  \label{eq:hatEco-hatEb-hatEli-beamsplitter-relation}
  \hat{E}_{c_{o}}(t)
  =
  \sqrt{\eta} \hat{E}_{b}(t) + \sqrt{1-\eta} \hat{E}_{l_{i}}(t)
  .
\end{eqnarray}
This relation is given in terms of the quadrature relation as
\begin{eqnarray}
  \label{eq:hatco-hatb-hatli-beamsplitter-relation}
  \hat{c}_{o}(\omega)
  =
  \sqrt{\eta} \hat{b}(\omega) + \sqrt{1-\eta} \hat{l}_{i}(\omega)
  .
\end{eqnarray}

%*******************************************************************

Here, we assign the states for the independent electric field
described in Fig.~\ref{fig:Simple_homodyne_detection}.
As mentioned above, the electric field from the local oscillator is in
the coherent state
(\ref{eq:hatli-coherent-state-vacuum-state-relation}).
In addition to this state, the electric fields associated with the
quadratures $\hat{d}_{i}$ and $\hat{c}_{i}$ are injected into the beam
splitter.
In this paper, we regard that the electric fields associated with the
quadratures $\hat{d}_{i}$ and $\hat{c}_{i}$ are in their vacua,
respectively.
The junction condition for the electric fields at the beam
splitter is given by
\begin{eqnarray}
  \label{eq:beam_splitter_junction_input_to_main}
  \hat{E}_{a}(t)
  =
  \sqrt{\eta} \hat{E}_{c_{i}}(t) - \sqrt{1-\eta}
  \hat{E}_{d_{i}}(t).
\end{eqnarray}
In terms of quadratures, this relation is given by
\begin{eqnarray}
  \label{eq:beam_splitter_junction_input_to_main_quadrature}
  \hat{a}(\omega)
  =
  \sqrt{\eta} \hat{c}_{i}(\omega) - \sqrt{1-\eta} \hat{d}_{i}(\omega).
\end{eqnarray}
Due to this relation
(\ref{eq:beam_splitter_junction_input_to_main_quadrature}), the state
associated with the quadrature $\hat{a}$ is described by the vacuum
states for the quadratures $\hat{d}_{i}$ and $\hat{c}_{i}$.

%*******************************************************************

Usually, the state associated with the quadrature $\hat{b}$ depends on
the state of the input field $\hat{E}_{a}$ into the main
interferometer and the other optical fields which inject to the main
interferometer~\cite{H.J.Kimble-Y.Levin-A.B.Matsko-K.S.Thorne-S.P.Vyatchanin-2001,H.Miao-PhDthesis-2010}.
Furthermore, in this paper, we consider the situation where the
output electric field $\hat{E}_{b}$ includes the information of
classical forces as in Eq.~(\ref{eq:hatbthehta-response}) and this
information are measured through the expectation value of the
operator $\hat{b}$.

%*******************************************************************

To evaluate the expectation value of the quadrature $\hat{b}$, we have
to specify the state of the total system.
Here, we assume that this state of the total system is given by
\begin{eqnarray}
  \label{eq:state_of_total_system}
  |\Psi\rangle
  =
  |\gamma\rangle_{l_{i}}\otimes|0\rangle_{c_{i}}\otimes|0\rangle_{d_{i}}\otimes|\psi\rangle_{\text{main}}
  ,
\end{eqnarray}
where the state $|\psi\rangle_{\text{main}}$ is the state for the
electric fields associated with the main interferometer, which is
independent of the states $|\gamma\rangle_{l_{i}}$,
$|0\rangle_{c_{i}}$, and $|0\rangle_{d_{i}}$.
As noted above, the output quadrature $\hat{b}$ may depends on the
input quadrature $\hat{a}$ and this input quadrature $\hat{a}$ is
related to the quadratures $\hat{c}_{i}$ and $\hat{d}_{i}$ through
Eq.~(\ref{eq:beam_splitter_junction_input_to_main_quadrature}).
Therefore, strictly speaking, the expectation value of the quadrature
$\hat{b}$ means that
\begin{eqnarray}
  \label{eq:expectation_value_of_b_def_original}
  \langle\hat{b}\rangle
  =
  \langle \psi|_{\text{main}}\otimes\langle 0|_{d_{i}}\otimes\langle 0|_{c_{i}}
  \hat{b}
  |0\rangle_{c_{i}}\otimes|0\rangle_{d_{i}}\otimes|\psi\rangle_{\text{main}},
\end{eqnarray}
but we denote simply
\begin{eqnarray}
  \label{eq:expectation_value_of_b_our_def}
  \langle\hat{b}\rangle
  :=
  \langle \Psi|
  \hat{b}
  |\Psi\rangle
  .
\end{eqnarray}

%*******************************************************************

PD in Fig.~\ref{fig:Simple_homodyne_detection} detects the photon
number of the electric field associated with the annihilation operator
$\hat{c}_{o}$.
So, we evaluate the expectation value of the photon number operator
$\hat{n}_{c_{o}}(\omega)
:=\hat{c}_{o}^{\dagger}(\omega)\hat{c}_{o}(\omega)$.
From the condition at the beam splitter, we obtain
\begin{eqnarray}
  \hat{n}_{c_{o}}(\omega)
  =
  \left(
    \sqrt{\eta} \hat{b}^{\dagger}(\omega)
    +
    \sqrt{1-\eta} \hat{l}_{i}^{\dagger}(\omega)
  \right)
  \left(
    \sqrt{\eta} \hat{b}(\omega)
    +
    \sqrt{1-\eta} \hat{l}_{i}(\omega)
  \right)
  .
  \label{eq:detected-photon-number-operator-simple-homodyne}
\end{eqnarray}
Through the definition of the coherent state
(\ref{eq:hatli-coherent-state-def}), the expectation value for this
number operator
(\ref{eq:detected-photon-number-operator-simple-homodyne}) under the
state (\ref{eq:state_of_total_system}) for the total system is given
by
\begin{eqnarray}
  \label{eq:PhotonNumberExpectation_SimpleHomodyne-def}
  \langle\hat{n}_{c_{o}}(\omega)\rangle
  &:=&
       \langle\Psi|\hat{n}_{c_{o}}(\omega)|\Psi\rangle
       \\
  \label{eq:PhotonNumberExpectation_SimpleHomodyne-result}
  &=&
      \eta
      \left\langle
      \hat{n}_{b}(\omega)
      \right\rangle
      +
      (1-\eta)
      \left|\gamma(\omega)\right|^{2}
      +
      \sqrt{\eta(1-\eta)}
      \left\langle
      \gamma^{*}(\omega)
      \hat{b}(\omega)
      +
      \gamma(\omega)
      \hat{b}^{\dagger}(\omega)
      \right\rangle
      ,
\end{eqnarray}
where we defined
\begin{eqnarray}
  &&
     \left\langle
     \hat{n}_{b}(\omega)
     \right\rangle
     =
      \left\langle
      \hat{b}^{\dagger}(\omega)
      \hat{b}(\omega)
      \right\rangle
      :=
       \langle\Psi|
       \hat{b}^{\dagger}(\omega)
       \hat{b}(\omega)
       |\Psi\rangle
       \label{eq:nb-SimpleHomodyne-def}
       ,
  \\
  &&
     \left\langle
     \gamma^{*}(\omega)
     \hat{b}(\omega)
     +
     \gamma(\omega)
     \hat{b}^{\dagger}(\omega)
     \right\rangle
     :=
     \langle\Psi|
     \left(
     \gamma^{*}(\omega)
     \hat{b}(\omega)
     +
     \gamma(\omega)
     \hat{b}^{\dagger}(\omega)
     \right)
     |\Psi\rangle
     .
     \label{eq:gammabdagger+bammaastb-SimpleHomodyne-def}
\end{eqnarray}

%*******************************************************************

As we assumed above, $\gamma(\omega)$ is a known complex function of
$\omega$.
Therefore, we can eliminate the term $(1-\eta)|\gamma(\omega)|^{2}$
from the expectation value $\langle\hat{n}_{c_{o}}(\omega)\rangle$,
which is the dominant term in
Eq.~(\ref{eq:PhotonNumberExpectation_SimpleHomodyne-result}) in the
situation where the absolute value of $\gamma(\omega)$ is very large.
In the same situation, we can neglect the first term
$\eta\left\langle\hat{n}_{b}(\omega)\right\rangle$ from the
expectation value
(\ref{eq:PhotonNumberExpectation_SimpleHomodyne-result}).
Then, we can measure the expectation value of the linear combination
$\gamma^{*}(\omega)\hat{b}(\omega)$ $+$
$\gamma(\omega)\hat{b}^{\dagger}(\omega)$
of the annihilation- and creation operators for the output field as
the subdominant contribution.
This is the simple homodyne detection in Fig.~\ref{fig:Simple_homodyne_detection}.
We note that the large $\gamma(\omega)$ is essential in this simple
homodyne detection.

%*******************************************************************

%%%%%%%%%%%%%%%%%%%%%%%%%%%%%%%%%%%%%%%%%%%%%%%%%%%%%%%
%%%%%%%%%%%%%%%%%%%%%%%%%%%%%%%%%%%%%%%%%%%%%%%%%%%%%%%
\subsection{Balanced homodyne detection}
\label{sec:balanced_homodyne_detection}
%%%%%%%%%%%%%%%%%%%%%%%%%%%%%%%%%%%%%%%%%%%%%%%%%%%%%%%
%%%%%%%%%%%%%%%%%%%%%%%%%%%%%%%%%%%%%%%%%%%%%%%%%%%%%%%

%*******************************************************************

\begin{figure}[ht]
  \centering
  \includegraphics[width=0.5\textwidth]{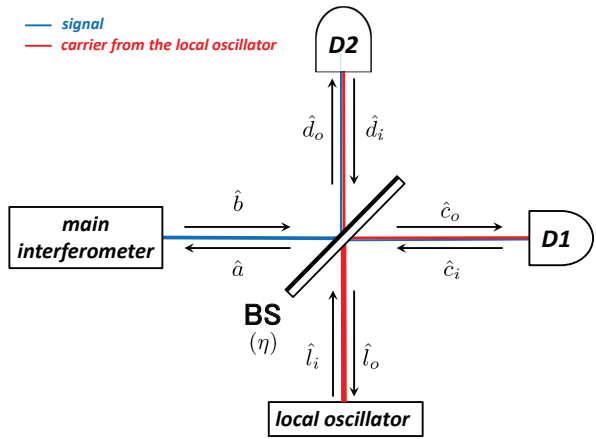}
  \caption{
    Configuration of the interferometer for the balanced homodyne
    detection.
    This notation of the quadratures are same as those in
    Fig.~\ref{fig:Simple_homodyne_detection}.
  }
  \label{fig:Balanced_homodyne_detection}
\end{figure}

The optical configuration of the balanced homodyne detection is almost
similar to the simple homodyne detection depicted in
Fig.~\ref{fig:Simple_homodyne_detection}, but the difference is in the
additional photon detection through the photodetector D2 in
Fig.~\ref{fig:Balanced_homodyne_detection} at another port of the
beam splitter.

%*******************************************************************

%*******************************************************************

To discuss the output of the balanced homodyne detection, we consider
the photon number operators detected by the photodetectors D1 and D2
in Fig.~\ref{fig:Balanced_homodyne_detection}, respectively.
At the photodetector D1,  we count the photon number which expressed
the operator
(\ref{eq:detected-photon-number-operator-simple-homodyne}) in
Sec.~\ref{sec:simple_homodyne_detection}.
Since the total state of the system is same as the state
(\ref{eq:state_of_total_system}) in the case of the simple homodyne
detector, the expectation value of the operator
(\ref{eq:detected-photon-number-operator-simple-homodyne}) is also
given by Eq.~(\ref{eq:PhotonNumberExpectation_SimpleHomodyne-result})
in Sec.~\ref{sec:simple_homodyne_detection}.
On the other hand, we also detect the photon number through the
photodetector D2 in the balanced homodyne detection.

%*******************************************************************

Following the notation of quadratures depicted in
Fig.~\ref{fig:Balanced_homodyne_detection}, the photon number operator
which is detected at the photodetector D2 is given by
$\hat{n}_{d_{o}}(\omega)
:=\hat{d}_{o}^{\dagger}(\omega)\hat{d}_{o}(\omega)$.
The output electric field
$\hat{E}_{d_{o}}(t)$ is produced through the mixing of the signal
electric field $\hat{E}_{b}(t)$ and the electric field
$\hat{E}_{l_{i}}(t)$ from the local oscillator at the beam splitter as
\begin{eqnarray}
  \label{eq:hatEdo-hatEb-hatEli-beamsplitter-relation-BHD}
  \hat{E}_{d_{o}}(t)
  =
  \sqrt{\eta} \hat{E}_{b}(t) - \sqrt{1-\eta} \hat{E}_{l_{i}}(t)
  .
\end{eqnarray}
This relation gives the quadrature relation as
\begin{eqnarray}
  \label{eq:hatdo-hatb-hatli-beamsplitter-relation-BHD}
  \hat{d}_{o}(\omega)
  =
  \sqrt{\eta} \hat{b}(\omega) - \sqrt{1-\eta} \hat{l}_{i}(\omega)
  .
\end{eqnarray}
Then, the photon number operator $\hat{n}_{d_{o}}(\omega)$ is given by
\begin{eqnarray}
  \hat{n}_{d_{o}}(\omega)
  =
  \left(
    \sqrt{\eta} \hat{l}_{i}(\omega)
    -
    \sqrt{1-\eta} \hat{b}(\omega)
  \right)^{\dagger}
  \left(
    \sqrt{\eta} \hat{l}_{i}(\omega)
    -
    \sqrt{1-\eta} \hat{b}(\omega)
  \right)
  .
  \label{eq:detected-photon-number-operator-D2-BHD}
\end{eqnarray}
The expectation value of the operator $\hat{n}_{d_{o}}(\omega)$
in the state (\ref{eq:state_of_total_system}) is given by
\begin{eqnarray}
  \langle\hat{n}_{d_{o}}(\omega)\rangle
  &:=&
       \langle\Psi|\hat{n}_{d_{o}}(\omega)|\Psi\rangle
       \nonumber\\
  &=&
      (1-\eta)
      \langle\hat{n}_{b}(\omega)\rangle
      +
      \eta |\gamma(\omega)|^{2}
      -
      \sqrt{\eta(1-\eta)}
      \left\langle
      \gamma^{*}(\omega) \hat{b}(\omega)
      +
      \gamma(\omega) \hat{b}^{\dagger}(\omega)
      \right\rangle
      ,
  \label{eq:PhotonNumberExpectation_D2_BHD}
\end{eqnarray}
where we used Eqs.~(\ref{eq:nb-SimpleHomodyne-def}) and
(\ref{eq:gammabdagger+bammaastb-SimpleHomodyne-def}).

%*******************************************************************

Taking the linear combination of
Eqs.~(\ref{eq:PhotonNumberExpectation_SimpleHomodyne-result}) and
(\ref{eq:PhotonNumberExpectation_D2_BHD}), we obtain
\begin{eqnarray}
  &&
  (1-\eta)\langle\hat{n}_{c_{o}}(\omega)\rangle
  -
  \eta\langle\hat{n}_{d_{o}}(\omega)\rangle
  \nonumber\\
  &=&
  \left(
    1 - 2 \eta
  \right)
  \left|\gamma(\omega)\right|^{2}
  +
  \sqrt{\eta(1-\eta)}
  \left\langle
    \gamma^{*}(\omega)
    \hat{b}(\omega)
    +
    \gamma(\omega)
    \hat{b}^{\dagger}(\omega)
  \right\rangle
  .
  \label{eq:BHD-combination}
\end{eqnarray}
Since $\gamma(\omega)$ is a known complex function, we can always
subtract the first term in Eq.~(\ref{eq:BHD-combination}) from the
output signal.
Then, we can unambiguously measure the linear combination of the
quadrature $\hat{b}(\omega)$ and $\hat{b}^{\dagger}(\omega)$ as
\begin{eqnarray}
  \left\langle
    \gamma^{*}(\omega)
    \hat{b}(\omega)
    +
    \gamma(\omega)
    \hat{b}^{\dagger}(\omega)
  \right\rangle
  &=&
      \frac{
      (1-\eta)\langle\hat{n}_{c_{o}}(\omega)\rangle
      -
      \eta\langle\hat{n}_{d_{o}}(\omega)\rangle
      -
      \left(
      1 - 2 \eta
      \right)
      \left|\gamma(\omega)\right|^{2}
      }{
      \sqrt{\eta(1-\eta)}
      }
      .
      \label{eq:BHD-output-expression}
\end{eqnarray}
This is the balanced homodyne detection.

%*******************************************************************

The right-hand side of Eq.~(\ref{eq:BHD-output-expression}) is also
regarded as the expectation value of the operator $\hat{s}(\omega)$
defined by
\begin{eqnarray}
  \label{eq:BHD-operator-def}
  \hat{s}(\omega)
  &:=&
       \frac{1}{(\eta(1-\eta))^{1/2}}
       \left\{
       (1-\eta)
       \hat{n}_{c_{o}}(\omega)
       -
       \eta
       \hat{n}_{d_{o}}(\omega)
       -
       \left(
       1 - 2 \eta
       \right)
       \left|\gamma(\omega)\right|^{2}
       \right\}
       ,
  \\
  \label{eq:BHD-operator-explicit}
  &=&
      \hat{l}_{i}^{\dagger}(\omega) \hat{b}(\omega)
      +
      \hat{b}^{\dagger}(\omega) \hat{l}_{i}(\omega)
      +
      \frac{1 - 2 \eta}{\sqrt{\eta(1-\eta)}}
      \left(
      \hat{l}_{i}^{\dagger}(\omega) \hat{l}_{i}(\omega)
      -
      \left|\gamma(\omega)\right|^{2}
      \right)
      ,
\end{eqnarray}
where we used
Eqs.~(\ref{eq:detected-photon-number-operator-simple-homodyne}) and (\ref{eq:detected-photon-number-operator-D2-BHD}).
Actually, the expectation value of the operator
(\ref{eq:BHD-operator-explicit}) under the state
(\ref{eq:state_of_total_system}) yields the left-hand side of
Eq.~(\ref{eq:BHD-output-expression}).
Here, we note that the operator $\hat{s}(\omega)$ is a self-adjoint
operator due to the fact that the photon number operators are
self-adjoint and $\hat{s}(\omega)$ is given by the linear combination
of these number operators with real coefficients.
Therefore, the expectation value $\langle\hat{s}(\omega)\rangle$ of
the operator $\hat{s}(\omega)$ must be real as seen in the right-hand
side Eq.~(\ref{eq:BHD-output-expression}).

\section{Balanced homodyne detection in the two-photon formulation}
\label{sec:BHD-two-photon}
%%%%%%%%%%%%%%%%%%%%%%%%%%%%%%%%%%%%%%%%%%%%%%%%%%%%%%%
%%%%%%%%%%%%%%%%%%%%%%%%%%%%%%%%%%%%%%%%%%%%%%%%%%%%%%%
%%%%%%%%%%%%%%%%%%%%%%%%%%%%%%%%%%%%%%%%%%%%%%%%%%%%%%%

%*******************************************************************

In the community of gravitational-wave detections, the two-photon
formulation~\cite{C.M.Caves-B.L.Schumaker-1985} is used under the
condition $\omega_{0}\gg\Omega$, where $\omega_{0}$ is the frequency
of the classical carrier field and $\Omega$ is its sideband frequency.
In Sec.~\ref{sec:ampl-quadr-phase}, we introduce the amplitude
quadrature $\hat{b}_{1}$ and phase-quadratures $\hat{b}_{2}$ in
Eq.~(\ref{eq:hatbthehta-def}).
In Sec.~\ref{sec:can_we_measure_coshatb1+sinhatb2}, we examine whether
or not we can measure the combination (\ref{eq:hatbthehta-def})
through the simple application of the balanced homodyne
detection reviewed in Sec.~\ref{sec:balanced_homodyne_detection} and
show that we cannot measure the expectation value of the combination
$\hat{b}_{\theta}$ in Eq.~(\ref{eq:hatbthehta-def}) by the simple
application of the balanced homodyne detection in
Sec.~\ref{sec:balanced_homodyne_detection}.
In Sec.~\ref{sec:examination-summary}, we describe some remarks of the
examination in Sec.~\ref{sec:can_we_measure_coshatb1+sinhatb2}.

%*******************************************************************

%%%%%%%%%%%%%%%%%%%%%%%%%%%%%%%%%%%%%%%%%%%%%%%%%%%%%%%
%%%%%%%%%%%%%%%%%%%%%%%%%%%%%%%%%%%%%%%%%%%%%%%%%%%%%%%
\subsection{Amplitude quadrature and phase quadrature }
\label{sec:ampl-quadr-phase}
%%%%%%%%%%%%%%%%%%%%%%%%%%%%%%%%%%%%%%%%%%%%%%%%%%%%%%%
%%%%%%%%%%%%%%%%%%%%%%%%%%%%%%%%%%%%%%%%%%%%%%%%%%%%%%%

%*******************************************************************

Keeping in our mind the above situation $\omega_{0}\gg\Omega$ of the
two-photon formulation, we introduce variables $\hat{b}_{\pm}$,
$\hat{l}_{\pm}$, and $\gamma_{\pm}$ as
\begin{eqnarray}
  \label{eq:hatbpm-hatlipm-gammapm-defs}
  \hat{b}_{\pm}(\Omega) := \hat{b}(\omega_{0}\pm\Omega), \quad
  \hat{l}_{i\pm}(\Omega) :=  \hat{l}_{i}(\omega_{0}\pm\Omega), \quad
  \gamma_{\pm}(\Omega) :=  \gamma(\omega_{0}\pm\Omega).
\end{eqnarray}
In this situation, the operator (\ref{eq:BHD-operator-explicit}) of
the homodyne detection is given by
\begin{eqnarray}
  \hat{s}_{\pm}
  &=&
  \hat{l}_{i\pm}^{\dagger} \hat{b}_{\pm}
  +
  \hat{b}_{\pm}^{\dagger} \hat{l}_{i\pm}
  +
  \frac{1 - 2 \eta}{\sqrt{\eta(1-\eta)}}
  \left(
    \hat{l}_{i\pm}^{\dagger} \hat{l}_{i\pm}
    -
    \left|\gamma_{\pm}\right|^{2}
  \right)
  .
  \label{eq:hatspm-def}
\end{eqnarray}
Furthermore, we introduce the amplitude and phase quadratures as
\begin{eqnarray}
  \label{eq:hatb1-hatb2-def}
  \hat{b}_{1}
  :=
  \frac{1}{\sqrt{2}}\left(\hat{b}_{+}+\hat{b}_{-}^{\dagger}\right)
  , \quad
  \hat{b}_{2}
  :=
  \frac{1}{\sqrt{2}i}\left(\hat{b}_{+}-\hat{b}_{-}^{\dagger}\right)
  .
\end{eqnarray}
Under the existence of the classical carrier field which proportional
to $\cos\omega_{0}t$, $\bar{b}_{1}$ and $\hat{b}_{2}$ are regarded as
the quadratures for the amplitude fluctuations and the phase
fluctuations, respectively.
The definitions (\ref{eq:hatb1-hatb2-def}) are equivalent to
\begin{eqnarray}
  \label{eq:hatb+hatb--hatb1-hatb2-relations}
  \hat{b}_{+}
  :=
  \frac{1}{\sqrt{2}} \left(
  \hat{b}_{1}
  +
  i \hat{b}_{2}
  \right)
  , \quad
  \hat{b}_{-}
  :=
  \frac{1}{\sqrt{2}} \left(
  \hat{b}_{1}^{\dagger}
  + i \hat{b}_{2}^{\dagger}
  \right)
  .
\end{eqnarray}

%*******************************************************************

As mentioned above, the aim of this section is to examine whether or
not the expectation-value measurement of the quadrature
(\ref{eq:hatbthehta-def}) with the homodyne angle $\theta$ is
possible through the simple application of the balanced homodyne
detection described in Sec.~\ref{sec:balanced_homodyne_detection}.
To carry out this examination, we have to consider the expectation values
of the operator $\hat{s}_{\pm}$ defined by  Eq.~(\ref{eq:hatspm-def})
as
\begin{eqnarray}
  \label{eq:hats+hats--exp}
  \left\langle
    \hat{s}_{+}
  \right\rangle
  =
  \left\langle
    \gamma_{+}^{*} \hat{b}_{+}
    +
    \hat{b}_{+}^{\dagger} \gamma_{+}
  \right\rangle
  , \quad
  \left\langle
    \hat{s}_{-}
  \right\rangle
  =
  \left\langle
    \gamma_{-}^{*} \hat{b}_{-}
    +
    \hat{b}_{-}^{\dagger} \gamma_{-}
  \right\rangle
  .
\end{eqnarray}
Substituting Eqs.~(\ref{eq:hatb+hatb--hatb1-hatb2-relations}) into
Eqs.~(\ref{eq:hats+hats--exp}) and taking linear combination of
$\langle\hat{s}_{+}\rangle$ and $\langle\hat{s}_{-}\rangle$, we obtain
\begin{eqnarray}
  \alpha \left\langle\hat{s}_{+}\right\rangle
  +
  \beta \left\langle\hat{s}_{-}\right\rangle
  &=&
      \frac{1}{\sqrt{2}}
      \left\langle
      \left(
      \alpha
      \gamma_{+}^{*}
      +
      \beta
      \gamma_{-}
      \right)
      \hat{b}_{1}
      +
      i
      \left(
      \alpha
      \gamma_{+}^{*}
      -
      \beta
      \gamma_{-}
      \right)
      \hat{b}_{2}
      \right.
      \nonumber\\
  && \quad\quad\quad\quad
     \left.
      +
      \left(
      \alpha
      \gamma_{+}
      +
      \beta
      \gamma_{-}^{*}
      \right)
      \hat{b}_{1}^{\dagger}
      +
      i
      \left(
      -
      \alpha
      \gamma_{+}
      +
      \beta
      \gamma_{-}^{*}
      \right)
      \hat{b}_{2}^{\dagger}
      \right\rangle
      .
  \label{eq:alphaexphats++betaexphats-}
\end{eqnarray}
The linear combination
(\ref{eq:alphaexphats++betaexphats-}) yields the possible
output signal by choosing the complex parameter $\alpha$,
$\beta$, and $\gamma_{\pm}$ which are completely
controllable.

%*******************************************************************

%%%%%%%%%%%%%%%%%%%%%%%%%%%%%%%%%%%%%%%%%%%%%%%%%%%%%%%
%%%%%%%%%%%%%%%%%%%%%%%%%%%%%%%%%%%%%%%%%%%%%%%%%%%%%%%
\subsection{Can we measure the expectation value of $\hat{b}_{\theta}$
  by the simple application of the balanced homodyne detection?}
\label{sec:can_we_measure_coshatb1+sinhatb2}
%%%%%%%%%%%%%%%%%%%%%%%%%%%%%%%%%%%%%%%%%%%%%%%%%%%%%%%
%%%%%%%%%%%%%%%%%%%%%%%%%%%%%%%%%%%%%%%%%%%%%%%%%%%%%%%

%*******************************************************************

To examine the problem whether or not we can measure the expectation
value of the linear combination defined by Eq.~(\ref{eq:hatbthehta-def})
through an appropriate choice of $\alpha$, $\beta$, and $\gamma_{\pm}$,
in this subsection, we consider the cases where the output signal
(\ref{eq:alphaexphats++betaexphats-}) yields the linear combination of
the two quadratures among $\hat{b}_{1}$, $\hat{b}_{1}^{\dagger}$,
$\hat{b}_{2}$, and $\hat{b}_{2}^{\dagger}$.
To do this, we treat following six cases.

%*******************************************************************

%%%%%%%%%%%%%%%%%%%%%%%%%%%%%%%%%%%%%%%%%%%%%%%%%%%%%%%
\subsubsection{Linear combination of $\hat{b}_{1}$ and $\hat{b}_{2}$}
\label{sec:hatb1-hatb2-case}
%%%%%%%%%%%%%%%%%%%%%%%%%%%%%%%%%%%%%%%%%%%%%%%%%%%%%%%

%*******************************************************************

Here, we check whether or not the linear combination
(\ref{eq:alphaexphats++betaexphats-}) is expressed by a linear combination
of $\hat{b}_{1}$ and $\hat{b}_{2}$, which is already given in
Ref.~\cite{K.Nakamura-M.-K.Fujimoto-2017b}.
This should be accomplished by the equation of the matrix form:
\begin{eqnarray}
  \label{eq:hatb1-hatb2-case-condition}
  \left(
    \begin{array}{cc}
        \gamma_{+} & \gamma_{-}^{*} \\
      - \gamma_{+} & \gamma_{-}^{*}
    \end{array}
  \right)
  \left(
    \begin{array}{c}
      \alpha \\
      \beta
    \end{array}
  \right)
  =
  \left(
    \begin{array}{c}
      0 \\
      0
    \end{array}
  \right)
  .
\end{eqnarray}
If a nontrivial solution $(\alpha,\beta)\neq(0,0)$ to
Eq.~(\ref{eq:hatb1-hatb2-case-condition}) exists, the determinant of
the matrix in Eq.~(\ref{eq:hatb1-hatb2-case-condition}) should vanish,
i.e., $2 \gamma_{+} \gamma_{-}^{*}=0$.
This implies that $\gamma_{+}=0$ or $\gamma_{-}=0$.
These cases are meaningless as balanced homodyne detections.
This fact implies that the linear combination of
$\langle\hat{s}_{\pm}\rangle$ does not give expectation values of
the linear combination of $\hat{b}_{1}$ and $\hat{b}_{2}$, by itself.
This result directly means that the output signal
(\ref{eq:alphaexphats++betaexphats-}) never yields the expectation
value of the operator (\ref{eq:hatbthehta-def}).

%*******************************************************************

%%%%%%%%%%%%%%%%%%%%%%%%%%%%%%%%%%%%%%%%%%%%%%%%%%%%%%%
\subsubsection{Linear combination of $\hat{b}_{1}$ and $\hat{b}_{1}^{\dagger}$}
\label{sec:hatb1-hatb1dagger-case}
%%%%%%%%%%%%%%%%%%%%%%%%%%%%%%%%%%%%%%%%%%%%%%%%%%%%%%%

%*******************************************************************

Here, we check whether or not
Eq.~(\ref{eq:alphaexphats++betaexphats-}) yields a linear combination
of $\hat{b}_{1}$ and $\hat{b}_{1}^{\dagger}$.
This situation should be accomplished by the equations of the matrix
form:
\begin{eqnarray}
  \label{eq:hatb1-hatb1dagger-case-condition}
  \left(
    \begin{array}{cc}
        \gamma_{+}^{*} & - \gamma_{-}     \\
      - \gamma_{+}     &   \gamma_{-}^{*}
    \end{array}
  \right)
  \left(
    \begin{array}{c}
      \alpha \\
      \beta
    \end{array}
  \right)
  =
  \left(
    \begin{array}{c}
      0 \\
      0
    \end{array}
  \right)
  .
\end{eqnarray}
If a nontrivial solution $(\alpha,\beta)\neq(0,0)$ to
Eq.~(\ref{eq:hatb1-hatb1dagger-case-condition}) exists, the
determinant of the matrix in
Eq.~(\ref{eq:hatb1-hatb1dagger-case-condition}) should vanish :
$(\gamma_{+}\gamma_{-})^{*}=\gamma_{+}\gamma_{-}$, i.e.,
$\gamma_{+}\gamma_{-}$ is real and we may choose so that
\begin{eqnarray}
  \label{eq:hatb1-hatb1dagger-case-vanishes-det-cond}
  \gamma_{+} = |\gamma_{+}|e^{i\theta}, \quad
  \gamma_{-} = |\gamma_{-}|e^{-i\theta}.
\end{eqnarray}
Substituting these expressions into
Eqs.~(\ref{eq:hatb1-hatb1dagger-case-condition}), we obtain
\begin{eqnarray}
  \label{eq:hatb1-hatb1dagger-case-alpha-beta-cond}
  \beta
  =
  \frac{|\gamma_{+}|}{|\gamma_{-}|}
  \alpha
  .
\end{eqnarray}
Furthermore, substituting
Eqs.~(\ref{eq:hatb1-hatb1dagger-case-vanishes-det-cond}) and
(\ref{eq:hatb1-hatb1dagger-case-alpha-beta-cond}) into
Eq.~(\ref{eq:alphaexphats++betaexphats-}) , we obtain
\begin{eqnarray}
  \frac{\left\langle\hat{s}_{+}\right\rangle}{|\gamma_{+}|}
  +
  \frac{\left\langle\hat{s}_{-}\right\rangle}{|\gamma_{-}|}
  =
  \sqrt{2}
  \left\langle
  \hat{b}_{1} e^{-i\theta}
  +
  \hat{b}_{1}^{\dagger} e^{+i\theta}
  \right\rangle
  .
  \label{eq:hatb1-hatb2-case-alphaexphats++betaexphats--result}
\end{eqnarray}

%*******************************************************************

%%%%%%%%%%%%%%%%%%%%%%%%%%%%%%%%%%%%%%%%%%%%%%%%%%%%%%%
\subsubsection{Linear combination of $\hat{b}_{1}$ and $\hat{b}_{2}^{\dagger}$}
\label{sec:hatb1-hatb2dagger-case}
%%%%%%%%%%%%%%%%%%%%%%%%%%%%%%%%%%%%%%%%%%%%%%%%%%%%%%%

%*******************************************************************

Here, we check whether or not
Eq.~(\ref{eq:alphaexphats++betaexphats-}) is expressed the linear
combination of $\hat{b}_{1}$ and $\hat{b}_{2}^{\dagger}$.
In this case, the following conditions should be satisfied:
\begin{eqnarray}
  \label{eq:hatb1-hatb2dagger-case-condition}
  \left(
    \begin{array}{cc}
      \gamma_{+}^{*} & - \gamma_{-} \\
      \gamma_{+}     &   \gamma_{-}^{*}
    \end{array}
  \right)
  \left(
    \begin{array}{c}
      \alpha \\
      \beta
    \end{array}
  \right)
  =
  \left(
    \begin{array}{c}
      0 \\
      0
    \end{array}
  \right)
  .
\end{eqnarray}
The condition for the existence of the nontrivial solution
$(\alpha,\beta)\neq(0,0)$ to
Eq.~(\ref{eq:hatb1-hatb2dagger-case-condition}) is
$(\gamma_{+}\gamma_{-})^{*}=-\gamma_{+}\gamma_{-}$, i.e.,
$\gamma_{+}\gamma_{-}$ is purely imaginary.
Therefore, we may choose $\gamma_{\pm}$ as
\begin{eqnarray}
  \label{eq:hatb1-hatb2dagger-case-vanishes-det-cond}
  \gamma_{+} = |\gamma_{+}| e^{+i\theta}, \quad
  \gamma_{-} = i |\gamma_{-}| e^{-i\theta}.
\end{eqnarray}
Substituting Eq.~(\ref{eq:hatb1-hatb2dagger-case-vanishes-det-cond})
into Eq.~(\ref{eq:hatb1-hatb2dagger-case-condition}), we obtain the
relation
\begin{eqnarray}
  \label{eq:hatb1-hatb2dagger-case-alpha-beta-cond}
  \beta
  =
  -
  i
  \frac{|\gamma_{+}|}{|\gamma_{-}|}
  \alpha
\end{eqnarray}
Furthermore, substituting
Eqs.~(\ref{eq:hatb1-hatb2dagger-case-vanishes-det-cond}) and
(\ref{eq:hatb1-hatb2dagger-case-alpha-beta-cond}) into
Eq.~(\ref{eq:alphaexphats++betaexphats-}), we obtain
\begin{eqnarray}
  \frac{\left\langle\hat{s}_{+}\right\rangle}{|\gamma_{+}|}
  -
  i
  \frac{\left\langle\hat{s}_{-}\right\rangle}{|\gamma_{-}|}
  =
  \sqrt{2}
  \left\langle
  e^{-i\theta}
  \hat{b}_{1}
  -
  i
  e^{+i\theta}
  \hat{b}_{2}^{\dagger}
  \right\rangle
  .
  \label{eq:hatb1-hatb2dagger-case-alphaexphats++betaexphats--result}
\end{eqnarray}

%*******************************************************************

%%%%%%%%%%%%%%%%%%%%%%%%%%%%%%%%%%%%%%%%%%%%%%%%%%%%%%%
\subsubsection{Linear combination of $\hat{b}_{1}^{\dagger}$ and $\hat{b}_{2}$}
\label{sec:hatb1dagger-hatb2-case}
%%%%%%%%%%%%%%%%%%%%%%%%%%%%%%%%%%%%%%%%%%%%%%%%%%%%%%%

%*******************************************************************

Here, we check whether or not
Eq.~(\ref{eq:alphaexphats++betaexphats-}) yields the linear
combination of $\hat{b}_{1}^{\dagger}$ and $\hat{b}_{2}$.
This should be accomplished by the equation of the matrix form:
\begin{eqnarray}
  \label{eq:hatb1dagger-hatb2-case-condition}
  \left(
    \begin{array}{cc}
        \gamma_{+}^{*} & \gamma_{-}      \\
      - \gamma_{+}       & \gamma_{-}^{*}
    \end{array}
  \right)
  \left(
    \begin{array}{c}
      \alpha \\
      \beta
    \end{array}
  \right)
  =
  \left(
    \begin{array}{c}
      0 \\
      0
    \end{array}
  \right)
  .
\end{eqnarray}
The existence of the nontrivial solution $(\alpha,\beta)\neq(0,0)$ to
Eq.~(\ref{eq:hatb1dagger-hatb2-case-condition}) is guaranteed by the
vanishing condition of the determinant of the matrix in
Eq.~(\ref{eq:hatb1dagger-hatb2-case-condition}).
This condition is given by
$(\gamma_{+}\gamma_{-})^{*}=-\gamma_{+}\gamma_{-}$, i.e.,
$\gamma_{+}\gamma_{-}$ is purely imaginary and we may choose
$\gamma_{\pm}$ as
\begin{eqnarray}
  \label{eq:hatb1dagger-hatb2-case-vanishes-det-cond}
  \gamma_{+} = |\gamma_{+}| e^{+i\theta}, \quad
  \gamma_{-} = i |\gamma_{-}| e^{-i\theta}.
\end{eqnarray}
Substituting Eq.~(\ref{eq:hatb1dagger-hatb2-case-vanishes-det-cond})
into Eq.~(\ref{eq:hatb1dagger-hatb2-case-condition}), we obtain
\begin{eqnarray}
  \label{eq:hatb1dagger-hatb2-case-alpha-beta-cond}
  \beta
  =
  i
  \frac{|\gamma_{+}|}{|\gamma_{-}|}
  \alpha
  .
\end{eqnarray}
Further, through
Eqs.~(\ref{eq:hatb1dagger-hatb2-case-vanishes-det-cond}) and
(\ref{eq:hatb1dagger-hatb2-case-alpha-beta-cond}), the linear
combination (\ref{eq:alphaexphats++betaexphats-}) is given by
\begin{eqnarray}
  \frac{\left\langle\hat{s}_{+}\right\rangle}{|\gamma_{+}|}
  +
  i
  \frac{\left\langle\hat{s}_{-}\right\rangle}{|\gamma_{-}|}
  =
  \sqrt{2}
  \left\langle
  i
  e^{-i\theta}
  \hat{b}_{2}
  +
  e^{+i\theta}
  \hat{b}_{1}^{\dagger}
  \right\rangle
  .
  \label{eq:hatb1dagger-hatb2-case-alphaexphats++betaexphats--result}
\end{eqnarray}

%*******************************************************************

%%%%%%%%%%%%%%%%%%%%%%%%%%%%%%%%%%%%%%%%%%%%%%%%%%%%%%%
\subsubsection{Linear combination of $\hat{b}_{1}^{\dagger}$ and $\hat{b}_{2}^{\dagger}$}
\label{sec:hatb1dagger-hatb2dagger-case}
%%%%%%%%%%%%%%%%%%%%%%%%%%%%%%%%%%%%%%%%%%%%%%%%%%%%%%%

%*******************************************************************

Here, we examine whether or not the linear combination
(\ref{eq:alphaexphats++betaexphats-}) is given by a linear
combination of $\hat{b}_{1}^{\dagger}$ and $\hat{b}_{2}^{\dagger}$.
This case is accomplished by the equation of the matrix form
\begin{eqnarray}
  \label{eq:hatb1dagger-hatb2dagger-case-condition}
  \left(
    \begin{array}{cc}
      \gamma_{+}^{*} &   \gamma_{-} \\
      \gamma_{+}^{*} & - \gamma_{-}
    \end{array}
  \right)
  \left(
    \begin{array}{c}
      \alpha \\
      \beta
    \end{array}
  \right)
  =
  \left(
    \begin{array}{c}
      0 \\
      0
    \end{array}
  \right)
  .
\end{eqnarray}
The existence of the nontrivial solution $(\alpha,\beta)\neq(0,0)$ to
Eq.~(\ref{eq:hatb1dagger-hatb2dagger-case-condition}) is given by
$\gamma_{+}^{*}\gamma_{-}$, which implies that $\gamma_{+}=0$ or
$\gamma_{-}=0$.
These are meaningless as balanced homodyne detections as in the case
in Sec.~\ref{sec:hatb1-hatb2-case}.
This fact implies that the linear combination
(\ref{eq:alphaexphats++betaexphats-}) does not give expectation values
of the linear combination of $\hat{b}_{1}^{\dagger}$ and
$\hat{b}_{2}^{\dagger}$.
This is consistent with the result in Sec.~\ref{sec:hatb1-hatb2-case}.

%*******************************************************************

%%%%%%%%%%%%%%%%%%%%%%%%%%%%%%%%%%%%%%%%%%%%%%%%%%%%%%%
\subsubsection{Linear combination of $\hat{b}_{2}$ and $\hat{b}_{2}^{\dagger}$}
\label{sec:hatb2-hatb2dagger-case}
%%%%%%%%%%%%%%%%%%%%%%%%%%%%%%%%%%%%%%%%%%%%%%%%%%%%%%%

%*******************************************************************

Finally, we check whether the linear combination
(\ref{eq:alphaexphats++betaexphats-}) is expressed by a linear
combination of $\hat{b}_{2}$ and $\hat{b}_{2}^{\dagger}$, or not.
This case is accomplished by the equation of the matrix form
\begin{eqnarray}
  \label{eq:hatb2-hatb2dagger-case-condition}
  \left(
    \begin{array}{cc}
      \gamma_{+}^{*} & \gamma_{-}     \\
      \gamma_{+}     & \gamma_{-}^{*}
    \end{array}
  \right)
  \left(
    \begin{array}{c}
      \alpha \\
      \beta
    \end{array}
  \right)
  =
  \left(
    \begin{array}{c}
      0 \\
      0
    \end{array}
  \right)
  .
\end{eqnarray}
The existence of nontrivial solutions $(\alpha,\beta)\neq(0,0)$ to
Eq.~(\ref{eq:hatb2-hatb2dagger-case-condition}) is guaranteed by the
vanishing determinant of the matrix in
Eq.~(\ref{eq:hatb2-hatb2dagger-case-condition}).
This condition is given by
$(\gamma_{+}\gamma_{-})^{*}=\gamma_{-}\gamma_{+}$, which means that
$\gamma_{+}\gamma_{-}$ is real.
Therefore, we may choose $\gamma_{\pm}$ so that
\begin{eqnarray}
  \label{eq:hatb2-hatb2dagger-case-vanishes-det-cond}
  \gamma_{+} = |\gamma_{+}| e^{+i\theta}, \quad
  \gamma_{-} = |\gamma_{-}| e^{-i\theta}.
\end{eqnarray}
Through Eqs.~(\ref{eq:hatb2-hatb2dagger-case-vanishes-det-cond}) and
(\ref{eq:hatb2-hatb2dagger-case-condition}), we obtain
\begin{eqnarray}
  \label{eq:hatb2-hatb2dagger-case-alpha-beta-cond}
  \beta
  =
  -
  \alpha
  \frac{|\gamma_{+}|}{|\gamma_{-}|}
  .
\end{eqnarray}
Substituting Eqs.~(\ref{eq:hatb2-hatb2dagger-case-vanishes-det-cond})
and (\ref{eq:hatb2-hatb2dagger-case-alpha-beta-cond}) into
Eq.~(\ref{eq:alphaexphats++betaexphats-}), we obtain
\begin{eqnarray}
  \label{eq:hatb2-hatb2dagger-case-alphaexphats++betaexphats--result}
  \frac{\left\langle\hat{s}_{+}\right\rangle}{|\gamma_{+}|}
  -
  \frac{\left\langle\hat{s}_{-}\right\rangle}{|\gamma_{-}|}
  =
  \sqrt{2} i
  \left\langle
  e^{-i\theta}
  \hat{b}_{2}
  -
  e^{+i\theta}
  \hat{b}_{2}^{\dagger}
  \right\rangle
  .
\end{eqnarray}

%*******************************************************************

%%%%%%%%%%%%%%%%%%%%%%%%%%%%%%%%%%%%%%%%%%%%%%%%%%%%%%%
%%%%%%%%%%%%%%%%%%%%%%%%%%%%%%%%%%%%%%%%%%%%%%%%%%%%%%%
\subsection{Remarks}
\label{sec:examination-summary}
%%%%%%%%%%%%%%%%%%%%%%%%%%%%%%%%%%%%%%%%%%%%%%%%%%%%%%%
%%%%%%%%%%%%%%%%%%%%%%%%%%%%%%%%%%%%%%%%%%%%%%%%%%%%%%%

%*******************************************************************

\begin{table}[t]
  \centering
  \begin{tabular}{cccccccc}
    \hline
    \hline
    Combination & Possible? & $\quad$ & $\gamma_{+}$ & $\quad$
    & $\gamma_{-}$  & $\quad$ &$\alpha$-$\beta$ relation \\
    \hline
    & & & & & & & \\
    $\displaystyle \hat{b}_{1}$ and $\displaystyle \hat{b}_{2}$
                                  & $\displaystyle \times$ & $\quad$ & --- & $\quad$ & --- & $\quad$ & --- \\
    & & & & & & & \\
    $\displaystyle \hat{b}_{1}$ and $\displaystyle \hat{b}_{1}^{\dagger}$
                & $\displaystyle \bigcirc$ & $\quad$
                                            & $\displaystyle |\gamma_{+}|e^{+i\theta}$
                                                           & $\quad$
    & $\displaystyle |\gamma_{-}|e^{-i\theta}$
                    & $\quad$
                              & $\displaystyle \beta=\frac{|\gamma_{+}|}{|\gamma_{-}|}\alpha$ \\
    & & & & & & & \\
    $\displaystyle \hat{b}_{1}$ and $\displaystyle \hat{b}_{2}^{\dagger}$
                & $\displaystyle \bigcirc$
                                  & $\quad$
                                            & $\displaystyle |\gamma_{+}|e^{+i\theta}$
                                                           & $\quad$
    & $\displaystyle i|\gamma_{-}|e^{-i\theta}$
                    & $\quad$
                              & $\displaystyle \beta= - i \frac{|\gamma_{+}|}{|\gamma_{-}|}\alpha$ \\
    & & & & & & & \\
    $\displaystyle \hat{b}_{1}^{\dagger}$ and $\displaystyle \hat{b}_{2}$
                & $\displaystyle \bigcirc$ & $\quad$
                                            & $\displaystyle |\gamma_{+}|e^{+i\theta}$
                                                           & $\quad$
    & $\displaystyle i|\gamma_{-}|e^{-i\theta}$
                    & $\quad$
                              & $\displaystyle \beta= + i \frac{|\gamma_{+}|}{|\gamma_{-}|}\alpha$ \\
    & & & & & & & \\
    $\displaystyle \hat{b}_{1}^{\dagger}$ and $\displaystyle \hat{b}_{2}^{\dagger}$
                & $\displaystyle \times$ & $\quad$  & --- & $\quad$  &
                                                                       --- & $\quad$  & --- \\
    & & & & & & & \\
    $\displaystyle \hat{b}_{2}$ and $\displaystyle \hat{b}_{2}^{\dagger}$
                & $\displaystyle \bigcirc$
                                  & $\quad$ & $\displaystyle |\gamma_{+}|e^{+i\theta}$
                                                           & $\quad$
    & $\displaystyle |\gamma_{-}|e^{-i\theta}$
                    & $\quad$
                              & $\displaystyle \beta= - \frac{|\gamma_{+}|}{|\gamma_{-}|}\alpha$ \\
    & & & & & & & \\
    \hline
    \hline
  \end{tabular}
  \caption{
    Summary of the problem whether or not the linear combination
    (\ref{eq:alphaexphats++betaexphats-}) can be reduced to the linear
    combination of two quadratures.
    In this table, the value of $\gamma_{+}$ is always arbitrary
    expect for the non-vanishing condition $\gamma_{+}\neq 0$ but the
    corresponding choice of $\gamma_{-}$ is nontrivial.
    This table indicates that the relative phase of $\gamma_{\pm}$ is
    important for the reduction of
    Eq.~(\ref{eq:alphaexphats++betaexphats-}) to the linear
    combination of two quadratures.
  }
  \label{tab:balanced-homodyne-combination-summary}
\end{table}

%*******************************************************************

We have shown that we cannot measure the expectation value of any linear
combination of $\hat{b}_{1}$ and $\hat{b}_{2}$ in
Sec.~\ref{sec:hatb1-hatb2-case} through the simple application of the
balanced homodyne detection.
Similarly, we cannot directly measure the linear combination of
$\hat{b}_{1}^{\dagger}$ and $\hat{b}_{2}^{\dagger}$ as we have seen in
Sec.~\ref{sec:hatb1dagger-hatb2dagger-case}.
These two cases are consistent with each other.
These also indicates that we cannot measure the expectation value of the
operator $\hat{b}_{\theta}$ through the simple application of the
balanced homodyne detection reviewed in
Sec.~\ref{sec:balanced_homodyne_detection}.

%*******************************************************************

On the other hand, as seen in Sec.~\ref{sec:hatb1-hatb1dagger-case},
we can measure the expectation value of the linear combination of
$\hat{b}_{1}$ and $\hat{b}_{1}^{\dagger}$ through the linear
combination of the expectation values
$\langle\hat{s}_{\pm}\rangle$ with real coefficients $\alpha$ and
$\beta$ as seen in
Eqs.~(\ref{eq:hatb1-hatb1dagger-case-alpha-beta-cond}) and
(\ref{eq:hatb1-hatb2-case-alphaexphats++betaexphats--result}).
To accomplish this, the phase of the complex amplitude $\gamma_{\pm}$
for the coherent state from the local oscillator should be opposite as
seen in Eq.~(\ref{eq:hatb1-hatb1dagger-case-vanishes-det-cond}).
This is an important requirement for the coherent state from the local
oscillator.

%*******************************************************************

Similarly, as seen in Sec.~\ref{sec:hatb2-hatb2dagger-case}, we can
also measure the expectation value of the linear combination of
$\hat{b}_{2}$ and $\hat{b}_{2}^{\dagger}$ through the linear
combination of the expectation values $\langle\hat{s}_{\pm}\rangle$
with real coefficients $\alpha$ and $\beta$ as seen in
Eqs.~(\ref{eq:hatb2-hatb2dagger-case-alpha-beta-cond}) and
(\ref{eq:hatb1-hatb2-case-alphaexphats++betaexphats--result}).
To accomplish this, the phase of the complex amplitude $\gamma_{\pm}$
for the coherent state from the local oscillator should be opposite as
seen in Eq.~(\ref{eq:hatb2-hatb2dagger-case-vanishes-det-cond}).
As in the case of Sec.~\ref{sec:hatb1-hatb1dagger-case}, the phase
condition for the local oscillator also is crucial.

%*******************************************************************

The results of the above arguments in
Sec.~\ref{sec:can_we_measure_coshatb1+sinhatb2} is summarized in
Table~\ref{tab:balanced-homodyne-combination-summary}.
In addition to the choice of the relation of the coefficients $\alpha$
and $\beta$, Table~\ref{tab:balanced-homodyne-combination-summary}
indicates that the relative phase of $\gamma_{\pm}$ of the local
oscillator is important for the reduction of
Eq.~(\ref{eq:alphaexphats++betaexphats-}) to the linear combination of
two quadratures from $\hat{b}_{1}$, $\hat{b}_{2}$,
$\hat{b}_{1}^{\dagger}$, and $\hat{b}_{2}^{\dagger}$.

%*******************************************************************

Two cases in Sec.~\ref{sec:hatb1-hatb1dagger-case} and
Sec.~\ref{sec:hatb2-hatb2dagger-case} are the cases of the real
coefficients $\alpha$ and $\beta$.
Since the operators $\hat{s}_{\pm}$ are self-adjoint operators, the
resulting operators are also self-adjoint.
Interestingly, the complex coefficients $\alpha$ and $\beta$ of the
linear combination of $\langle\hat{s}_{\pm}\rangle$ is also possible
as seen in Sec.~\ref{sec:hatb1-hatb2dagger-case} and in
Sec.~\ref{sec:hatb1dagger-hatb2-case}.
In these cases, the results are no longer self-adjoint operators in
spite that the operators $\hat{s}_{\pm}$ are self-adjoint.
Therefore, we can calculate an expectation value of non-self-adjoint
operators from the expectation values of the self-adjoint operators
which can be directly calculable from the expectation values of photon
number operators in the balanced homodyne detection reviewed in
Sec.~\ref{sec:balanced_homodyne_detection}.

%*******************************************************************

Finally, we also note that in any possible cases in
Sec.~\ref{sec:can_we_measure_coshatb1+sinhatb2}, the phase control of
the upper- and lower-sidebands is necessary.
Namely, from the local oscillator, we have to introduce the coherent
state with the complex amplitudes
$\gamma_{\pm}=\gamma(\omega_{0}\pm\Omega)$ for upper- and
lower-sideband have the opposite phase as in
Eqs.~(\ref{eq:hatb1-hatb1dagger-case-vanishes-det-cond}),
(\ref{eq:hatb1-hatb2dagger-case-vanishes-det-cond}),
(\ref{eq:hatb1dagger-hatb2-case-vanishes-det-cond}), and
(\ref{eq:hatb2-hatb2dagger-case-vanishes-det-cond}).
Before these important requirements for the phase of $\gamma_{\pm}$,
$\gamma_{\pm}$ must have their non-vanishing support at the frequency
$\omega_{0}\pm\Omega$.
These requirements for the coherent state from the local oscillator
are also important results from our examination in this section.

%*******************************************************************

%%%%%%%%%%%%%%%%%%%%%%%%%%%%%%%%%%%%%%%%%%%%%%%%%%%%%%%
%%%%%%%%%%%%%%%%%%%%%%%%%%%%%%%%%%%%%%%%%%%%%%%%%%%%%%%
%%%%%%%%%%%%%%%%%%%%%%%%%%%%%%%%%%%%%%%%%%%%%%%%%%%%%%%
\section{Double balanced homodyne detection}
\label{sec:Double_balanced_homodyne_detection}
%%%%%%%%%%%%%%%%%%%%%%%%%%%%%%%%%%%%%%%%%%%%%%%%%%%%%%%
%%%%%%%%%%%%%%%%%%%%%%%%%%%%%%%%%%%%%%%%%%%%%%%%%%%%%%%
%%%%%%%%%%%%%%%%%%%%%%%%%%%%%%%%%%%%%%%%%%%%%%%%%%%%%%%

%*******************************************************************

Through the understanding of the balanced homodyne detections based on
the examinations in Sec.~\ref{sec:BHD-two-photon}, we propose the
``double balanced homodyne detection'' which enable us
to measure the expectation values for the photon annihilation and creation
operators themselves.
In Sec.~\ref{sec:theoretical_proposal}, we describe our theoretical
proposal of a double balanced homodyne detection.
In Sec.~\ref{sec:realization-of-DBHD}, we propose a realization of our
double balanced homodyne detection through the interferometer setup.
In Sec.~\ref{sec:input-vaccum-to-the-main-inteterfemeter}, we
discuss the commutation relations between each electric field in the
interferometer with the input field to the main interferometer.
In Sec.~\ref{sec:measurement_of-hatb-or-hatbdagger}, we show the
double balanced homodyne detection through our interferometer setup
enable us to measure the expectation values of the annihilation and
creation operators for the output quadrature from the main
interferometer.
Finally, in Sec.~\ref{sec:DBHD_in_two_photon_formulation}, we show
that the expectation value of the operator (\ref{eq:hatbthehta-def})
of the output quadrature from the main interferometer can be measured
through our interferometer setup.
In this section, we assume $|\gamma_{\pm}|=:|\gamma|$, for simplicity.
The generalization to the case where $|\gamma_{+}|\neq|\gamma_{-}|$ is
always possible and straightforward as seen in
Appendix~\ref{sec:Deriv_of_hatt+_hatt-_operators}.

%*******************************************************************

%%%%%%%%%%%%%%%%%%%%%%%%%%%%%%%%%%%%%%%%%%%%%%%%%%%%%%%
%%%%%%%%%%%%%%%%%%%%%%%%%%%%%%%%%%%%%%%%%%%%%%%%%%%%%%%
\subsection{Theoretical proposal}
\label{sec:theoretical_proposal}
%%%%%%%%%%%%%%%%%%%%%%%%%%%%%%%%%%%%%%%%%%%%%%%%%%%%%%%
%%%%%%%%%%%%%%%%%%%%%%%%%%%%%%%%%%%%%%%%%%%%%%%%%%%%%%%

%*******************************************************************

In this subsection, we describe our theoretical proposal of the double
balanced homodyne detection.
In Sec.~\ref{sec:exp_val_b1_b1dagger_b2_b2dagger}, we describe how to
realize the expectation values of $\hat{b}_{1}$,
$\hat{b}_{1}^{\dagger}$, $\hat{b}_{2}$, or $\hat{b}_{2}^{\dagger}$ in
the calculations on papers.
The idea in Sec.~\ref{sec:exp_val_b1_b1dagger_b2_b2dagger} is extended
in Sec.~\ref{sec:exp_val_costhetab1+sinthetab2} to the measurement of
the linear combination (\ref{eq:hatbthehta-def}).

%*******************************************************************

%%%%%%%%%%%%%%%%%%%%%%%%%%%%%%%%%%%%%%%%%%%%%%%%%%%%%%%
\subsubsection{Expectation values of $\hat{b}_{1}$,
  $\hat{b}_{1}^{\dagger}$, $\hat{b}_{2}$, and $\hat{b}_{2}^{\dagger}$}
\label{sec:exp_val_b1_b1dagger_b2_b2dagger}
%%%%%%%%%%%%%%%%%%%%%%%%%%%%%%%%%%%%%%%%%%%%%%%%%%%%%%%

%*******************************************************************

Here again, we consider the case in
Sec.~\ref{sec:hatb1-hatb1dagger-case}.
In this case, the linear combination
(\ref{eq:alphaexphats++betaexphats-}) is given by
\begin{eqnarray}
  \frac{1}{\sqrt{2}|\gamma|}
  \left(
  \left\langle\hat{s}_{+}\right\rangle
  +
  \left\langle\hat{s}_{-}\right\rangle
  \right)
  =
  \langle
  \hat{b}_{1}
  \rangle
  e^{-i\theta}
  +
  \langle
  \hat{b}_{1}^{\dagger}
  \rangle
  e^{+i\theta}
  .
  \label{eq:hatb1-hatb1dagger-case-result-gammapm=gamma}
\end{eqnarray}
Since the complex amplitude $\gamma$ including its phase is completely
controllable, we may choose $\theta=0$ or $\theta=\pi/2$, for example.
In the cases where $\theta=0$ and $\theta=\pi/2$,
Eq.~(\ref{eq:hatb1-hatb1dagger-case-result-gammapm=gamma}) are
given by
\begin{eqnarray}
  &&
     \left.
     \frac{1}{\sqrt{2}|\gamma|}
     \left(
     \left\langle\hat{s}_{+}\right\rangle
     +
     \left\langle\hat{s}_{-}\right\rangle
     \right)
     \right|_{\theta=0}
     =
     \langle
     \hat{b}_{1}
     \rangle
     +
     \langle
     \hat{b}_{1}^{\dagger}
     \rangle
     ,
     \label{eq:hatb1-hatb1dagger-case-result-gammapm=gamma-theta0}
  \\
  &&
     \left.
     \frac{1}{\sqrt{2}|\gamma|}
     \left(
     \left\langle\hat{s}_{+}\right\rangle
     +
     \left\langle\hat{s}_{-}\right\rangle
     \right)
     \right|_{\theta=\pi/2}
     =
     -
     i
     \langle
     \hat{b}_{1}
     \rangle
     +
     i
     \langle
     \hat{b}_{1}^{\dagger}
     \rangle
     .
     \label{eq:hatb1-hatb1dagger-case-result-gammapm=gamma-thetapi/2}
\end{eqnarray}
If we can produce the results
Eqs.~(\ref{eq:hatb1-hatb1dagger-case-result-gammapm=gamma-theta0}) and
(\ref{eq:hatb1-hatb1dagger-case-result-gammapm=gamma-thetapi/2}) as
the measurement results at the same time, we can calculate
\begin{eqnarray}
  \frac{1}{2\sqrt{2}|\gamma|}
  \left\{
  \left.
  \left(
  \left\langle\hat{s}_{+}\right\rangle
  +
  \left\langle\hat{s}_{-}\right\rangle
  \right)
  \right|_{\theta=0}
  +
  \left.
  i
  \left(
  \left\langle\hat{s}_{+}\right\rangle
  +
  \left\langle\hat{s}_{-}\right\rangle
  \right)
  \right|_{\theta=\pi/2}
  \right\}
  &=&
      \langle\hat{b}_{1}\rangle
      ,
      \label{eq:hatb1-hatb1dagger-case-hatb1-expect-combination}
  \\
  \frac{1}{2\sqrt{2}|\gamma|}
  \left\{
  \left.
  \left(
  \left\langle\hat{s}_{+}\right\rangle
  +
  \left\langle\hat{s}_{-}\right\rangle
  \right)
  \right|_{\theta=0}
  -
  \left.
  i
  \left(
  \left\langle\hat{s}_{+}\right\rangle
  +
  \left\langle\hat{s}_{-}\right\rangle
  \right)
  \right|_{\theta=\pi/2}
  \right\}
  &=&
      \langle\hat{b}_{1}^{\dagger}\rangle
      .
      \label{eq:hatb1-hatb1dagger-case-hatb1dagger-expect-combination}
\end{eqnarray}
Note that $\langle\hat{s}_{\pm}\rangle$ are measured from the
photon-number expectation values through the balanced homodyne detection.
Since $\gamma$ is controllable,
Eq.~(\ref{eq:hatb1-hatb1dagger-case-hatb1-expect-combination}) and
(\ref{eq:hatb1-hatb1dagger-case-hatb1dagger-expect-combination}) imply
that the expectation values $\langle\hat{b}_{1}\rangle$ and
$\langle\hat{b}_{1}^{\dagger}\rangle$ of the quadrature is indirectly
calculable from these measurement, respectively, though the
operators $\hat{b}_{1}$ is not self-adjoint operator.
This situation is similar to the cases in
Sec.~\ref{sec:hatb1-hatb2dagger-case} and
Sec.~\ref{sec:hatb1dagger-hatb2-case}.

%*******************************************************************

Similar analysis is possible even in the case in
Sec.~\ref{sec:hatb2-hatb2dagger-case}.
The results yield that we can calculate the expectation values of the
quadrature $\langle\hat{b}_{2}\rangle$ and
$\langle\hat{b}_{2}^{\dagger}\rangle$ from the expectation values
$\langle\hat{s}_{\pm}\rangle$ which can be calculated from the
expectation values of photon number operators in the balanced homodyne
detection.
Actually, the result
(\ref{eq:hatb2-hatb2dagger-case-alphaexphats++betaexphats--result}) in
Sec.~\ref{sec:hatb2-hatb2dagger-case} is given by
\begin{eqnarray}
  \label{eq:hatb2-hatb2dagger-case-result-gammapm=gamma}
  \frac{1}{\sqrt{2}i|\gamma|}
  \left(
  \langle\hat{s}_{+}\rangle
  -
  \langle\hat{s}_{-}\rangle
  \right)
  =
  e^{-i\theta}
  \langle
  \hat{b}_{2}
  \rangle
  -
  e^{+i\theta}
  \langle
  \hat{b}_{2}^{\dagger}
  \rangle
  .
\end{eqnarray}
As in the case of
Eq.~(\ref{eq:hatb1-hatb1dagger-case-result-gammapm=gamma}), we obtain
the expectation values $\langle\hat{b}_{2}\rangle$ and
$\langle\hat{b}_{2}^{\dagger}\rangle$ as follows:
\begin{eqnarray}
  \label{eq:hatb2-hatb2dagger-case-hatb2-expect-combination}
  \frac{1}{2\sqrt{2}|\gamma|}
  \left\{
  \left.
  - i
  \left(
  \langle\hat{s}_{+}\rangle
  -
  \langle\hat{s}_{-}\rangle
  \right)
  \right|_{\theta=0}
  +
  \left.
  \left(
  \langle\hat{s}_{+}\rangle
  -
  \langle\hat{s}_{-}\rangle
  \right)
  \right|_{\theta=\pi/2}
  \right\}
  &=&
      \langle \hat{b}_{2} \rangle
      ,
  \\
  \frac{1}{2\sqrt{2}|\gamma|}
  \left\{
  i
  \left.
  \left(
  \langle\hat{s}_{+}\rangle
  -
  \langle\hat{s}_{-}\rangle
  \right)
  \right|_{\theta=0}
  +
  \left.
  \left(
  \langle\hat{s}_{+}\rangle
  -
  \langle\hat{s}_{-}\rangle
  \right)
  \right|_{\theta=\pi/2}
  \right\}
  &=&
      \langle \hat{b}_{2}^{\dagger} \rangle
      .
      \label{eq:hatb2-hatb2dagger-case-hatb2dagger-expect-combination}
\end{eqnarray}
We note that
Eq.~(\ref{eq:hatb2-hatb2dagger-case-hatb2dagger-expect-combination})
is important in the gravitational-wave detection, because the phase
quadrature $\hat{b}_{2}$ includes gravitational wave signal in
many conventional interferometers.

%*******************************************************************

%%%%%%%%%%%%%%%%%%%%%%%%%%%%%%%%%%%%%%%%%%%%%%%%%%%%%%%
\subsubsection{Expectation value of $\hat{b}_{\theta}=\cos\theta\hat{b}_{1}+\sin\theta\hat{b}_{2}$}
\label{sec:exp_val_costhetab1+sinthetab2}
%%%%%%%%%%%%%%%%%%%%%%%%%%%%%%%%%%%%%%%%%%%%%%%%%%%%%%%

%*******************************************************************

Inspecting analyses in the previous section and
Appendix~\ref{sec:Deriv_of_hatt+_hatt-_operators}, we introduce the
operators $\hat{t}_{\pm}$ which are defined by
\begin{eqnarray}
  \label{eq:hatt+-def}
  \hat{t}_{+}
  :=
  \frac{1}{\sqrt{2}|\gamma|} \left(
  \hat{s}_{+}
  +
  \hat{s}_{-}
  \right)
  ,
  \quad
  \hat{t}_{-}
  :=
  \frac{1}{\sqrt{2}i|\gamma|}
  \left(
  \hat{s}_{+}
  -
  \hat{s}_{-}
  \right)
\end{eqnarray}
to consider the expectation value of the operator $\hat{b}_{\theta}$.
We require $\theta_{\pm}=\theta$ as the requirement for the
phase of the complex amplitudes $\gamma_{\pm}$ as discussed in
Appendix~\ref{sec:Deriv_of_hatt+_hatt-_operators}.
In addition, we assumed $|\gamma_{\pm}|=|\gamma|$ for simplicity.
We note the operator $\hat{t}_{+}$ is self-adjoint and the operator
$\hat{t}_{-}$ is anti-self-adjoint.

%*******************************************************************

From the expressions (\ref{eq:hatspm-def})
of the operators $\hat{s}_{\pm}$, we obtain
\begin{eqnarray}
  \label{eq:hatspmovergammapm}
  \frac{\hat{s}_{\pm}}{|\gamma|}
  &=&
  \frac{\hat{l}_{i\pm}^{\dagger}}{|\gamma|} \hat{b}_{\pm}
  +
  \frac{\hat{l}_{i\pm}}{|\gamma|} \hat{b}_{\pm}^{\dagger}
  +
  \frac{1 - 2 \eta}{\eta^{1/2}(1-\eta)^{1/2}}
  |\gamma|
  \left(
    \frac{\hat{l}_{i\pm}^{\dagger}}{|\gamma|}
    \frac{\hat{l}_{i\pm}}{|\gamma|}
    -
    1
  \right)
\end{eqnarray}
and the above operators $\hat{t}_{\pm}$ are given by
\begin{eqnarray}
  \hat{t}_{+}
  &=&
      \frac{1}{\sqrt{2}} \left\{
      \frac{\hat{l}_{i+}^{\dagger}}{|\gamma|} \hat{b}_{+}
      +
      \frac{\hat{l}_{i+}}{|\gamma|} \hat{b}_{+}^{\dagger}
      +
      \frac{1 - 2 \eta}{\eta^{1/2}(1-\eta)^{1/2}}
      |\gamma|
      \left(
      \frac{\hat{l}_{i+}^{\dagger}}{|\gamma|}
      \frac{\hat{l}_{i+}}{|\gamma|}
      -
      1
      \right)
      \right.
      \nonumber\\
  && \quad\quad\quad
     \left.
     +
     \frac{\hat{l}_{i-}^{\dagger}}{|\gamma|} \hat{b}_{-}
     +
     \frac{\hat{l}_{i-}}{|\gamma|} \hat{b}_{-}^{\dagger}
     +
     \frac{1 - 2 \eta}{\eta^{1/2}(1-\eta)^{1/2}}
     |\gamma|
     \left(
     \frac{\hat{l}_{i-}^{\dagger}}{|\gamma|}
     \frac{\hat{l}_{i-}}{|\gamma|}
     -
     1
     \right)
     \right\}
     ,
     \label{eq:hatt+explicitform}
\end{eqnarray}
and
\begin{eqnarray}
  \hat{t}_{-}
  &=&
      \frac{1}{\sqrt{2}i}
      \left\{
      \frac{\hat{l}_{i+}^{\dagger}}{|\gamma|} \hat{b}_{+}
      +
      \frac{\hat{l}_{i+}}{|\gamma|} \hat{b}_{+}^{\dagger}
      +
      \frac{1 - 2 \eta}{\eta^{1/2}(1-\eta)^{1/2}}
      |\gamma|
      \left(
      \frac{\hat{l}_{i+}^{\dagger}}{|\gamma|}
      \frac{\hat{l}_{i+}}{|\gamma|}
      -
      1
      \right)
      \right.
      \nonumber\\
  && \quad\quad\quad
     \left.
     -
     \frac{\hat{l}_{i-}^{\dagger}}{|\gamma|} \hat{b}_{-}
     -
     \frac{\hat{l}_{i-}}{|\gamma|} \hat{b}_{-}^{\dagger}
     -
     \frac{1 - 2 \eta}{\eta^{1/2}(1-\eta)^{1/2}}
     |\gamma|
     \left(
     \frac{\hat{l}_{i-}^{\dagger}}{|\gamma|}
     \frac{\hat{l}_{i-}}{|\gamma|}
     -
     1
     \right)
     \right\}
     .
     \label{eq:hatt-explicitform}
\end{eqnarray}
The expectation value of the operators
(\ref{eq:hatt+explicitform}) and (\ref{eq:hatt-explicitform}) with the
requirement $\theta_{\pm}=\theta$ are given by
\begin{eqnarray}
  \left.
  \left\langle
  \hat{t}_{+}
  \right\rangle
  \right|_{\theta_{\pm}=\theta}
  &=&
      \frac{1}{\sqrt{2}}
      \left\langle
      e^{-i\theta} \hat{b}_{+}
      +
      e^{+i\theta} \hat{b}_{+}^{\dagger}
      +
      e^{-i\theta} \hat{b}_{-}
      +
      e^{+i\theta} \hat{b}_{-}^{\dagger}
      \right\rangle
      \label{eq:exp_val_hatt+thetapm=theta}
      \\
  &=&
      \left\langle
      \left(
      \cos\theta
      \hat{b}_{1}
      +
      \sin\theta
      \hat{b}_{2}
      \right)
      +
      \left(
      \cos\theta
      \hat{b}_{1}
      +
      \sin\theta
      \hat{b}_{2}
      \right)^{\dagger}
      \right\rangle
      ,
      \label{eq:exp_val_hatt+thetapm=theta-twophoton}
\end{eqnarray}
and
\begin{eqnarray}
  \left.
  \left\langle
  \hat{t}_{-}
  \right\rangle
  \right|_{\theta_{\pm}=\theta}
  &=&
      \frac{1}{\sqrt{2}i}
      \left\langle
      e^{-i\theta} \hat{b}_{+}
      +
      e^{+i\theta} \hat{b}_{+}^{\dagger}
      -
      e^{-i\theta} \hat{b}_{-}
      -
      e^{+i\theta} \hat{b}_{-}^{\dagger}
      \right\rangle
      \label{eq:exp_val_hatt-thetapm=theta}
      \\
  &=&
      \left\langle
      \left(
      -
      \sin\theta
      \hat{b}_{1}
      +
      \cos\theta
      \hat{b}_{2}
      \right)
      -
      \left(
      -
      \sin\theta
      \hat{b}_{1}
      +
      \cos\theta
      \hat{b}_{2}
      \right)^{\dagger}
      \right\rangle
      .
      \label{eq:exp_val_hatt-thetapm=theta-twophoton}
\end{eqnarray}
Here, we used the fact that the states for the quadratures
$\hat{l}_{\pm}$ are in the coherent state $|\gamma\rangle_{l_{\pm}}$,
i.e., $\hat{l}_{\pm}|\gamma\rangle_{l_{i}}$ $=$
$\gamma_{\pm}|\gamma\rangle_{l_{\pm}}$ $=$
$|\gamma|e^{+i\theta}|\gamma\rangle_{l_{\pm}}$ and
Eqs.~(\ref{eq:hatb+hatb--hatb1-hatb2-relations}).
We note that Eq.~(\ref{eq:exp_val_hatt-thetapm=theta-twophoton})
indicates that
\begin{eqnarray}
  \left.
  \left\langle
  \hat{t}_{-}
  \right\rangle
  \right|_{\theta_{\pm}=\theta+\pi/2}
  =
  \left\langle
  \left(
  \cos\theta
  \hat{b}_{1}
  +
  \sin\theta
  \hat{b}_{2}
  \right)
  -
  \left(
  \cos\theta
  \hat{b}_{1}
  +
  \sin\theta
  \hat{b}_{2}
  \right)^{\dagger}
  \right\rangle
  .
  \label{eq:exp_val_hatt-thetapm=theta+pi/2-twophoton}
\end{eqnarray}

%*******************************************************************

From (\ref{eq:exp_val_hatt+thetapm=theta-twophoton}) and
(\ref{eq:exp_val_hatt-thetapm=theta+pi/2-twophoton}), we obtain
\begin{eqnarray}
  \frac{1}{2}
  \left\{
  \left.
  \left\langle
  \hat{t}_{+}
  \right\rangle
  \right|_{\theta_{\pm}=\theta}
  +
  \left.
  \left\langle
  \hat{t}_{-}
  \right\rangle
  \right|_{\theta_{\pm}=\theta+\pi/2}
  \right\}
  =
  \left\langle
  \cos\theta \hat{b}_{1}
  +
  \sin\theta \hat{b}_{2}
  \right\rangle
  .
  \label{eq:conventional_homodyne_detection_result}
\end{eqnarray}
The expectation values of the operator $\hat{s}_{\pm}$ are
calculated from the expectation values of photon-number
 operators through balanced homodyne detections.
The complex amplitudes $\gamma_{\pm}$ are completely controllable
including their phases.
Therefore, the expectation value in the left hand side of
Eq.~(\ref{eq:conventional_homodyne_detection_result})
is also calculable if we can measure the first- and the second terms at
the same time.
This implies that the expectation value
$\langle\hat{b}_{\theta}\rangle$ is also calculable from measurable
quantities under the same situation.

%*******************************************************************

%%%%%%%%%%%%%%%%%%%%%%%%%%%%%%%%%%%%%%%%%%%%%%%%%%%%%%%
%%%%%%%%%%%%%%%%%%%%%%%%%%%%%%%%%%%%%%%%%%%%%%%%%%%%%%%
\subsection{A realization of our double balanced homodyne detection}
\label{sec:realization-of-DBHD}
%%%%%%%%%%%%%%%%%%%%%%%%%%%%%%%%%%%%%%%%%%%%%%%%%%%%%%%
%%%%%%%%%%%%%%%%%%%%%%%%%%%%%%%%%%%%%%%%%%%%%%%%%%%%%%%

%*******************************************************************

\begin{figure}[ht]
  \centering
  \includegraphics[width=0.8\textwidth]{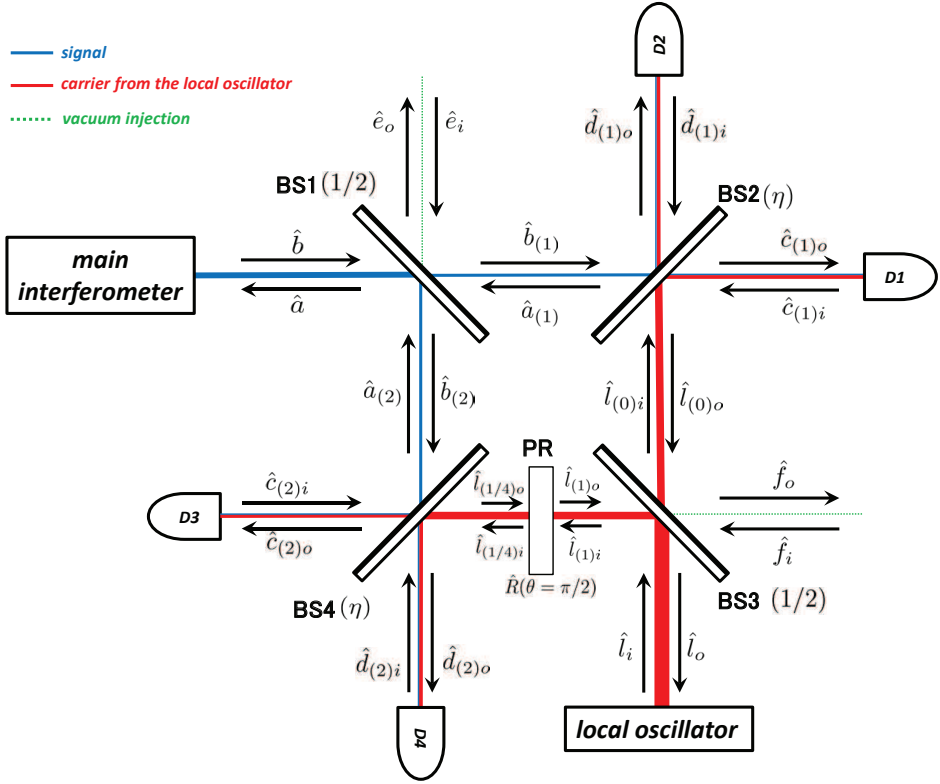}
  \caption{
    A relaization of the meaurement process of the quadrature
    $\langle\hat{b}_{\theta}\rangle$.
    ``BS'' is the beam splitter and ``PR'' is the phase rotator.
    To measure the expectation values
    (\ref{eq:exp_val_hatt+thetapm=theta-twophoton}) and
    (\ref{eq:exp_val_hatt-thetapm=theta+pi/2-twophoton}) at the same
    time, we introduce two balanced homodyne detection.
    To carry out these two balanced homodyne detections, we separate
    the signal photon field associated with the quadrature $\hat{b}$
    from the main interferometer and the photon field associated with
    the quadrature $\hat{l}_{i}$ from the local oscillator through the
    beam splitters BS1 and BS3, respectively.
    One of these two paths is used for the usual balanced homodyne
    detection through the beam splitter BS2 and the photodetectors D1
    and D2.
    Along the other path, we introduce the phase rotator PR to
    introduce $\pi/2$-phase difference in the photon field from the
    local oscillator and detect by the usual balanced homodyne
    detection through the beam splitter B4 and the photodetector D3
    and D2 after this phase addition.
    We describe the paths of the signal photon field and the field
    from the local oscillator.
    The notation of the quadratures for the photon fields are also
    described in this figure.
    We consider the case where the transmittance $\eta$ for BS2 and
    BS4 is not $1/2$ as a simple model of imperfection of the
    interferometer setup.
  }
  \label{fig:DoubleBalancedHomodyneDetection-configuration}
\end{figure}

In this subsection, we consider the realization of the measurement of
the expectation value $\langle\hat{b}_{\theta}\rangle$ of the operator
defined by Eq.~(\ref{eq:hatbthehta-def}) through an interferometer
setup.
Our proposal of the interferometer configuration is depicted in
Fig.~\ref{fig:DoubleBalancedHomodyneDetection-configuration}.
This interferometer setup is so called the ``eight-port homodyne
detection''~\cite{N.G.Walker-etal-1986-J.W.Noh-etal-1991-1993-M.G.Raymer-etal-1993}.
We call our use of the eight-port homodyne detection in this paper as
``double balanced homodyne detection.''
In Sec.~\ref{sec:outline_of_DBHD}, we describe the outline of the
interferometer setup for our double balanced homodyne detection in
Fig.~\ref{fig:DoubleBalancedHomodyneDetection-configuration}.
In Sec.~\ref{sec:separatioin_of_the_signal_field}, we explain about
the separation of the signal field from the main interferometer through
the beam splitter BS1.
In Sec.~\ref{sec:separation_of_the_field _from_the_local_oscillator},
we explain about the separation of the coherent state from the local
oscillator through the beam splitter BS3.
In Sec.~\ref{sec:balanced_homodyne_detection_BS2}, we explain about
the balanced homodyne detection through the beam splitter BS2.
Finally, in Sec.~\ref{sec:balanced_homodyne_detection_BS4}, we explain
about the balanced homodyne detection through the beam splitter BS4.
The results derived in this subsection lead to our main results in
this paper.

%*******************************************************************

%*******************************************************************

%%%%%%%%%%%%%%%%%%%%%%%%%%%%%%%%%%%%%%%%%%%%%%%%%%%%%%%
\subsubsection{Outline of double balanced homodyne detection}
\label{sec:outline_of_DBHD}
%%%%%%%%%%%%%%%%%%%%%%%%%%%%%%%%%%%%%%%%%%%%%%%%%%%%%%%

%*******************************************************************

To measure the expectation values
(\ref{eq:exp_val_hatt+thetapm=theta-twophoton}) and
(\ref{eq:exp_val_hatt-thetapm=theta+pi/2-twophoton}) at the same time,
we introduce two balanced homodyne detections.
To carry out these two balanced homodyne detection, through the beam
splitter BS1, we separate the signal photon field associated with the
quadrature $\hat{b}$ from the main interferometer.
Furthermore, through the beam splitter BS3, we also separate the
photon field associated with the quadrature $\hat{l}_{i}$ from the
local oscillator.
One of the these separated set of the signal field associated with the
quadrature $\hat{b}_{(1)}$ and the field associated with the
quadrature $\hat{l}_{(0)}$ from the local oscillator are mixed at the
beam splitter BS2 and this mixed fields are detected by photodetectors
D1 and D2 to carry out the usual balanced homodyne detection.
In another set of these separated operator, the field from the local
oscillator associated with the quadrature $\hat{l}_{(1)i}$ goes
through the phase rotator PR and gain the additional $\pi/2$-phase.
We denote the quadrature for this phase-added field by
$\hat{l}_{(1/4)i}$.
This field and the separated signal field associated with the
quadrature $\hat{b}_{(2)}$ are mixed the beam splitter BS4 and then
detected by the photodetectors  D3 and D4 to carry out the usual
balanced homodyne detection.

%*******************************************************************

%%%%%%%%%%%%%%%%%%%%%%%%%%%%%%%%%%%%%%%%%%%%%%%%%%%%%%%
\subsubsection{Separation of the signal field}
\label{sec:separatioin_of_the_signal_field}
%%%%%%%%%%%%%%%%%%%%%%%%%%%%%%%%%%%%%%%%%%%%%%%%%%%%%%%

%*******************************************************************

From here, we analyze the photon fields according to the
interferometer setup depicted in
Fig.~\ref{fig:DoubleBalancedHomodyneDetection-configuration}, in detail.

%*******************************************************************

First, at the 50:50 beam splitter 1 (BS1), the output signal
$\hat{b}$ from the main interferometer is separated into two parts,
which we denote $\hat{b}_{(1)}$ and $\hat{b}_{(2)}$, respectively.
In addition to the output quadrature $\hat{b}$, the additional
noise source is inserted, which is denoted $\hat{e}_{i}$ in
Fig.~\ref{fig:DoubleBalancedHomodyneDetection-configuration}.
Then, at BS1, the quadratures $\hat{b}$, $\hat{b}_{(1)}$,
$\hat{b}_{(2)}$, and $\hat{e}_{i}$ are related as
\begin{eqnarray}
  \label{eq:DBHD-BS1-junction-hatb(1)-hatb(2)}
  \hat{b}_{(1)}
  =
  \frac{1}{\sqrt{2}} \left(
  \hat{b} - \hat{e}_{i}
  \right)
  ,
  \quad
  \hat{b}_{(2)}
  =
  \frac{1}{\sqrt{2}} \left(
  \hat{b} + \hat{e}_{i}
  \right)
\end{eqnarray}
through the field junction conditions at the BS1.
We assume the state associated with the quadrature $\hat{e}_{i}$ is
vacuum.

%*******************************************************************

%%%%%%%%%%%%%%%%%%%%%%%%%%%%%%%%%%%%%%%%%%%%%%%%%%%%%%%
\subsubsection{Separation of the field from the local oscillator}
\label{sec:separation_of_the_field _from_the_local_oscillator}
%%%%%%%%%%%%%%%%%%%%%%%%%%%%%%%%%%%%%%%%%%%%%%%%%%%%%%%

%*******************************************************************

At the 50:50 beam splitter 3 (BS3), the incident electric field from
the local oscillator is in the coherent state and its quadrature is
denoted by $\hat{l}_{i}$.
Furthermore, from the configuration depicted in
Fig.~\ref{fig:DoubleBalancedHomodyneDetection-configuration}, another
incident field to BS3, which we denote its quadrature by
$\hat{f}_{i}$, should be taken into account.
On the other hand, the beam splitter BS3 separate the electric field
into two paths.
The electric field along one of these two paths goes from BS3 to BS2.
We denote the quadrature associated with this electric field by
$\hat{l}_{(0)i}$.
On the other hand, the electric field along another path from BS3 is
towards to the beam splitter 4 (BS4).
We denote the quadrature associated with this electric field by
$\hat{l}_{(1)i}$.
By the beam splitter condition, quadratures $\hat{l}_{(0)i}$ and
$\hat{l}_{(1)i}$ are determined by the equation
\begin{eqnarray}
  \label{eq:hatl(0)i-hatl(1)ihatli-hatfi-rel}
  \hat{l}_{(0)i}
  =
  \frac{1}{\sqrt{2}} \left(
    \hat{l}_{i} - \hat{f}_{i}
  \right)
  , \quad
  \hat{l}_{(1)i}
  =
  \frac{1}{\sqrt{2}}\left(
    \hat{l}_{i} + \hat{f}_{i}
  \right)
  .
\end{eqnarray}
We also assume that the state associated with the quadrature
$\hat{f}_{i}$ is also vacuum.
The field associated with the quadrature $\hat{l}_{(0)i}$ is
used in the balanced homodyne detection through the beam splitter
2 (BS2).

%*******************************************************************

To make $\pi/2$-phase difference in the electric field from the local
oscillator as discussed in Sec.~\ref{sec:theoretical_proposal}, we
introduce the phase rotator (PR) between BS3 and BS4.
As shown in Appendix A in
Ref.~\cite{H.J.Kimble-Y.Levin-A.B.Matsko-K.S.Thorne-S.P.Vyatchanin-2001},
the phase rotator operates the rotation operator $\hat{R}(\theta)$
defined by
\begin{eqnarray}
  \label{eq:phase_rotation_operator}
  \hat{R}(\theta)
  =
  \exp\left\{
    - i \theta
    \int_{0}^{+\infty} \frac{d\Omega}{2\pi}
    \left(
      \hat{l}_{(1)i+}^{\dagger}\hat{l}_{(1)i+}
      +
      \hat{l}_{(1)i-}^{\dagger}\hat{l}_{(1)i-}
    \right)
  \right\}
  .
\end{eqnarray}
The phase rotator (PR) in
Fig.~\ref{fig:DoubleBalancedHomodyneDetection-configuration} is
defined as a device which operates to the quadrature $\hat{l}_{(1)i}$
and produce the quantum field associated with the quadrature
$\hat{l}_{(1/4)i}$ as the output field as
\begin{eqnarray}
  \label{eq:phase_rotation_operator_operation}
  \hat{l}_{(1/4)i}
  =
  i
  \hat{l}_{(1)i}
  .
\end{eqnarray}
This quadrature $\hat{l}_{(1/4)i}$ is used the balanced homodyne
detection through BS4.

%*******************************************************************

%%%%%%%%%%%%%%%%%%%%%%%%%%%%%%%%%%%%%%%%%%%%%%%%%%%%%%%
\subsubsection{Balanced homodyne detection through BS2}
\label{sec:balanced_homodyne_detection_BS2}
%%%%%%%%%%%%%%%%%%%%%%%%%%%%%%%%%%%%%%%%%%%%%%%%%%%%%%%

%*******************************************************************

Now, we consider the balanced homodyne detection through the beam
splitter BS2.
Here, we assume that the transmittance of the beamsplitter
BS2 is $\eta$ as a simple model of the imperfections of the
interferometer configurations.
The important incident fields to the beam splitter BS2 are
$\hat{b}_{(1)}$ and $\hat{l}_{(0)i}$ and the important outgoing
fields from BS2 are $\hat{c}_{(1)o}$ and $\hat{d}_{(1)o}$.
At the BS2, these fields are related as
\begin{eqnarray}
  \label{eq:DBHD-D1-D2-output}
  \hat{c}_{(1)o}
  =
  \sqrt{\eta} \hat{b}_{(1)}
  +
  \sqrt{1-\eta} \hat{l}_{(0)i}
  ,
  \quad
  \hat{d}_{(1)o}
  =
  \sqrt{\eta} \hat{l}_{(0)i}
  -
  \sqrt{1-\eta} \hat{b}_{(1)}
  .
\end{eqnarray}

%*******************************************************************

The quadrature $\hat{b}_{(1)}$ is related to the direct output
quadrature $\hat{b}$ through the first equation in
Eqs.~(\ref{eq:DBHD-BS1-junction-hatb(1)-hatb(2)}) and the quadrature
$\hat{l}_{(0)i}$ is related to the quadrature $\hat{l}_{i}$ of the
incident field from the local oscillator through
the first equation in
Eqs.~(\ref{eq:hatl(0)i-hatl(1)ihatli-hatfi-rel}).
Then, Eqs.~(\ref{eq:DBHD-D1-D2-output}) are given by
\begin{eqnarray}
  \hat{c}_{(1)o}
  &=&
  \frac{1}{\sqrt{2}}
  \left(
    \sqrt{\eta}
    \hat{b}
    +
    \sqrt{1-\eta}
    \hat{l}_{i}
  \right)
  -
  \frac{1}{\sqrt{2}}
  \left(
    \sqrt{\eta}
    \hat{e}_{i}
    +
    \sqrt{1-\eta}
    \hat{f}_{i}
  \right)
  \label{eq:hatc(1)o=hatb-hatli-hatei-hatfi-relation}
  , \\
  \hat{d}_{(1)o}
  &=&
  \frac{1}{\sqrt{2}}
  \left(
    \sqrt{\eta}
    \hat{l}_{i}
    -
    \sqrt{1-\eta}
    \hat{b}
  \right)
  +
  \frac{1}{\sqrt{2}}
  \left(
    \sqrt{1-\eta}
    \hat{e}_{i}
    -
    \sqrt{\eta}
    \hat{f}_{i}
  \right)
  \label{eq:hatd(1)o=hatli-hatb-hatei-hatfi-relation}
  .
\end{eqnarray}
Here, we note that the second terms in
Eqs.~(\ref{eq:hatc(1)o=hatb-hatli-hatei-hatfi-relation}) and
(\ref{eq:hatd(1)o=hatli-hatb-hatei-hatfi-relation}) are the
contribution of the incident vacuum field due to the interferometer
configuration depicted in
Fig.~\ref{fig:DoubleBalancedHomodyneDetection-configuration},
respectively.

%*******************************************************************

The photon numbers of the output fields associated with the
quadratures $\hat{c}_{(1)o}$ and $\hat{d}_{(1)o}$ are detected
through the photodetector D1 and D2 in
Fig.~\ref{fig:DoubleBalancedHomodyneDetection-configuration},
respectively.
These photon-number operators are defined by
\begin{eqnarray}
  \label{eq:hatnc(1)o-hatnd(1)o-defs}
  \hat{n}_{c_{(1)o}}
  :=
  \hat{c}_{(1)o}^{\dagger}\hat{c}_{(1)o}
  ,
  \quad
  \hat{n}_{d_{(1)o}}
  :=
  \hat{d}_{(1)o}^{\dagger}\hat{d}_{(1)o}
  .
\end{eqnarray}
Through Eqs.~(\ref{eq:hatc(1)o=hatb-hatli-hatei-hatfi-relation}) and
(\ref{eq:hatd(1)o=hatli-hatb-hatei-hatfi-relation}), these operators
are given in terms of the quadratures $\hat{b}$,
$\hat{l}_{i}$, $\hat{e}_{i}$ and $\hat{f}_{i}$.
The expectation values of the photon-number operators
$\hat{n}_{c_{(1)o}}$ and $\hat{n}_{d_{(1)o}}$, which is detected by
the photodetector D1 and D2, respectively, are given by
\begin{eqnarray}
  \langle\hat{n}_{c_{(1)o}}\rangle
  &=&
      \frac{\eta}{2}
      \left\langle
      \hat{n}_{b}
      \right\rangle
      +
      \frac{\sqrt{\eta(1-\eta)}}{2}
      \left\langle
      \gamma^{*}
      \hat{b}
      +
      \gamma
      \hat{b}^{\dagger}
      \right\rangle
      +
      \frac{(1-\eta)}{2}
      |\gamma|^{2}
      ,
  \\
  \label{eq:DBHD-output-D1-photon-number-expect}
  \langle\hat{n}_{d_{(1)o}}\rangle
  &=&
      \frac{\eta}{2}
      |\gamma|^{2}
      -
      \frac{\sqrt{\eta(1-\eta)}}{2}
      \left\langle
      \gamma^{*}
      \hat{b}
      +
      \gamma
      \hat{b}^{\dagger}
      \right\rangle
      +
      \frac{1-\eta}{2}
      \langle\hat{n}_{b}\rangle
      .
      \label{eq:DBHD-output-D2-photon-number-expect}
\end{eqnarray}
In the derivation of
Eqs.~(\ref{eq:DBHD-output-D1-photon-number-expect}) and
(\ref{eq:DBHD-output-D2-photon-number-expect}), we used the field
associated with the quadrature $\hat{l}_{i}$ is in the coherent state
with the complex amplitude $\gamma$ and the fields associated with the
quadratures $\hat{e}_{i}$ and $\hat{f}_{i}$ are in vacua, respectively.
As in the case of the conventional balanced homodyne detection in
Sec.~\ref{sec:balanced_homodyne_detection}, we obtain
\begin{eqnarray}
  \label{eq:BHD-D1-D2-result}
  \frac{2}{\sqrt{\eta(1-\eta)}}
  \left(
  (1-\eta)\langle\hat{n}_{c_{(1)o}}\rangle
  -
  \eta\langle\hat{n}_{d_{(1)o}}\rangle
  -
  \frac{1-2\eta}{2}
  |\gamma|^{2}
  \right)
  =
  \left\langle
  \gamma^{*}
  \hat{b}
  +
  \gamma
  \hat{b}^{\dagger}
  \right\rangle
\end{eqnarray}
from Eqs.~(\ref{eq:DBHD-output-D1-photon-number-expect}) and
(\ref{eq:DBHD-output-D2-photon-number-expect}).
Thus, the expectation value of the self-adjoint operator
$\gamma^{*}\hat{b}+\gamma\hat{b}^{\dagger}$ can be calculated through
the photon-number expectation value $\langle\hat{n}_{c_{(1)o}}\rangle$
and $\langle\hat{n}_{d_{(1)o}}\rangle$, if the non-vanishing complex
amplitude $\gamma$ of the coherent state from the local oscillator and
the transmittance $\eta$ of the beam splitter are given.

%*******************************************************************

The left-hand side of Eq.~(\ref{eq:BHD-D1-D2-result}) is regarded as
the expectation value of the operator $\hat{s}_{D1D2}$ defined by
\begin{eqnarray}
  \label{eq:hats-D1D2-operator-original-def}
  \hat{s}_{D1D2}
  :=
  \frac{2}{\sqrt{\eta(1-\eta)}}
  \left(
  (1-\eta)\hat{n}_{c_{(1)o}}
  -
  \eta\hat{n}_{d_{(1)o}}
  -
  \frac{1-2\eta}{2}
  |\gamma|^{2}
  \right)
  .
\end{eqnarray}
Through Eqs.~(\ref{eq:hatc(1)o=hatb-hatli-hatei-hatfi-relation}),
(\ref{eq:hatd(1)o=hatli-hatb-hatei-hatfi-relation}) and
(\ref{eq:hatnc(1)o-hatnd(1)o-defs}), in terms of the quadrature
$\hat{b}$, $\hat{l}_{i}$, $\hat{e}_{i}$, and $\hat{f}_{i}$, the
operator $\hat{s}_{D1D2}$ defined by
Eq.~(\ref{eq:hats-D1D2-operator-original-def}) is also expressed as
\begin{eqnarray}
  \hat{s}_{D1D2}
  &=&
      \hat{b} \hat{l}_{i}^{\dagger}
      + \hat{b}^{\dagger} \hat{l}_{i}
      +
      \frac{1-2\eta}{\sqrt{\eta(1-\eta)}}
      \left(
      \hat{l}_{i}^{\dagger} \hat{l}_{i}
      -
      |\gamma|^{2}
      \right)
      \nonumber\\
  &&
     - \hat{b} \hat{f}_{i}^{\dagger}
     - \hat{b}^{\dagger} \hat{f}_{i}
     + \hat{e}_{i} \hat{f}_{i}^{\dagger}
     + \hat{e}_{i}^{\dagger} \hat{f}_{i}
     - \hat{e}_{i} \hat{l}_{i}^{\dagger}
     - \hat{e}_{i}^{\dagger} \hat{l}_{i}
     \nonumber\\
  &&
     +
     \frac{1-2\eta}{\sqrt{\eta(1-\eta)}}
     \left(
     \hat{f}_{i}^{\dagger} \hat{f}_{i}
     - \hat{f}_{i}^{\dagger} \hat{l}_{i}
     - \hat{l}_{i}^{\dagger} \hat{f}_{i}
     \right)
     .
     \label{eq:hats-D1D2-operator-expression-2}
\end{eqnarray}
In the right-hand side of
Eq.~(\ref{eq:hats-D1D2-operator-expression-2}), the first line gives
the expectation value (\ref{eq:BHD-D1-D2-result}), the second-line is
the contributions from the vacuum inputs, and the final line is the
vacuum contribution due to the imperfection of the beam splitter BS2.

%*******************************************************************

%%%%%%%%%%%%%%%%%%%%%%%%%%%%%%%%%%%%%%%%%%%%%%%%%%%%%%%
\subsubsection{Balanced homodyne detection through BS4}
\label{sec:balanced_homodyne_detection_BS4}
%%%%%%%%%%%%%%%%%%%%%%%%%%%%%%%%%%%%%%%%%%%%%%%%%%%%%%%

%*******************************************************************

Next, we consider the balanced homodyne detection through the beam
splitter BS4.
Here, we again assume that the transmittance of the beamsplitter BS4
is $\eta$, as a simple model of the imperfections of the
interferometer configurations.
The important incident fields to the beam splitter BS4 is
$\hat{b}_{(2)}$ and $\hat{l}_{(1/4)i}$ and the important
outgoing fields from BS4 is $\hat{c}_{(2)o}$ and
$\hat{d}_{(2)o}$.
At the BS4, these fields are related as
\begin{eqnarray}
  \label{eq:DBHD-D3-D4-output}
  \hat{c}_{(2)o}
  =
  \sqrt{\eta} \hat{l}_{(1/4)i}
  -
  \sqrt{1-\eta} \hat{b}_{(2)}
  ,
  \quad
  \hat{d}_{(2)o}
  =
  \sqrt{\eta} \hat{b}_{(2)}
  +
  \sqrt{1-\eta} \hat{l}_{(1/4)i}
  .
\end{eqnarray}
Here, we consider the case where the transmittance of the beam
splitter BS4 is also $\eta$, which equal to the transmittance of the
beam splitter BS2, as a simple model of imperfections of the
interferometer setup.

%*******************************************************************

The quadrature $\hat{b}_{(2)}$ is related to the direct output
quadrature $\hat{b}$ through the second equation in
Eqs.~(\ref{eq:DBHD-BS1-junction-hatb(1)-hatb(2)}).
On the other hand, the quadrature $\hat{l}_{(1/4)i}$ is related to the
quadrature $\hat{l}_{i}$ of the incident field from the local
oscillator through the second equation of
Eqs.~(\ref{eq:hatl(0)i-hatl(1)ihatli-hatfi-rel}) and the effect of the
phase rotator (\ref{eq:phase_rotation_operator_operation}).
Then, Eqs.~(\ref{eq:DBHD-D3-D4-output}) are given by
\begin{eqnarray}
  \hat{c}_{(2)o}
  &=&
      \frac{1}{\sqrt{2}}
      \left(
      i
      \sqrt{\eta}
      \hat{l}_{i}
      -
      \sqrt{1-\eta}
      \hat{b}
      \right)
      +
      \frac{1}{\sqrt{2}}
      \left(
      i
      \sqrt{\eta}
      \hat{f}_{i}
      -
      \sqrt{1-\eta}
      \hat{e}_{i}
      \right)
      ,
      \label{eq:hatc(2)o=hatb-hatli-hatfi-hatei-relation}
      \\
  \hat{d}_{(2)o}
  &=&
      \frac{1}{\sqrt{2}}
      \left(
      \sqrt{\eta}
      \hat{b}
      +
      i \sqrt{1-\eta} \hat{l}_{i}
      \right)
      +
      \frac{1}{\sqrt{2}}
      \left(
      \sqrt{\eta}
      \hat{e}_{i}
      +
      i \sqrt{1-\eta} \hat{f}_{i}
      \right)
      .
      \label{eq:hatd(2)o=hatb-hatli-hatei-hatfi-relation}
\end{eqnarray}
Here, we note that the second terms in
Eqs.~(\ref{eq:hatc(2)o=hatb-hatli-hatfi-hatei-relation}) and
(\ref{eq:hatd(2)o=hatb-hatli-hatei-hatfi-relation}) are the
contribution of the incident vacuum field due to the interferometer
configuration depicted in
Fig.~\ref{fig:DoubleBalancedHomodyneDetection-configuration},
respectively.

%*******************************************************************

The photon numbers of the output fields associated with the
quadratures $\hat{c}_{(2)o}$ and $\hat{d}_{(2)o}$ are detected
through the photodetector D3 and D4 in
Fig.~\ref{fig:DoubleBalancedHomodyneDetection-configuration},
respectively.
These photon-number operators are defined by
\begin{eqnarray}
  \label{eq:hatnc(2)o-hatnd(2)o-defs}
  \hat{n}_{c_{(2)o}}
  :=
  \hat{c}_{(2)o}^{\dagger}\hat{c}_{(2)o}
  , \quad
  \hat{n}_{d_{(2)o}}
  :=
  \hat{d}_{(2)o}^{\dagger}\hat{d}_{(2)o}
  .
\end{eqnarray}
Through Eqs.~(\ref{eq:hatc(2)o=hatb-hatli-hatfi-hatei-relation}) and
(\ref{eq:hatd(2)o=hatb-hatli-hatei-hatfi-relation}), these operators
are given in terms of the quadratures $\hat{b}$, $\hat{l}_{i}$,
$\hat{e}_{i}$ and $\hat{f}_{i}$.
The expectation values of the photon number operators
$\hat{n}_{c_{(2)}o}$ and $\hat{n}_{d_{(2)o}}$, which are detected
by the photodetector D3 and D4, respectively, are given by
\begin{eqnarray}
  \langle\hat{n}_{c_{(2)o}}\rangle
  &=&
      \frac{\eta}{2}
      |\gamma|^{2}
      +
      \frac{i\sqrt{\eta(1-\eta)}}{2}
      \left\langle
      \gamma^{*}
      \hat{b}
      -
      \gamma
      \hat{b}^{\dagger}
      \right\rangle
      +
      \frac{1-\eta}{2}
      \left\langle
      \hat{n}_{b}
      \right\rangle
      ,
      \label{eq:DBHD-output-D3-photon-number-expect}
  \\
  \langle\hat{n}_{d_{(2)o}}\rangle
  &=&
      \frac{\eta}{2}
      \left\langle
      \hat{n}_{b}
      \right\rangle
      -
      \frac{i\sqrt{\eta(1-\eta)}}{2}
      \left\langle
      \gamma^{*}
      \hat{b}
      -
      \gamma
      \hat{b}^{\dagger}
      \right\rangle
      +
      \frac{1-\eta}{2}
      |\gamma|^{2}
      .
      \label{eq:DBHD-output-D4-photon-number-expect}
\end{eqnarray}
In the derivation of
Eqs.~(\ref{eq:DBHD-output-D3-photon-number-expect}) and
(\ref{eq:DBHD-output-D4-photon-number-expect}), we used the field
associated with the quadrature $\hat{l}_{i}$ is in the coherent state
with the complex amplitude $\gamma$ and the fields associated with the
quadratures $\hat{e}_{i}$ and $\hat{f}_{i}$ are in vacua.
As in the case of the conventional balanced homodyne detection in
Sec.~\ref{sec:balanced_homodyne_detection}, we obtain
\begin{eqnarray}
  \label{eq:BHD-D3-D4-result}
  -
  \frac{2i}{\sqrt{\eta(1-\eta)}}
  \left(
  \eta\langle\hat{n}_{c_{(2)o}}\rangle
  -
  (1-\eta)\langle\hat{n}_{d_{(2)o}}\rangle
  + \frac{1-2\eta}{2} |\gamma|^{2}
  \right)
  =
  \left\langle
  \gamma^{*}
  \hat{b}
  -
  \gamma
  \hat{b}^{\dagger}
  \right\rangle
  .
\end{eqnarray}
from Eqs.~(\ref{eq:DBHD-output-D3-photon-number-expect}) and
(\ref{eq:DBHD-output-D4-photon-number-expect}).
Thus, the expectation value of the anti-self-adjoint operator
$\gamma^{*}\hat{b}-\gamma\hat{b}^{\dagger}$ can be calculated through
the photon number expectation value $\langle\hat{n}_{c_{(2)o}}\rangle$ and
$\langle\hat{n}_{d_{(2)o}}\rangle$, if the complex amplitude $\gamma$
of the coherent state from the local oscillator and the transmittance
$\eta$ of the beam splitters B2 and B4 are given.

%*******************************************************************

The left-hand side of Eq.~(\ref{eq:BHD-D3-D4-result}) is regarded as
the expectation value of the operator $\hat{s}_{D3D4}(\omega)$ defined
by
\begin{eqnarray}
  \label{eq:hats-D3D4-operator-original-def}
  \hat{s}_{D3D4}
  :=
  \frac{2i}{\sqrt{\eta(1-\eta)}}
  \left(
  (1-\eta)\hat{n}_{d_{(2)o}}
  -
  \eta\hat{n}_{c_{(2)o}}
  -
  \frac{1-2\eta}{2} |\gamma|^{2}
  \right)
  .
\end{eqnarray}
Through Eqs.~(\ref{eq:hatc(2)o=hatb-hatli-hatfi-hatei-relation}),
(\ref{eq:hatd(2)o=hatb-hatli-hatei-hatfi-relation}), and
(\ref{eq:hatnc(2)o-hatnd(2)o-defs}), in terms of the quadrature
$\hat{b}$, $\hat{l}_{i}$, $\hat{e}_{i}$, and $\hat{f}_{i}$, the
operator $\hat{s}_{D3D4}$ defined by
Eq.~(\ref{eq:hats-D3D4-operator-original-def}) is also expressed as
\begin{eqnarray}
  \hat{s}_{D3D4}
  &=&
      \hat{l}_{i}^{\dagger}
      \hat{b}
      -
      \hat{b}^{\dagger}
      \hat{l}_{i}
      +
      \frac{i(1-2\eta)}{\sqrt{\eta(1-\eta)}}
      \left(
      \hat{l}_{i}^{\dagger}
      \hat{l}_{i}
      - |\gamma|^{2}
      \right)
      \nonumber\\
  &&
     - \hat{b}^{\dagger} \hat{f}_{i}
     + \hat{b} \hat{f}_{i}^{\dagger}
     + \hat{l}_{i}^{\dagger} \hat{e}_{i}
     + \hat{f}_{i}^{\dagger} \hat{e}_{i}
     - \hat{e}_{i}^{\dagger} \hat{l}_{i}
     - \hat{e}_{i}^{\dagger} \hat{f}_{i}
     \nonumber\\
  &&
     +
     \frac{i(1-2\eta)}{\sqrt{\eta(1-\eta)}}
     \left(
     \hat{f}_{i}^{\dagger}
     \hat{l}_{i}
     +
     \hat{l}_{i}^{\dagger}
     \hat{f}_{i}
     +
     \hat{f}_{i}^{\dagger}
     \hat{f}_{i}
     \right)
     .
  \label{eq:hats-D3D4-operator-expression-2}
\end{eqnarray}
In the right-hand side of
Eq.~(\ref{eq:hats-D3D4-operator-expression-2}), the first line gives
the expectation value (\ref{eq:BHD-D3-D4-result}), the second-line is
the contributions from the vacuum inputs, and the final line is the
vacuum contribution due to the unbalance of the beam splitter BS4.

%*******************************************************************

We note that the overall factor in the left-hand side of
Eq.~(\ref{eq:BHD-D3-D4-result}) and the right-hand side of the
definition (\ref{eq:hats-D3D4-operator-original-def}) are purely
imaginary.
These factors breaks the self-adjointness of our results.

%*******************************************************************

%%%%%%%%%%%%%%%%%%%%%%%%%%%%%%%%%%%%%%%%%%%%%%%%%%%%%%%
%%%%%%%%%%%%%%%%%%%%%%%%%%%%%%%%%%%%%%%%%%%%%%%%%%%%%%%
\subsection{Input vacuum to the main interferometer}
\label{sec:input-vaccum-to-the-main-inteterfemeter}
%%%%%%%%%%%%%%%%%%%%%%%%%%%%%%%%%%%%%%%%%%%%%%%%%%%%%%%
%%%%%%%%%%%%%%%%%%%%%%%%%%%%%%%%%%%%%%%%%%%%%%%%%%%%%%%

%*******************************************************************

The output operators (\ref{eq:hats-D1D2-operator-expression-2}) and
(\ref{eq:hats-D3D4-operator-expression-2}) to measure the operator
$\hat{b}$ or $\hat{b}^{\dagger}$ are described by the quadratures
$\hat{b}$, $\hat{l}_{i}$, $\hat{e}_{i}$, and $\hat{f}_{i}$.
As depicted in
Fig.~\ref{fig:DoubleBalancedHomodyneDetection-configuration},
$\hat{e}_{i}$, $\hat{f}_{i}$, and $\hat{l}_{i}$ are quadratures of the
independent fields.
Then, we should regard that these operators commute with each other:
\begin{eqnarray}
  \label{eq:ei-fi-li-independence}
  \left[
  \hat{e}_{i}, \hat{f}_{i}
  \right]
  =
  \left[
  \hat{e}_{i}, \hat{f}_{i}^{\dagger}
  \right]
  =
  \left[
  \hat{f}_{i}, \hat{l}_{i}
  \right]
  =
  \left[
  \hat{f}_{i}, \hat{l}_{i}^{\dagger}
  \right]
  =
  \left[
  \hat{l}_{i}, \hat{e}_{i}
  \right]
  =
  \left[
  \hat{l}_{i}, \hat{e}_{i}^{\dagger}
  \right]
  =
  0
  .
\end{eqnarray}
The main purpose of this subsection is to check whether or not we may
regard the output quadrature $\hat{b}$ is also independent of
$\hat{e}_{i}$, $\hat{f}_{i}$, and $\hat{l}_{i}$.

%*******************************************************************

To check this independence, we have to remind that the operator $\hat{b}$
is the output quadrature from the main interferometer and we have the input
quadrature $\hat{a}$ to the main interferometer.
In general, the output quadrature $\hat{b}$ may depend on this input
quadrature $\hat{a}$.
In many situation, we regard that the field associated with the
quadrature $\hat{a}$ is in vacuum.
Here, we check this vacuum state for the quadrature $\hat{a}$ is an
independent vacuum from the states of the photon fields associated
with the quadratures $\hat{e}_{i}$, $\hat{f}_{i}$, and $\hat{l}_{i}$
through the interferometer configuration depicted in
Fig.~\ref{fig:DoubleBalancedHomodyneDetection-configuration}.

%*******************************************************************

At the BS1, the quadrature $\hat{a}$ is given by
\begin{eqnarray}
  \hat{a}
  =
  \frac{1}{\sqrt{2}} \left(
    \hat{a}_{(1)}
    +
    \hat{a}_{(2)}
  \right)
  ,
  \label{eq:hata-input-to-main-interferometer}
\end{eqnarray}
and, at the BS2, the quadrature $\hat{a}_{(1)}$ is given by
\begin{eqnarray}
  \hat{a}_{(1)}
  =
  \sqrt{\eta} \hat{c}_{(1)i}
  -
  \sqrt{1-\eta} \hat{d}_{(1)i}
  .
  \label{eq:hata(1)-def}
\end{eqnarray}
The quadratures $\hat{c}_{(1)i}$ and $\hat{d}_{(1)i}$ are
the quadratures of the incident fields from detectors D1 and D2
to the BS2, respectively.
On the other hand, through the junction at the BS4, the
quadrature $\hat{a}_{(2)}$ is given by
\begin{eqnarray}
  \hat{a}_{(2)}
  =
  \sqrt{\eta} \hat{d}_{(2)i}
  -
  \sqrt{1-\eta} \hat{c}_{(2)i}
  .
  \label{eq:hata(2)-def}
\end{eqnarray}
The quadratures $\hat{c}_{(2)i}$ and $\hat{d}_{(2)i}$ are
the quadratures of the incident fields from detectors D3 and D4
to the BS4, respectively.

%*******************************************************************

Substituting Eqs.~(\ref{eq:hata(1)-def}) and (\ref{eq:hata(2)-def})
into Eq.~(\ref{eq:hata-input-to-main-interferometer}), we obtain
\begin{eqnarray}
  \hat{a}
  =
  \sqrt{\frac{\eta}{2}}
  \left(
  \hat{c}_{(1)i}
  +
  \hat{d}_{(2)i}
  \right)
  -
  \sqrt{\frac{1-\eta}{2}}
  \left(
  \hat{d}_{(1)i}
  +
  \hat{c}_{(2)i}
  \right)
  .
  \label{eq:hata-hatc(1)i-hatd(2)i-hatd(1)i-hatc(2)i-relation}
\end{eqnarray}
We assume that the states associated with the quadratures
$\hat{c}_{(1)i}$, $\hat{c}_{(2)i}$, $\hat{d}_{(1)i}$, and
$\hat{d}_{(2)i}$ from the detectors D1, D3, D2, and D4 are
vacua.
Due to this situation, the state associated with the input quadrature
$\hat{a}$ is also regarded as a vacuum.
Furthermore,
Eq.~(\ref{eq:hata-hatc(1)i-hatd(2)i-hatd(1)i-hatc(2)i-relation})
indicates that the quadrature $\hat{a}$ is independent of the
quadratures $\hat{e}_{i}$, $\hat{f}_{i}$, and $\hat{l}_{i}$.
Therefore, we may regard that
\begin{eqnarray}
  \label{eq:hata-hatei-hatfi-hatli-independence}
  \left[
  \hat{a}, \hat{e}_{i}
  \right]
  =
  \left[
  \hat{a}, \hat{e}_{i}^{\dagger}
  \right]
  =
  \left[
  \hat{a}, \hat{f}_{i}
  \right]
  =
  \left[
  \hat{a}, \hat{f}_{i}^{\dagger}
  \right]
  =
  \left[
  \hat{a}, \hat{l}_{i}
  \right]
  =
  \left[
  \hat{a}, \hat{l}_{i}^{\dagger}
  \right]
  =
  0
  ,
\end{eqnarray}
and then, we may regard that
\begin{eqnarray}
  \label{eq:hatb-hatei-hatfi-hatli-independence}
  \left[
  \hat{b}, \hat{e}_{i}
  \right]
  =
  \left[
  \hat{b}, \hat{e}_{i}^{\dagger}
  \right]
  =
  \left[
  \hat{b}, \hat{f}_{i}
  \right]
  =
  \left[
  \hat{b}, \hat{f}_{i}^{\dagger}
  \right]
  =
  \left[
  \hat{b}, \hat{l}_{i}
  \right]
  =
  \left[
  \hat{b}, \hat{l}_{i}^{\dagger}
  \right]
  =
  0
  ,
\end{eqnarray}
even if the output quadrature $\hat{b}$ depends on the input
quadrature $\hat{a}$.

%*******************************************************************

We use the commutation relations (\ref{eq:ei-fi-li-independence}) and
(\ref{eq:hatb-hatei-hatfi-hatli-independence}) when we evaluate the
fluctuations in the measurements discussed in
Sec.~\ref{sec:measurement_of-hatb-or-hatbdagger} and
Sec.~\ref{sec:DBHD_in_two_photon_formulation}.

%*******************************************************************

%%%%%%%%%%%%%%%%%%%%%%%%%%%%%%%%%%%%%%%%%%%%%%%%%%%%%%%
%%%%%%%%%%%%%%%%%%%%%%%%%%%%%%%%%%%%%%%%%%%%%%%%%%%%%%%
\subsection{Measurements of an annihilation operator $\hat{b}$ and
  a creation operator $\hat{b}^{\dagger}$}
\label{sec:measurement_of-hatb-or-hatbdagger}
%%%%%%%%%%%%%%%%%%%%%%%%%%%%%%%%%%%%%%%%%%%%%%%%%%%%%%%
%%%%%%%%%%%%%%%%%%%%%%%%%%%%%%%%%%%%%%%%%%%%%%%%%%%%%%%

%*******************************************************************

In Sec.~\ref{sec:realization-of-DBHD}, the expectation value of the operators
$\hat{s}_{D1D2}$ and $\hat{s}_{D3D4}$ can be calculated through the
photon-number expectation values at the photodetectors D1, D2, D3, and D4.
In this subsection, we note that we can calculate the expectation
values of the annihilation and creation operators of photon field from
the expectation values of these operators.
The simplified version of the ingredients of this subsection is
already explained in Ref.~\cite{K.Nakamura-M.-K.Fujimoto-2017b}.

%*******************************************************************

Note that the field associated with the quadrature $\hat{l}_{i}$ is in
the coherent state with the complex amplitude $\gamma$ and the fields
associated with the quadratures $\hat{f}_{i}$ and $\hat{e}_{i}$
are in their vacuum states in the derivation of the form of the
operators $\hat{s}_{D1D2}$ and $\hat{s}_{D3D4}$.
The guiding principle of the construction of the operators
$\hat{s}_{D1D2}$ and $\hat{s}_{D3D4}$ are just their expectation
values which are given by (\ref{eq:BHD-D1-D2-result}) and
(\ref{eq:BHD-D3-D4-result}).
From these expectation values, the expectation values of
$\langle\hat{b}\rangle$ and $\langle\hat{b}^{\dagger}\rangle$ are
given by
\begin{eqnarray}
  &&
     \label{eq:expecation_valu_of_hatb-result}
     \frac{1}{2\gamma^{*}}
     \left(
     \left\langle
     \hat{s}_{D1D2}
     \right\rangle
     +
     \left\langle
     \hat{s}_{D3D4}
     \right\rangle
     \right)
     =
     \left\langle\hat{b}\right\rangle
     ,
  \\
  &&
     \label{eq:expecation_valu_of_hatbdagger-result}
     \frac{1}{2\gamma}
     \left(
     \left\langle
     \hat{s}_{D1D2}
     \right\rangle
     -
     \left\langle
     \hat{s}_{D3D4}
     \right\rangle
     \right)
     =
     \left\langle\hat{b}^{\dagger}\right\rangle
     .
\end{eqnarray}
These are assertions which is pointed out in
Ref.~\cite{K.Nakamura-M.-K.Fujimoto-2017b} and are trivial result from
the construction of the operators $\hat{s}_{D1D2}$ and
$\hat{s}_{D3D4}$.
We have to emphasize that the expectation values of the operators
$\hat{s}_{D1D2}$ and $\hat{s}_{D3D4}$ are calculable through the
expectation values of photon number operators which are measured
at  the photodetector D1, D2, D3, and D4, the complex amplitude
$\gamma$ for the coherent state from the local oscillator, and the
transmittance $\eta$ of the beam splitters B2 and B4.
A similar assertion was also reported in
Ref.~\cite{E.Shchukin-Th.Richter-W.Vogel-2005} in the context
of the nonclassicality criteria of quantum system.

%*******************************************************************

Here, we also note that the direct calculation from
Eqs.~(\ref{eq:hats-D1D2-operator-expression-2}) and
(\ref{eq:hats-D3D4-operator-expression-2}) yields
\begin{eqnarray}
  \hat{t}_{b+}
  &:=&
      \frac{1}{2\gamma^{*}}
      \left(
      \hat{s}_{D1D2}
      +
      \hat{s}_{D3D4}
      \right)
      \nonumber\\
  &=&
      \hat{b} \frac{\hat{l}_{i}^{\dagger}}{\gamma^{*}}
      +
      \frac{(1+i)(1-2\eta)}{2\sqrt{\eta(1-\eta)}}
      \gamma
      \left(
      \frac{\hat{l}_{i}^{\dagger} \hat{l}_{i}}{|\gamma|^{2}}
      -
      1
      \right)
      \nonumber\\
  &&
     +
     \frac{1}{\gamma^{*}}
     \left(
     - \hat{b}^{\dagger} \hat{f}_{i}
     + \hat{e}_{i} \hat{f}_{i}^{\dagger}
     - \hat{e}_{i}^{\dagger} \hat{l}_{i}
     \right)
     \nonumber\\
  &&
     +
     \frac{(1+i)(1-2\eta)}{2\sqrt{\eta(1-\eta)}\gamma^{*}}
     \left(
     + i \hat{f}_{i}^{\dagger} \hat{f}_{i}
     -   \hat{f}_{i}^{\dagger} \hat{l}_{i}
     -   \hat{l}_{i}^{\dagger} \hat{f}_{i}
     \right)
     .
     \label{eq:hatb-measure-operator-result}
\end{eqnarray}
and
\begin{eqnarray}
  \hat{t}_{b-}
  &:=&
       \frac{1}{2\gamma}
       \left(
       \hat{s}_{D1D2}
       -
       \hat{s}_{D3D4}
       \right)
       \nonumber\\
  &=&
      \hat{b}^{\dagger} \frac{\hat{l}_{i}}{\gamma}
      +
      \frac{(1-i)(1-2\eta)}{2\sqrt{\eta(1-\eta)}}
      \gamma^{*}
      \left(
      \frac{\hat{l}_{i}^{\dagger} \hat{l}_{i}}{|\gamma|^{2}}
      -
      1
      \right)
      \nonumber\\
  &&
     +
     \frac{1}{\gamma}
     \left(
     - \hat{b} \hat{f}_{i}^{\dagger}
     + \hat{e}_{i}^{\dagger} \hat{f}_{i}
     - \hat{e}_{i} \hat{l}_{i}^{\dagger}
     \right)
  \nonumber\\
  &&
     +
     \frac{(1-i)(1-2\eta)}{2\sqrt{\eta(1-\eta)}\gamma}
     \left(
     - i \hat{f}_{i}^{\dagger} \hat{f}_{i}
     -   \hat{l}_{i}^{\dagger} \hat{f}_{i}
     -   \hat{f}_{i}^{\dagger} \hat{l}_{i}
     \right)
     .
     \label{eq:hatbdagger-measure-operator-result}
\end{eqnarray}
We also note that
\begin{eqnarray}
  \label{eq:hattbplusdagger_is_hattbminus}
  \hat{t}_{b+}^{\dagger}
  =
  \hat{t}_{b-}
\end{eqnarray}
from the explicit expression (\ref{eq:hatb-measure-operator-result})
and (\ref{eq:hatbdagger-measure-operator-result}).

%*******************************************************************

Through the explicit expression of the operators
(\ref{eq:hatb-measure-operator-result}) and
(\ref{eq:hatbdagger-measure-operator-result}), we evaluate the
fluctuations in the measurement of the expectation values
(\ref{eq:expecation_valu_of_hatb-result}) and
(\ref{eq:expecation_valu_of_hatbdagger-result}).
Since the operators $\hat{t}_{b\pm}$ are not self-adjoint as in
Eq.~(\ref{eq:hattbplusdagger_is_hattbminus}), there is no guiding
principle to evaluate their fluctuations, in general.
In this paper, we evaluate the fluctuations through the expectation
value
\begin{eqnarray}
  \label{eq:hattbplushattbplusdagger+hattbplusdaggerhattbdagger}
  \left\langle
  \frac{1}{2} \left(
  \hat{t}_{b\pm}(\omega)\hat{t}_{b\pm}^{\dagger}(\omega')
  +
  \hat{t}_{b\pm}^{\dagger}(\omega')\hat{t}_{b\pm}(\omega)
  \right)
  \right\rangle
  .
\end{eqnarray}
The evaluation through
Eq.~(\ref{eq:hattbplushattbplusdagger+hattbplusdaggerhattbdagger}) is
commonly used to evaluate spectral
densities of gravitational-wave
detectors~\cite{H.J.Kimble-Y.Levin-A.B.Matsko-K.S.Thorne-S.P.Vyatchanin-2001,H.Miao-PhDthesis-2010}.
Furthermore, the operator $\hat{t}_{b\pm}$ has the property
(\ref{eq:hattbplusdagger_is_hattbminus}), the expectation value
Eq.~(\ref{eq:hattbplushattbplusdagger+hattbplusdaggerhattbdagger}) is
evaluated as
\begin{eqnarray}
  \left\langle
  \frac{1}{2} \left(
  \hat{t}_{b\pm}(\omega)\hat{t}_{b\pm}^{\dagger}(\omega')
  +
  \hat{t}_{b\pm}^{\dagger}(\omega')\hat{t}_{b\pm}(\omega)
  \right)
  \right\rangle
  =
  \frac{1}{2}
  \left(
  \langle\hat{t}_{b\pm}(\omega)|\hat{t}_{b\mp}(\omega')\rangle
  +
  \langle\hat{t}_{b\mp}(\omega')|\hat{t}_{b\pm}(\omega)\rangle
  \right)
  ,
  \label{eq:flucutation_evaluation-hatbmp-hatbpm}
\end{eqnarray}
where we defined
\begin{eqnarray}
  |\hat{t}_{b\pm}(\omega)\rangle
  :=
  \hat{t}_{b\pm}(\omega)|\Psi\rangle
  .
  \label{eq:hattbpm-ket}
\end{eqnarray}
Furthermore, we note that
\begin{eqnarray}
  \label{eq:hattbminushattbplusast=hattbplushattbminus-exp}
  \hat{t}_{b+}(\omega')\hat{t}_{b+}^{\dagger}(\omega)
  +
  \hat{t}_{b+}^{\dagger}(\omega)\hat{t}_{b+}(\omega')
  &=&
      \hat{t}_{b+}(\omega')\hat{t}_{b-}(\omega)
      +
      \hat{t}_{b-}(\omega)\hat{t}_{b+}(\omega')
      \nonumber\\
  &=&
      \hat{t}_{b-}^{\dagger}(\omega')\hat{t}_{b-}(\omega)
      +
      \hat{t}_{b-}(\omega)\hat{t}_{b-}^{\dagger}(\omega')
      .
\end{eqnarray}
This implies that the fluctuations in the measurement of the operator
$\hat{t}_{b-}$ through
Eq.~(\ref{eq:hattbplushattbplusdagger+hattbplusdaggerhattbdagger}) is
given by the fluctuations in the measurement of the operator
$\hat{t}_{b-}$ through
Eq.~(\ref{eq:hattbplushattbplusdagger+hattbplusdaggerhattbdagger}) and
the replacement $\omega\leftrightarrow\omega'$.
For this reason, we only show the results of the fluctuations in the
measurement of the expectation value of the operator $\hat{t}_{b+}$:
\begin{eqnarray}
%  &&
%     \!\!\!\!\!\!\!\!\!\!
     \frac{1}{2}
     \left\langle
     \hat{t}_{b+} \hat{t}_{b+}^{'\dagger}
     +
     \hat{t}_{b+}^{'\dagger} \hat{t}_{b+}
     \right\rangle
%     \nonumber\\
  &=&
      \left\langle
      \frac{1}{2}
      \left(
      \hat{b}^{'\dagger}
      \hat{b}
      +
      \hat{b}
      \hat{b}^{'\dagger}
      \right)
      \right\rangle
      \nonumber\\
  &&
     +
     \left(
     \frac{\langle\hat{n}_{b}\rangle}{|\gamma|^{2}}
     +
     \frac{1}{2}
     \right)
     2 \pi \delta(\omega-\omega')
     \nonumber\\
  &&
     +
     \frac{1-2\eta}{4\sqrt{\eta(1-\eta)}|\gamma|^{2}}
     \left\langle
     \gamma^{*}
     (\gamma^{*}+1)
     \hat{b}
     +
     \gamma
     (\gamma + 1)
     \hat{b}^{\dagger}
     \right\rangle
     2 \pi \delta(\omega-\omega')
     \nonumber\\
  &&
     +
     i
     \frac{1-2\eta}{4\sqrt{\eta(1-\eta)}|\gamma|^{2}}
     \left\langle
     \gamma^{*} (\gamma^{*} - 1)
     \hat{b}
     -
     \gamma (\gamma - 1)
     \hat{b}^{\dagger}
     \right\rangle
     2 \pi \delta(\omega-\omega')
     \nonumber\\
  &&
     +
     \frac{(1-2\eta)^{2}}{2\eta(1-\eta)}
     \left(
     1
     +
     |\gamma|^{2}
     \right)
     2 \pi \delta(\omega-\omega')
     .
  \label{eq:hattbplushattbminus+hattbminushattbplus-2}
\end{eqnarray}
Here, we note that the second line of the light-hand side of
Eq.~(\ref{eq:hattbplushattbminus+hattbminushattbplus-2}) comes from
the shot noise contribution of the additional input vacua in the
interferometer and the remaining lines of the right-hand side is  due to
the imperfection of the beam splitters BS2 and BS4 from 50:50.

%*******************************************************************

Since we are concentrating on the case where the operators
$\hat{t}_{b\pm}$, $\hat{b}$, and $\hat{b}^{\dagger}$ have the
nontrivial expectation values
$\langle\hat{t}_{b+}\rangle=\langle\hat{b}\rangle$ and
$\langle\hat{t}_{b-}\rangle=\langle\hat{b}^{\dagger}\rangle$,
Eq.~(\ref{eq:hattbplushattbminus+hattbminushattbplus-2}) includes not
only the information of the fluctuations but also the information of
the correlation of these expectation values.
Therefore, to consider the fluctuations in the measurements of
$\hat{t}_{b\pm}$, we have to eliminate the information of the
correlation of the expectation values
$\langle\hat{t}_{b+}\rangle=\langle\hat{b}\rangle$ and
$\langle\hat{t}_{b-}\rangle=\langle\hat{b}^{\dagger}\rangle$ from Eq.~(\ref{eq:hattbplushattbminus+hattbminushattbplus-2}).
Actually, if we define the noise operator $\hat{t}_{b\pm}^{(n)}$ and
$\hat{b}^{(n)}$ so that
\begin{eqnarray}
  \label{eq:hattbplus-noise-operators}
  \hat{t}_{b+}
  &=:&
       \langle\hat{b}\rangle + \hat{t}_{b+}^{(n)},
       \quad
       \langle\hat{t}_{b+}^{(n)}\rangle=0,
  \\
  \label{eq:hattbminus-noise-operators}
  \hat{t}_{b-}
  &=:&
       \langle\hat{b}^{\dagger}\rangle + \hat{t}_{b-}^{(n)},
       \quad
       \langle\hat{t}_{b-}^{(n)}\rangle=0,
  \\
  \label{eq:hatb-noise-operators}
  \hat{b}
  &=:&
       \langle\hat{b}\rangle + \hat{b}^{(n)},
       \quad
       \langle\hat{b}^{(n)}\rangle=0
       ,
\end{eqnarray}
the left-hand side of
Eq.~(\ref{eq:hattbplushattbminus+hattbminushattbplus-2}) and the first
term in the right-hand side of
Eq.~(\ref{eq:hattbplushattbminus+hattbminushattbplus-2}) yield
\begin{eqnarray}
  \label{eq:hattbplushattbminus+hattbminushattbplus-noise}
  \frac{1}{2} \left\langle
  \hat{t}_{b+}\hat{t}_{b+}^{'\dagger}
  +
  \hat{t}_{b+}^{'\dagger}\hat{t}_{b+}
  \right\rangle
  &=&
      \frac{1}{2} \left\langle
      \hat{t}_{b+}^{(n)}
      \hat{t}_{b+}^{(n)'\dagger}
      +
      \hat{t}_{b+}^{(n)'\dagger}
      \hat{t}_{b+}^{(n)}
      \right\rangle
      +
      \langle\hat{b}'\rangle^{*}
      \langle\hat{b}\rangle
      ,
  \\
  \label{eq:hatbhatbdagger+hatbdaggerhatb-noise}
  \frac{1}{2} \left\langle
  \hat{b}\hat{b}^{'\dagger}
  +
  \hat{b}^{'\dagger}\hat{b}
  \right\rangle
  &=&
      \frac{1}{2} \left\langle
      \hat{b}^{(n)}
      \hat{b}^{(n)'\dagger}
      +
      \hat{b}^{(n)'\dagger}
      \hat{b}^{(n)}
      \right\rangle
      +
      \langle\hat{b}'\rangle^{*}
      \langle\hat{b}\rangle
      .
\end{eqnarray}
In the right-hand side of
Eqs.~(\ref{eq:hattbplushattbminus+hattbminushattbplus-noise}) and
(\ref{eq:hatbhatbdagger+hatbdaggerhatb-noise}), the first terms
corresponds to the noise correlation in the measurements of operators
$\hat{b}$ and $\hat{b}^{\dagger}$.

%*******************************************************************

For the quantum operator $\hat{Q}$ with its expectation value
$\langle\hat{Q}\rangle=0$, we introduce the noise spectral density as
in
Refs.~\cite{H.J.Kimble-Y.Levin-A.B.Matsko-K.S.Thorne-S.P.Vyatchanin-2001,H.Miao-PhDthesis-2010}
by
\begin{eqnarray}
  \label{eq:spectral_density_def}
  \frac{1}{2} S_{\hat{Q}}(\omega) 2 \pi \delta(\omega-\omega')
  :=
  \frac{1}{2} \left(
  \hat{Q}(\omega)\hat{Q}^{\dagger}(\omega')
  +
  \hat{Q}^{\dagger}(\omega')\hat{Q}(\omega)
  \right)
  .
\end{eqnarray}
Then, Eq.~(\ref{eq:hattbplushattbminus+hattbminushattbplus-2}) yields
the relation between the noise spectral densities as
\begin{eqnarray}
  S_{\hat{t}_{b+}^{(n)}}(\omega)
  &=&
      S_{\hat{b}^{(n)}}(\omega)
      +
      \frac{2\langle\hat{n}_{b}\rangle}{|\gamma|^{2}}
      +
      1
     \nonumber\\
  &&
     +
     \frac{1-2\eta}{2\sqrt{\eta(1-\eta)}|\gamma|^{2}}
     \left\langle
     \gamma^{*}
     (\gamma^{*}+1)
     \hat{b}
     +
     \gamma
     (\gamma + 1)
     \hat{b}^{\dagger}
     \right\rangle
     \nonumber\\
  &&
     +
     i
     \frac{1-2\eta}{2\sqrt{\eta(1-\eta)}|\gamma|^{2}}
     \left\langle
     \gamma^{*} (\gamma^{*} - 1)
     \hat{b}
     -
     \gamma (\gamma - 1)
     \hat{b}^{\dagger}
     \right\rangle
     \nonumber\\
  &&
     +
     \frac{(1-2\eta)^{2}}{\eta(1-\eta)}
     \left(
     1
     +
     |\gamma|^{2}
     \right)
     .
  \label{eq:spectral-density-relation-in-DBHD}
\end{eqnarray}

%*******************************************************************

Equation (\ref{eq:spectral-density-relation-in-DBHD}) indicates that
the imperfection $\eta$ of the beam splitter gives additional noise.
On the other hand, in the case where the beam splitters BS2 and BS4 is
50:50, i.e., $\eta=1/2$, equation
(\ref{eq:spectral-density-relation-in-DBHD}) gives
\begin{eqnarray}
  S_{\hat{t}_{b+}^{(n)}}(\omega)
  &=&
      S_{\hat{b}^{(n)}}(\omega)
      +
      \frac{2}{|\gamma|^{2}}
      \langle
      \hat{n}_{b}
      \rangle
      +
      1
     .
  \label{eq:spectral-density-relation-in-DBHD-eta=1/2}
\end{eqnarray}
This is the result which was shown in
Ref.~\cite{K.Nakamura-M.-K.Fujimoto-2017b} and the similar result is
previously reported~\cite{E.Shchukin-Th.Richter-W.Vogel-2005} in the
context of the characterization of the non-classicality for a quantum system.
Equation (\ref{eq:spectral-density-relation-in-DBHD-eta=1/2})
indicates that in addition to the noise spectral density
$S_{\hat{b}^{(n)}}(\omega)$, we have the additional fluctuations which
described by $2\langle\hat{n}_{b}\rangle/|\gamma^{2}|+1$ in the double
balanced homodyne detection to measure $\hat{b}$ through the
measurement of $\hat{t}_{b+}$.
We note that the term $2\langle\hat{n}_{b}\rangle/|\gamma^{2}|$ will be
negligible if the absolute value of the complex amplitude $\gamma$ is
much larger than the expectation value of the output photon number
$\langle\hat{n}_{b}\rangle$, i.e., $\langle\hat{n}_{b}\rangle \ll |\gamma^{2}|$.
On the other hand, the term $1$ in
$2\langle\hat{n}_{b}\rangle/|\gamma^{2}|+1$, which comes from the shot
noise from the additional input vacuum fields in the interferometer
for the double balanced homodyne detection, is not controllable.
Furthermore, the comparison of
Eqs.~(\ref{eq:spectral-density-relation-in-DBHD}) and
(\ref{eq:spectral-density-relation-in-DBHD-eta=1/2}) indicates that
the the behavior of the noise spectrum density
$S_{\hat{t}_{b+}^{(n)}}(\omega)$ is sensitive to the transmittance
$\eta$.
In particular, the last term in
Eq.~(\ref{eq:spectral-density-relation-in-DBHD}) may become dominant
when $|\gamma|\gg 1$.

%*******************************************************************

%%%%%%%%%%%%%%%%%%%%%%%%%%%%%%%%%%%%%%%%%%%%%%%%%%%%%%%
\subsection{Double balanced homodyne detection in two-photon description}
\label{sec:DBHD_in_two_photon_formulation}
%%%%%%%%%%%%%%%%%%%%%%%%%%%%%%%%%%%%%%%%%%%%%%%%%%%%%%%

%*******************************************************************

Here, we consider the two-photon description and show that we can
calculate the expectation value of the operator $\hat{b}_{\theta}$
through the interferometer setup depicted in
Fig.~\ref{fig:DoubleBalancedHomodyneDetection-configuration}.

%*******************************************************************

As discussed in Sec.~\ref{sec:BHD-two-photon}, we consider the
sideband with the frequencies $\omega_{0}\pm\Omega$.
The output through the balanced homodyne detection using D1 and D2
has the information of the operators
$\hat{s}_{D1D2}(\omega_{0}\pm\Omega)=\hat{s}_{D1D2\pm}$ defined by
\begin{eqnarray}
  \hat{s}_{D1D2\pm}
  &=&
      \hat{b}_{\pm} \hat{l}_{i\pm}^{\dagger}
      + \hat{b}_{\pm}^{\dagger} \hat{l}_{i\pm}
      +
      \frac{1-2\eta}{\sqrt{\eta(1-\eta)}}
      \left(
      \hat{l}_{i\pm}^{\dagger} \hat{l}_{i\pm}
      -
      |\gamma_{\pm}|^{2}
      \right)
      \nonumber\\
  &&
     - \hat{b}_{\pm} \hat{f}_{i\pm}^{\dagger}
     - \hat{b}_{\pm}^{\dagger} \hat{f}_{i\pm}
     + \hat{e}_{i\pm} \hat{f}_{i\pm}^{\dagger}
     + \hat{e}_{i\pm}^{\dagger} \hat{f}_{i\pm}
     - \hat{e}_{i\pm} \hat{l}_{i\pm}^{\dagger}
     - \hat{e}_{i\pm}^{\dagger} \hat{l}_{i\pm}
     \nonumber\\
  &&
     +
     \frac{1-2\eta}{\sqrt{\eta(1-\eta)}}
     \left(
     \hat{f}_{i\pm}^{\dagger} \hat{f}_{i\pm}
     - \hat{f}_{i\pm}^{\dagger} \hat{l}_{i\pm}
     - \hat{l}_{i\pm}^{\dagger} \hat{f}_{i\pm}
     \right)
     \label{eq:hatsD1D2-two-photon-upper-lower-defs}
\end{eqnarray}
from Eq.~(\ref{eq:hats-D1D2-operator-expression-2}).
Following the discussion in
Sec.~\ref{sec:exp_val_costhetab1+sinthetab2}, we introduce the
operators $\hat{t}_{D1D2+}$ defined by
\begin{eqnarray}
  \hat{t}_{D1D2+}
  :=
  \frac{1}{\sqrt{2}}
  \left(
    \frac{
      \hat{s}_{D1D2+}
    }{
      |\gamma_{+}|
    }
    +
    \frac{
      \hat{s}_{D1D2-}
    }{
      |\gamma_{-}|
    }
  \right)
  ,
  \label{eq:hattD1D2plus-two-photon-def}
\end{eqnarray}
where the electric field assoicated with the quadratures
$\hat{l}_{i\pm}$ are in the coherent state with the complex amplitude
$\gamma_{\pm}$ and the phase $\theta_{\pm}$ of $\gamma_{\pm}$ is
chosen so that $\theta_{\pm}$ $=$ $\theta$ and $|\gamma_{+}|$ $=$
$|\gamma_{-}|$  $=:$  $|\gamma|$.
This choice of the phase is essential for the measurement of the expectation value
$\langle\hat{b}_{\theta}\rangle$.
Under this situation, we evaluate the operator $\hat{t}_{D1D2+}$
defined by Eqs.~(\ref{eq:hattD1D2plus-two-photon-def}).
Substituting Eqs.~(\ref{eq:hatsD1D2-two-photon-upper-lower-defs}) into
Eq.~(\ref{eq:hattD1D2plus-two-photon-def}) and using the definitions
(\ref{eq:hatb1-hatb2-def}) of the operators $\hat{b}_{1}$ and
$\hat{b}_{2}$, we obtain
\begin{eqnarray}
  \hat{t}_{D1D2+}
  &=&
      \frac{1}{2|\gamma|}
      \left(
      \hat{l}_{i-}
      +
      \hat{l}_{i+}^{\dagger}
      \right)
      \hat{b}_{1}
      +
      \frac{1}{2i|\gamma|}
      \left(
      \hat{l}_{i-}
      -
      \hat{l}_{i+}^{\dagger}
      \right)
      \hat{b}_{2}
      \nonumber\\
  &&
      +
      \frac{1}{2|\gamma|}
      \left(
      \hat{l}_{i+}
      +
      \hat{l}_{i-}^{\dagger}
      \right)
      \hat{b}_{1}^{\dagger}
      +
      \frac{1}{2i|\gamma|}
      \left(
      \hat{l}_{i+}
      -
      \hat{l}_{i-}^{\dagger}
      \right)
      \hat{b}_{2}^{\dagger}
      \nonumber\\
  &&
     +
     \frac{1-2\eta}{\sqrt{2\eta(1-\eta)}|\gamma|}
     \left(
     \hat{l}_{i+}^{\dagger}
     \hat{l}_{i+}
     +
     \hat{l}_{i-}^{\dagger}
     \hat{l}_{i-}
     -
     2
     |\gamma|^{2}
     \right)
     \nonumber\\
  &&
     +
     \frac{1}{\sqrt{2}|\gamma|}
     \left(
     - \hat{b}_{+} \hat{f}_{i+}^{\dagger}
     - \hat{b}_{+}^{\dagger} \hat{f}_{i+}
     + \hat{e}_{i+} \hat{f}_{i+}^{\dagger}
     + \hat{e}_{i+}^{\dagger} \hat{f}_{i+}
     - \hat{e}_{i+} \hat{l}_{i+}^{\dagger}
     - \hat{e}_{i+}^{\dagger} \hat{l}_{i+}
     \right)
     \nonumber\\
  &&
     +
     \frac{1}{\sqrt{2}|\gamma|}
     \left(
     - \hat{b}_{-} \hat{f}_{i-}^{\dagger}
     - \hat{b}_{-}^{\dagger} \hat{f}_{i-}
     + \hat{e}_{i-} \hat{f}_{i-}^{\dagger}
     + \hat{e}_{i-}^{\dagger} \hat{f}_{i-}
     - \hat{e}_{i-} \hat{l}_{i-}^{\dagger}
     - \hat{e}_{i-}^{\dagger} \hat{l}_{i-}
     \right)
     \nonumber\\
  &&
     +
     \frac{1-2\eta}{\sqrt{2\eta(1-\eta)}}
     \frac{1}{|\gamma|}
     \left(
     \hat{f}_{i+}^{\dagger} \hat{f}_{i+}
     - \hat{f}_{i+}^{\dagger} \hat{l}_{i+}
     - \hat{l}_{i+}^{\dagger} \hat{f}_{i+}
     \right.
     \nonumber\\
  && \quad\quad\quad\quad\quad\quad\quad\quad\quad
     \left.
     + \hat{f}_{i-}^{\dagger} \hat{f}_{i-}
     - \hat{f}_{i-}^{\dagger} \hat{l}_{i-}
     - \hat{l}_{i-}^{\dagger} \hat{f}_{i-}
     \right)
     .
     \label{eq:hattD1D2plus-resultsofDBHD}
\end{eqnarray}
The expectation values of the operator $\hat{t}_{D1D2+}$ is
derived from the expectation values of the operator $\hat{s}_{D1D2\pm}$,
which are given by the definition
(\ref{eq:hats-D1D2-operator-original-def}) and
Eq.~(\ref{eq:BHD-D1-D2-result}).
From these expectation values, we can check that the expectation value
of the operator (\ref{eq:hattD1D2plus-resultsofDBHD}) is given by
\begin{eqnarray}
  \left\langle
  \hat{t}_{D1D2+}
  \right\rangle
  =
  \left\langle
  \cos\theta
  \hat{b}_{1}
  +
  \sin\theta
  \hat{b}_{2}
  +
  \cos\theta
  \hat{b}_{1}^{\dagger}
  +
  \sin\theta
  \hat{b}_{2}^{\dagger}
  \right\rangle
  \label{eq:hattD1D2plus-resultsofDBHD-exp-value}
\end{eqnarray}

%*******************************************************************

On the other hand, the output through the balanced homodyne detection
using D3 and D4 has the information of the operators
$\hat{s}_{D3D4}(\omega_{0}\pm\Omega)=\hat{s}_{D3D4\pm}$ as
\begin{eqnarray}
  \hat{s}_{D3D4\pm}
  &=&
      \hat{l}_{i\pm}^{\dagger}
      \hat{b}_{\pm}
      -
      \hat{b}_{\pm}^{\dagger}
      \hat{l}_{i\pm}
      +
      \frac{i(1-2\eta)}{\sqrt{\eta(1-\eta)}}
      \left(
      \hat{l}_{i\pm}^{\dagger}
      \hat{l}_{i\pm}
      - |\gamma_{\pm}|^{2}
      \right)
      \nonumber\\
  &&
     - \hat{b}_{\pm}^{\dagger} \hat{f}_{i\pm}
     + \hat{b}_{\pm} \hat{f}_{i\pm}^{\dagger}
     + \hat{l}_{i\pm}^{\dagger} \hat{e}_{i\pm}
     + \hat{f}_{i\pm}^{\dagger} \hat{e}_{i\pm}
     - \hat{e}_{i\pm}^{\dagger} \hat{l}_{i\pm}
     - \hat{e}_{i\pm}^{\dagger} \hat{f}_{i\pm}
     \nonumber\\
  &&
     +
     \frac{i(1-2\eta)}{\sqrt{\eta(1-\eta)}}
     \left(
     \hat{f}_{i\pm}^{\dagger}
     \hat{l}_{i\pm}
     +
     \hat{l}_{i\pm}^{\dagger}
     \hat{f}_{i\pm}
     +
     \hat{f}_{i\pm}^{\dagger}
     \hat{f}_{i\pm}
     \right)
     .
  \label{eq:hatsD3D4-two-photon-upper-lower-defs}
\end{eqnarray}
from Eq.~(\ref{eq:hats-D3D4-operator-expression-2}).
Following the discussion in
Sec.~\ref{sec:exp_val_costhetab1+sinthetab2}, we introduce the
operator $\hat{t}_{D3D4-}$ defined by
\begin{eqnarray}
  \label{eq:hattD3D4muinus-def-2}
  \hat{t}_{D3D4-}
  :=
  \frac{1}{\sqrt{2}}
  \left(
    \frac{
      \hat{s}_{D3D4+}
    }{
      |\gamma_{+}|
    }
    -
    \frac{
      \hat{s}_{D3D4-}
    }{
      |\gamma_{-}|
    }
  \right)
  =
  \frac{1}{\sqrt{2}|\gamma|}
  \left(
  \hat{s}_{D3D4+}
  -
  \hat{s}_{D3D4-}
  \right)
  .
\end{eqnarray}
Here, we used our situation where $\theta_{\pm}=\theta$ and
$|\gamma_{\pm}|=|\gamma|$.
Substituting Eqs.~(\ref{eq:hatsD3D4-two-photon-upper-lower-defs}) into
Eq.~(\ref{eq:hattD3D4muinus-def-2}) and using the definitions
(\ref{eq:hatb1-hatb2-def}) of the operators $\hat{b}_{1}$ and
$\hat{b}_{2}$, we obtain
\begin{eqnarray}
  \hat{t}_{D3D4-}
  &=&
      \frac{1}{2|\gamma|}
      \left(
      \hat{l}_{i-}
      +
      \hat{l}_{i+}^{\dagger}
      \right)
      \hat{b}_{1}
      +
      \frac{1}{2i|\gamma|}
      \left(
      \hat{l}_{i-}
      -
      \hat{l}_{i+}^{\dagger}
      \right)
      \hat{b}_{2}
      \nonumber\\
  &&
      -
      \frac{1}{2|\gamma|}
      \left(
      \hat{l}_{i+}
      +
      \hat{l}_{i-}^{\dagger}
      \right)
      \hat{b}_{1}^{\dagger}
      -
      \frac{1}{2i|\gamma|}
      \left(
      \hat{l}_{i+}
      -
      \hat{l}_{i-}^{\dagger}
      \right)
      \hat{b}_{2}^{\dagger}
      \nonumber\\
  &&
     +
     \frac{i(1-2\eta)}{|\gamma|\sqrt{2\eta(1-\eta)}}
     \left(
     + \hat{l}_{i+}^{\dagger} \hat{l}_{i+}
     - \hat{l}_{i-}^{\dagger} \hat{l}_{i-}
     \right)
     \nonumber\\
  &&
     +
     \frac{1}{\sqrt{2}|\gamma|}
     \left(
     - \hat{b}_{+}^{\dagger} \hat{f}_{i+}
     + \hat{b}_{+} \hat{f}_{i+}^{\dagger}
     + \hat{l}_{i+}^{\dagger} \hat{e}_{i+}
     + \hat{f}_{i+}^{\dagger} \hat{e}_{i+}
     - \hat{e}_{i+}^{\dagger} \hat{l}_{i+}
     - \hat{e}_{i+}^{\dagger} \hat{f}_{i+}
     \right)
     \nonumber\\
  &&
     +
     \frac{1}{\sqrt{2}|\gamma|}
     \left(
     + \hat{b}_{-}^{\dagger} \hat{f}_{i-}
     - \hat{b}_{-} \hat{f}_{i-}^{\dagger}
     - \hat{l}_{i-}^{\dagger} \hat{e}_{i-}
     - \hat{f}_{i-}^{\dagger} \hat{e}_{i-}
     + \hat{e}_{i-}^{\dagger} \hat{l}_{i-}
     + \hat{e}_{i-}^{\dagger} \hat{f}_{i-}
     \right)
  \nonumber\\
  && 
  +
  \frac{i(1-2\eta)}{|\gamma|\sqrt{2\eta(1-\eta)}}
  \left(
    + \hat{f}_{i+}^{\dagger} \hat{l}_{i+}
    + \hat{l}_{i+}^{\dagger} \hat{f}_{i+}
    + \hat{f}_{i+}^{\dagger} \hat{f}_{i+}
     \right.
     \nonumber\\
  && \quad\quad\quad\quad\quad\quad\quad\quad\quad
     \left.
    - \hat{f}_{i-}^{\dagger} \hat{l}_{i-}
    - \hat{l}_{i-}^{\dagger} \hat{f}_{i-}
    - \hat{f}_{i-}^{\dagger} \hat{f}_{i-}
  \right)
  .
  \label{eq:hattD3D4minus-resultsofDBHD}
\end{eqnarray}
As in the case of the operator $\hat{t}_{D1D2+}$, the expectation
values of the operators $\hat{s}_{D3D4\pm}$ are given by the
definition (\ref{eq:hats-D3D4-operator-original-def}) and
Eq.~(\ref{eq:BHD-D3-D4-result}).
From these expectation values, the expectation values of the operator
$\hat{t}_{D3D4-}$ defined by (\ref{eq:hattD3D4minus-resultsofDBHD}) is
given by
\begin{eqnarray}
  \left\langle
  \hat{t}_{D3D4-}
  \right\rangle
  =
  \left\langle
  \cos\theta
  \hat{b}_{1}
  +
  \sin\theta
  \hat{b}_{2}
  -
  \cos\theta
  \hat{b}_{1}^{\dagger}
  -
  \sin\theta
  \hat{b}_{2}^{\dagger}
  \right\rangle
  .
  \label{eq:hattD3D4minus-resultsofDBHD-exp-value}
\end{eqnarray}

%*******************************************************************

From the expectation values
(\ref{eq:hattD1D2plus-resultsofDBHD-exp-value}) and
(\ref{eq:hattD3D4minus-resultsofDBHD-exp-value}), we obtain the
expected result
\begin{eqnarray}
  \left\langle
  \frac{1}{2}
  \left(
  \hat{t}_{D1D2+}
  +
  \hat{t}_{D3D4-}
  \right)
  \right\rangle
  =
  \left\langle
  \cos\theta
  \hat{b}_{1}
  +
  \sin\theta
  \hat{b}_{2}
  \right\rangle
  \label{eq:halfofhattD1D2plus+hattD3D4-result}
\end{eqnarray}
Thus, we have derived that the operator whose expectation value yields
$\langle\cos\theta\hat{b}_{1}+\sin\theta\hat{b}_{2}\rangle$ is given
by
\begin{eqnarray}
  \hat{t}_{\theta}
  &:=&
       \frac{1}{2}
       \left(
       \hat{t}_{D1D2+}
       +
       \hat{t}_{D3D4-}
       \right)
       \nonumber\\
  &=&
      \frac{1}{2\sqrt{2}|\gamma|}
      \left(
      \hat{s}_{D1D2+}
      +
      \hat{s}_{D1D2-}
      +
      \hat{s}_{D3D4+}
      -
      \hat{s}_{D3D4-}
      \right)
      \nonumber\\
  &=&
      \frac{1}{2|\gamma|}
      \left(
      \hat{l}_{i-}
      +
      \hat{l}_{i+}^{\dagger}
      \right)
      \hat{b}_{1}
      +
      \frac{1}{2i|\gamma|}
      \left(
      \hat{l}_{i-}
      -
      \hat{l}_{i+}^{\dagger}
      \right)
      \hat{b}_{2}
      \nonumber\\
  &&
     +
     \frac{1-2\eta}{2\sqrt{2\eta(1-\eta)}|\gamma|}
     \left\{
     (1+i)
     \hat{l}_{i+}^{\dagger}
     \hat{l}_{i+}
     +
     (1-i)
     \hat{l}_{i-}^{\dagger}
     \hat{l}_{i-}
     -
     2
     |\gamma|^{2}
     \right\}
     \nonumber\\
  &&
     +
     \frac{1}{\sqrt{2}|\gamma|}
     \left\{
     - \hat{b}_{+}^{\dagger} \hat{f}_{i+}
     + \hat{e}_{i+} \hat{f}_{i+}^{\dagger}
     - \hat{e}_{i+}^{\dagger} \hat{l}_{i+}
     - \hat{b}_{-} \hat{f}_{i-}^{\dagger}
     - \hat{e}_{i-} \hat{l}_{i-}^{\dagger}
     + \hat{e}_{i-}^{\dagger} \hat{f}_{i-}
     \right\}
     \nonumber\\
  &&
     +
     \frac{(1+i)(1-2\eta)}{2\sqrt{2\eta(1-\eta)}|\gamma|}
     \left\{
     \hat{f}_{i+}^{\dagger} \hat{f}_{i+}
     + i \hat{f}_{i+}^{\dagger} \hat{l}_{i+}
     + i \hat{l}_{i+}^{\dagger} \hat{f}_{i+}
     - i \hat{f}_{i-}^{\dagger} \hat{f}_{i-}
     \right.
     \nonumber\\
  && \quad\quad\quad\quad\quad\quad\quad\quad\quad
     \left.
     -   \hat{f}_{i-}^{\dagger} \hat{l}_{i-}
     -   \hat{l}_{i-}^{\dagger} \hat{f}_{i-}
     \right\}
     .
     \label{eq:hatttheta-def}
\end{eqnarray}

%*******************************************************************

As in the case of the measurement of $\langle\hat{b}\rangle$ or
$\langle\hat{b}^{\dagger}\rangle$ in
Sec.~\ref{sec:measurement_of-hatb-or-hatbdagger}, we evaluate the
fluctuations in the measurement of the expectation value
(\ref{eq:halfofhattD1D2plus+hattD3D4-result}) through the explicit
expression of the operators (\ref{eq:hatttheta-def}).
We evaluate of the fluctuations of the measurement of the operator
$\hat{t}_{\theta}$ through the expectation value
$\left\langle\left(\hat{t}_{\theta}\hat{t}_{\theta}^{'\dagger} + \hat{t}_{\theta}^{'\dagger}\hat{t}_{\theta}\right)/2\right\rangle$
as in the case of Sec.~\ref{sec:measurement_of-hatb-or-hatbdagger},
where we introduced the notation $\hat{Q}' := \hat{Q}(\Omega')$ for
the frequency-dependent operator $\hat{Q}=\hat{Q}(\Omega)$.
Tedious but straightforward calculations leads to the result
\begin{eqnarray}
%  &&
%     \!\!\!\!\!\!\!\!\!\!
     \left\langle
     \frac{1}{2}
     \left(
     \hat{t}_{\theta}\hat{t}_{\theta}^{'\dagger}
     +
     \hat{t}_{\theta}^{'\dagger}\hat{t}_{\theta}
     \right)
     \right\rangle
%     \nonumber\\
  &=&
      \left\langle
      \frac{1}{2}
      \left(
      \hat{b}_{\theta}
      \hat{b}_{\theta}^{'\dagger}
      +
      \hat{b}_{\theta}^{'\dagger}
      \hat{b}_{\theta}
      \right)
      \right\rangle
%      \nonumber\\
%  &&
     +
     \frac{1}{2}
     \left(
     1
     +
     \frac{\langle\hat{n}_{b-}+\hat{n}_{b+}\rangle}{|\gamma|^{2}}
     \right)
     2 \pi \delta(\Omega-\Omega')
     \nonumber\\
  &&
     +
     \frac{1-2\eta}{2\sqrt{2\eta(1-\eta)}|\gamma|}
     \left\{
     \left( \cos\theta - \sin\theta \right)
     \langle \hat{b}_{1} + \hat{b}_{1}^{\dagger} \rangle
     \right.
     \nonumber\\
  && \quad\quad\quad\quad\quad\quad\quad\quad\quad
     \left.
     +
     \left( \cos\theta + \sin\theta \right)
     \langle \hat{b}_{2} + \hat{b}_{2}^{\dagger} \rangle
     \right\}
     2 \pi \delta(\Omega-\Omega')
     \nonumber\\
  &&
     +
     \frac{(1-2\eta)^{2}}{\eta(1-\eta)}
     2 \pi \delta(\Omega-\Omega')
     .
  \label{eq:hattthetahattthetadagger+hattthetadaggerhatttheta-result}
\end{eqnarray}
The second term in the first line of the right-hand side of
Eq.~(\ref{eq:hattthetahattthetadagger+hattthetadaggerhatttheta-result})
comes from the shot noise contribution of the additional input vacua
in the interferometer of the double balanced homodyne detection and
the second-, third-, and fourth-lines of
Eq.~(\ref{eq:hattthetahattthetadagger+hattthetadaggerhatttheta-result})
is due to the imperfection of the beam splitters BS2 and BS4 from
50:50 as in the case of Sec.~\ref{sec:measurement_of-hatb-or-hatbdagger}.

%*******************************************************************

The left-hand side and the first term in the right-hand side of
Eq.~(\ref{eq:hattthetahattthetadagger+hattthetadaggerhatttheta-result})
includes not only the information of the fluctuations in the
measurement of the operator $\hat{t}_{\theta}$ but also the
information of the expectation value
$\langle\hat{t}_{\theta}\rangle=\langle\hat{b}_{\theta}\rangle$.
Therefore, we  have to eliminate the information of the expectation
value of the operator $\hat{t}_{\theta}$ from
Eq.~(\ref{eq:hattthetahattthetadagger+hattthetadaggerhatttheta-result}).
To carry out this elimination, as in the case in
Sec.~\ref{sec:measurement_of-hatb-or-hatbdagger}, we introduce the
noise operators $\hat{t}_{\theta}^{(n)}$ and $\hat{b}_{\theta}^{(n)}$
by
\begin{eqnarray}
  \label{eq:hatttheta_noise_def}
  \hat{t}_{\theta}
  &=:&
      \langle\hat{b}_{\theta}\rangle + \hat{t}_{\theta}^{(n)}
      , \quad
       \langle\hat{t}_{\theta}^{(n)}\rangle = 0,
  \\
  \label{eq:hatbtheta_noise_def}
  \hat{b}_{\theta}
  &=:&
      \langle\hat{b}_{\theta}\rangle + \hat{b}_{\theta}^{(n)}
      , \quad
      \langle\hat{b}_{\theta}^{(n)}\rangle = 0.
\end{eqnarray}
In terms of these noise operators $\hat{t}_{\theta}^{(n)}$ and
$\hat{b}_{\theta}^{(n)}$, the left-hand side of
Eq.~(\ref{eq:hattthetahattthetadagger+hattthetadaggerhatttheta-result})
is given by
\begin{eqnarray}
  \label{eq:hattthetahattthetadagger+hattthetadaggerhatttheta-left}
  \left\langle
  \frac{1}{2} \left(
  \hat{t}_{\theta} \hat{t}_{\theta}^{'\dagger}
  +
  \hat{t}_{\theta}^{'\dagger} \hat{t}_{\theta}
  \right)
  \right\rangle
  =
  \frac{1}{2}
  \left\langle
  \hat{t}_{\theta}^{(n)}
  \hat{t}_{\theta}^{(n)'\dagger}
  +
  \hat{t}_{\theta}^{(n)'\dagger}
  \hat{t}_{\theta}^{(n)}
  \right\rangle
  +
  \langle\hat{b}_{\theta}\rangle
  \langle\hat{b}_{\theta}'\rangle^{*}
  .
\end{eqnarray}
Similarly, the first term in the right-hand side of
Eq.~(\ref{eq:hattthetahattthetadagger+hattthetadaggerhatttheta-result})
is also given by
\begin{eqnarray}
  \label{eq:hattthetahattthetadagger+hattthetadaggerhatttheta-right-first}
  \frac{1}{2}
  \left\langle
  \hat{b}_{\theta} \hat{b}_{\theta}^{'\dagger}
  +
  \hat{b}_{\theta}^{'\dagger} \hat{b}_{\theta}
  \right\rangle
  &=&
      \frac{1}{2}
      \left\langle
      \hat{b}_{\theta}^{(n)}
      \hat{b}_{\theta}^{(n)'\dagger}
      +
      \hat{b}_{\theta}^{(n)}
      \hat{b}_{\theta}^{(n)'\dagger}
      \right\rangle
      +
      \langle\hat{b}_{\theta}\rangle
      \langle\hat{b}_{\theta}'\rangle^{*}
      .
\end{eqnarray}
In
Eqs.~(\ref{eq:hattthetahattthetadagger+hattthetadaggerhatttheta-left})
and
(\ref{eq:hattthetahattthetadagger+hattthetadaggerhatttheta-right-first}),
the first terms are corresponds to the noise correlation and the
second term is the correlation of the expectation values.

%*******************************************************************

Here, following Eq.~(\ref{eq:spectral_density_def}), we introduce the
noise spectral densities for the noise operators
$\hat{t}_{\theta}^{(n)}$ and $\hat{b}_{\theta}^{(n)}$.
Then, in terms of the noise spectral densities,
Eq.~(\ref{eq:hattthetahattthetadagger+hattthetadaggerhatttheta-result})
is given by
\begin{eqnarray}
  S_{\hat{t}_{\theta}^{(n)}}(\Omega)
  &=&
      S_{\hat{b}_{\theta}^{(n)}}(\Omega)
      +
      \frac{\langle\hat{n}_{b-}+\hat{n}_{b+}\rangle}{|\gamma|^{2}}
      +
      1
      \nonumber\\
  &&
     +
     \frac{1-2\eta}{\sqrt{2\eta(1-\eta)}|\gamma|}
     \left\{
     \left( \cos\theta - \sin\theta \right)
     \langle \hat{b}_{1} + \hat{b}_{1}^{\dagger} \rangle
     +
     \left( \cos\theta + \sin\theta \right)
     \langle \hat{b}_{2} + \hat{b}_{2}^{\dagger} \rangle
     \right\}
     \nonumber\\
  &&
     +
     \frac{(1-2\eta)^{2}}{2\eta(1-\eta)}
     .
  \label{eq:hatttheta-hatbtheta-spectral-density-relation}
\end{eqnarray}

%*******************************************************************

In the case where the beam splitters BS2 and BS4 is 50:50, i.e.,
$\eta=1/2$, equation (\ref{eq:spectral-density-relation-in-DBHD})
gives
\begin{eqnarray}
  S_{\hat{t}_{\theta}^{(n)}}(\Omega)
  =
  S_{\hat{b}_{\theta}^{(n)}}(\Omega)
  +
  \frac{\langle\hat{n}_{b-}+\hat{n}_{b+}\rangle}{|\gamma|^{2}}
  +
  1
  .
  \label{eq:hatttheta-hatbtheta-spectral-density-relation-eta=1/2}
\end{eqnarray}
As Eq.~(\ref{eq:spectral-density-relation-in-DBHD-eta=1/2}), equation
(\ref{eq:hatttheta-hatbtheta-spectral-density-relation-eta=1/2})
indicates that in addition to the spectral density
$S_{\hat{b}_{\theta}}(\omega)$, we have the additional fluctuations which
described by $\langle\hat{n}_{b+}+\hat{n}_{b-}\rangle/|\gamma^{2}|+1$
in the double balanced homodyne detection to measure the expectation
value of the operator $\hat{b}_{\theta}$ through the
measurement of $\hat{t}_{\theta}$.
We note that the term
$\langle\hat{n}_{b+}+\hat{n}_{b-}\rangle/|\gamma^{2}|$ will be
negligible if the absolute value of the complex amplitude $\gamma$ is
much larger than the expectation value of the total output photon
number $\langle\hat{n}_{b+}+\hat{n}_{b-}\rangle$ from the main
interferometer as in the case of
Sec.~\ref{sec:measurement_of-hatb-or-hatbdagger}.
On the other hand, the term $1$ in
$2\langle\hat{n}_{b+}+\hat{n}_{b-}\rangle/|\gamma^{2}|+1$, which is
comes from the shot noise from the additional input vacuum fields in the
interferometer for the double balanced homodyne detection, is not
controllable.
This situation is same as that in
Sec.~\ref{sec:measurement_of-hatb-or-hatbdagger}.

%*******************************************************************

%%%%%%%%%%%%%%%%%%%%%%%%%%%%%%%%%%%%%%%%%%%%%%%%%%%%%%%
%%%%%%%%%%%%%%%%%%%%%%%%%%%%%%%%%%%%%%%%%%%%%%%%%%%%%%%
%%%%%%%%%%%%%%%%%%%%%%%%%%%%%%%%%%%%%%%%%%%%%%%%%%%%%%%
\section{Noise spectrum of the gravitational-wave detector using the
  double balanced homodyne detection}
\label{sec:Noise_spectrum_GW_detector}
%%%%%%%%%%%%%%%%%%%%%%%%%%%%%%%%%%%%%%%%%%%%%%%%%%%%%%%
%%%%%%%%%%%%%%%%%%%%%%%%%%%%%%%%%%%%%%%%%%%%%%%%%%%%%%%
%%%%%%%%%%%%%%%%%%%%%%%%%%%%%%%%%%%%%%%%%%%%%%%%%%%%%%%

%*******************************************************************

Here, we comment on the homodyne detection in gravitational-wave
detectors.
In this section, we concentrate only on the case $\eta=1/2$.

%*******************************************************************

The input-output relation for the main interferometer to detect
gravitational waves is formally given by
Eq.~(\ref{eq:hatbthehta-response}).
The gravitational-wave signal $h(\Omega)$ is a classical signal which
proportional to the identity operator in the sense of quantum theory.
On the other hand, the noise operator $\hat{h}_{n}(\Omega)$ satisfies
the property
\begin{eqnarray}
  \label{eq:noise_operator_property}
  \langle\hat{h}_{n}(\Omega)\rangle = 0.
\end{eqnarray}
Furthermore, the response function $R(\Omega)$ is a complex function.

%*******************************************************************

Substituting Eq.~(\ref{eq:hatbthehta-response}) into the
expression (\ref{eq:halfofhattD1D2plus+hattD3D4-result}) of the
expectation value $\langle\hat{t}_{\theta}\rangle$, we obtain
\begin{eqnarray}
  \label{eq:GWD-DBHD-expect}
  \langle\hat{t}_{\theta}(\Omega)\rangle
  =
  R(\Omega) h(\Omega)
  .
\end{eqnarray}
where we used the definition (\ref{eq:hatttheta-def}) of the operator
$\hat{t}_{\theta}$ and the property of the noise operator
$\hat{h}_{n}$.
Therefore, the expectation value of the operator
$\langle\hat{t}_{\theta}(\Omega)\rangle/R(\Omega)$ yields the
gravitational-wave signal $h(\Omega)$ in the frequency domain.

%*******************************************************************

In the case of $h(\Omega)\neq 0$, the fluctuation directly evaluated
through
$\left\langle\left(\hat{t}_{\theta}\hat{t}_{\theta}^{'\dagger} + \hat{t}_{\theta}^{'\dagger}\hat{t}_{\theta}\right)/2\right\rangle$
also includes the information of the signal $h(\Omega)$ as in the case
of Secs.~\ref{sec:measurement_of-hatb-or-hatbdagger} and
\ref{sec:DBHD_in_two_photon_formulation}.
This can be easily seen through
Eq.~(\ref{eq:hattthetahattthetadagger+hattthetadaggerhatttheta-result})
as
\begin{eqnarray}
  \left\langle
  \frac{1}{2}
  \left(
  \hat{t}_{\theta}\hat{t}_{\theta}^{'\dagger}
  +
  \hat{t}_{\theta}^{'\dagger}\hat{t}_{\theta}
  \right)
  \right\rangle
  &=&
      R(\Omega)
      R^{*}(\Omega')
      \left\langle
      \frac{1}{2}
      \left(
      \hat{h}_{n}(\Omega)
      \hat{h}_{n}^{\dagger}(\Omega')
      +
      \hat{h}_{n}^{\dagger}(\Omega')
      \hat{h}_{n}(\Omega)
      \right)
      \right\rangle
      \nonumber\\
  &&
      +
      R(\Omega)
      R^{*}(\Omega')
      h^{*}(\Omega')
      h(\Omega)
      \nonumber\\
  &&
      +
      \frac{1}{2}
      \left(
      1
      +
      \frac{\langle\hat{n}_{b-}+\hat{n}_{b+}\rangle}{|\gamma|^{2}}
      \right)
      2 \pi \delta(\Omega-\Omega')
     .
  \label{eq:GWD-DBHD-symmetrized-two-point-function}
\end{eqnarray}
The second-line of the right-hand side in
Eq.~(\ref{eq:GWD-DBHD-symmetrized-two-point-function}) corresponds to
the signal correlation.
Since $\langle\hat{h}_{n}\rangle=0$, we may apply the definition
(\ref{eq:spectral_density_def}) of the noise spectral density.
Then, we can easily obtain the definition of the signal referred noise
spectral density in the measurement of the operator $\hat{t}_{\theta}$
as
\begin{eqnarray}
  \label{eq:GWD-DBHD-ttheta-noise-spectral-density-def}
  \frac{1}{2R(\Omega)R^{*}(\Omega')} S_{\hat{t}_{\theta}^{(n)}}(\Omega) 2 \pi \delta(\Omega-\Omega')
  :=
  \left\langle
  \frac{1}{R(\Omega)R^{*}(\Omega')}
  \frac{1}{2}
  \left(
  \hat{t}_{\theta}\hat{t}_{\theta}^{'\dagger}
  +
  \hat{t}_{\theta}^{'\dagger}\hat{t}_{\theta}
  \right)
  \right\rangle
  -
  h^{*}(\Omega')
  h(\Omega)
  .
\end{eqnarray}
We can also define the signal referred noise spectral density for the
main interferometer as the noise spectral density for the operator
$\hat{h}_{n}$ through Eq.~(\ref{eq:spectral_density_def}) :
\begin{eqnarray}
  \frac{1}{2} S_{\hat{h}_{n}}(\Omega) 2 \pi \delta(\Omega-\Omega')
  :=
  \left\langle
  \frac{1}{2}
  \left(
  \hat{h}_{n}(\Omega)
  \hat{h}_{n}^{\dagger}(\Omega')
  +
  \hat{h}_{n}^{\dagger}(\Omega')
  \hat{h}_{n}(\Omega)
  \right)
  \right\rangle
  .
\end{eqnarray}
Then, we obtain the relation of the noise spectral densities by
\begin{eqnarray}
  \label{eq:GWD-DBHD-signal-referred-noise-spectral-desnity}
 \frac{1}{|R(\Omega)|^{2}} S_{\hat{t}_{\theta}^{(n)}}(\Omega)
  =
  S_{\hat{h}_{n}}(\Omega)
  +
  \frac{1}{|R(\Omega)|^{2}}
  \left(
  1
  +
  \frac{\langle\hat{n}_{b-}(\Omega)+\hat{n}_{b+}(\Omega)\rangle}{|\gamma(\Omega)|^{2}}
  \right)
  .
\end{eqnarray}
It is reasonable to regard the noise spectral density
(\ref{eq:GWD-DBHD-signal-referred-noise-spectral-desnity}) as the total
signal referred noise spectral density through our double balanced
homodyne detection in the situation where gravitational-wave signal exist.

%*******************************************************************

From Eq.~(\ref{eq:GWD-DBHD-signal-referred-noise-spectral-desnity}),
we may say that in the situation
$\langle\hat{n}_{b-}(\Omega)+\hat{n}_{b+}(\Omega)\rangle\ll
|\gamma(\Omega)|^{2}$, the last term in the bracket of the second term
of the right-hand side of
Eq.~(\ref{eq:GWD-DBHD-signal-referred-noise-spectral-desnity}) becomes
harmless but the ``1'' of the first term in the same bracket cannot be
controllable as in the cases of
Secs.~\ref{sec:measurement_of-hatb-or-hatbdagger} and
\ref{sec:DBHD_in_two_photon_formulation}.
However, the situation of
Eq.~(\ref{eq:GWD-DBHD-signal-referred-noise-spectral-desnity}) is
different from those in Secs.~\ref{sec:measurement_of-hatb-or-hatbdagger} and
\ref{sec:DBHD_in_two_photon_formulation}.
In
Eq.~(\ref{eq:GWD-DBHD-signal-referred-noise-spectral-desnity}),
we have the response function in the front of the additional noise due
to our double balanced homodyne detection.

%*******************************************************************

Finally, it is instructive to show an example of the conventional
gravitational-wave detectors.
We consider the input-output relation for a Fabry-P\'erot
gravitational-wave detector discussed in
Ref.~\cite{H.J.Kimble-Y.Levin-A.B.Matsko-K.S.Thorne-S.P.Vyatchanin-2001}
\begin{eqnarray}
  \label{eq:Kimble-16}
  \hat{b}_{1} = \hat{a}_{1} e^{2i\beta}, \quad
  \hat{b}_{2} = (\hat{a}_{2}-{\cal K}\hat{a}_{1}) e^{2i\beta}
  +
  \sqrt{2{\cal K}} \frac{h}{h_{SQL}} e^{i\beta}.
\end{eqnarray}
In this case, the operator $\hat{b}_{\theta}$ defined by
Eq.~(\ref{eq:hatbthehta-def}) is given by
\begin{eqnarray}
  \label{eq:Kimble-hatbtheta}
  \hat{b}_{\theta}
  =
  \frac{
  e^{i\beta}
  \sin\theta
  \sqrt{2{\cal K}}
  }{
  h_{SQL}
  }
  \left(
  \frac{
  e^{i\beta}
  h_{SQL}
  }{
  \sqrt{2{\cal K}}
  }
  \left(
  (\cot\theta - {\cal K}) \hat{a}_{1}
  +
  \hat{a}_{2}
  \right)
  +
  h
  \right)
  .
\end{eqnarray}
Identifying this input-output relation (\ref{eq:Kimble-hatbtheta})
with Eq.~(\ref{eq:hatbthehta-response}), we obtain
\begin{eqnarray}
  \label{eq:Kimble-hn-R}
  \hat{h}_{n}
  =
  \frac{
  e^{i\beta}
  h_{SQL}
  }{
  \sqrt{2{\cal K}}
  }
  \left(
  (\cot\theta - {\cal K}) \hat{a}_{1}
  +
  \hat{a}_{2}
  \right)
  ,
  \quad
  R(\Omega)
  =
  \frac{
  e^{i\beta}
  \sin\theta
  \sqrt{2{\cal K}}
  }{
  h_{SQL}
  }
  .
\end{eqnarray}

%*******************************************************************

Although we did not discuss the generation of the frequency-dependent
homodyne angle within this paper, we assume that we can generate the
homodyne angle $\theta$ with the dependence on the sideband frequency
$\Omega$ appropriately.
Under this assumption, we consider the case where
$|\gamma|^{2}\gg\langle\hat{n}_{b-}+\hat{n}_{b+}\rangle$.
In this case, the signal referred noise spectral density
(\ref{eq:GWD-DBHD-signal-referred-noise-spectral-desnity}) for our
double balanced homodyne detection
\begin{eqnarray}
  \frac{1}{|R(\Omega)|^{2}} S_{\hat{t}_{\theta}^{(n)}}(\Omega)
  &\gtrsim&
            \frac{h_{SQL}^{2}}{2{\cal K}}
            \left[
            2 \left(
            \cot\theta
            -
            \frac{{\cal K}}{2}
            \right)^{2}
            +
            \frac{{\cal K}^{2}}{2}
            +
            2
            \right]
            \nonumber\\
  &\geq&
         \frac{h_{SQL}^{2}}{2{\cal K}}
         \left(
         \frac{{\cal K}^{2}}{2}
         +
         2
         \right)
  \geq
         h_{SQL}^{2}
         .
         \label{eq:Kimble-GWD-DBHD-signal-referred-noise-spectral-desnity-1}
\end{eqnarray}
Here, we chose the frequency-dependent homodyne angle
$\cot\theta={\cal K}/2$ in
Eq.~(\ref{eq:Kimble-GWD-DBHD-signal-referred-noise-spectral-desnity-1}),
where ${\cal K}$ depends on the frequency $\Omega$~\cite{H.J.Kimble-Y.Levin-A.B.Matsko-K.S.Thorne-S.P.Vyatchanin-2001}.

%*******************************************************************

The noise spectral density
(\ref{eq:Kimble-GWD-DBHD-signal-referred-noise-spectral-desnity-1})
indicates that the additional noise from the vacuum fluctuation in our
realization of the double homodyne detection breaks the advantage of
the frequency-dependent homodyne detection discussed in
Ref.~\cite{H.J.Kimble-Y.Levin-A.B.Matsko-K.S.Thorne-S.P.Vyatchanin-2001}.
One of conclusions in
Ref.~\cite{H.J.Kimble-Y.Levin-A.B.Matsko-K.S.Thorne-S.P.Vyatchanin-2001}
is that we can completely eliminate the radiation-pressure noise
through the frequency-dependent homodyne detection.
We have to emphasize that we should not parallelly compare the noise
spectral density
(\ref{eq:Kimble-GWD-DBHD-signal-referred-noise-spectral-desnity-1})
with those in
Ref.~\cite{H.J.Kimble-Y.Levin-A.B.Matsko-K.S.Thorne-S.P.Vyatchanin-2001},
because
Eq.~(\ref{eq:Kimble-GWD-DBHD-signal-referred-noise-spectral-desnity-1})
is already taking into account of the readout scheme, while this effect
is not taken into account in
Ref.~\cite{H.J.Kimble-Y.Levin-A.B.Matsko-K.S.Thorne-S.P.Vyatchanin-2001}.
Since the total quantum measurement process even for classical
gravitational waves is completed through the inclusion of its readout scheme,
we have to discuss the signal-noise trade off relation through this
total quantum measurement process including its readout scheme.

%*******************************************************************

%%%%%%%%%%%%%%%%%%%%%%%%%%%%%%%%%%%%%%%%%%%%%%%%%%%%%%%
%%%%%%%%%%%%%%%%%%%%%%%%%%%%%%%%%%%%%%%%%%%%%%%%%%%%%%%
%%%%%%%%%%%%%%%%%%%%%%%%%%%%%%%%%%%%%%%%%%%%%%%%%%%%%%%
\section{Summary}
\label{sec:Summary}
%%%%%%%%%%%%%%%%%%%%%%%%%%%%%%%%%%%%%%%%%%%%%%%%%%%%%%%
%%%%%%%%%%%%%%%%%%%%%%%%%%%%%%%%%%%%%%%%%%%%%%%%%%%%%%%
%%%%%%%%%%%%%%%%%%%%%%%%%%%%%%%%%%%%%%%%%%%%%%%%%%%%%%%

%********************************************************************

In summary, in Sec.~\ref{sec:conventional-Homodyne-detections}, we first reviewed
the simple and the balanced homodyne detection through the understanding in the
references~\cite{H.M.Wiseman-G.J.Milburn-1993,H.M.Wiseman-G.J.Milburn-book-2009}
to exclude ambiguities of the understanding of ``homodyne detections''.
In this review, we use the Heisenberg picture in quantum theories and
this picture enable us to make the explanations of the homodyne
detections as compact as possible.
Based on this understanding of the homodyne detection, in
Sec.~\ref{sec:BHD-two-photon}, we examine the balanced homodyne
detection in the two-photon
formulation~\cite{C.M.Caves-B.L.Schumaker-1985}.
Our target is the statement, which states ``{\it the measurement of
  the quadrature
  $\hat{b}_{\theta}:=\cos\theta\hat{b}_{1}+\sin\theta\hat{b}_{2}$ is
  carried out by the balanced homodyne
  detection~\cite{S.P.Vyatchanin-A.B.Matsko-1993}.}''
To examine whether this statement is correct, or not, we consider the
unambiguous statement, ``{\it the expectation value of the operator
  $\hat{b}_{\theta}$ can be measured as the linear combination of the
  upper- and lower-sidebands from the output of the balanced homodyne
  detection.}''
Then, we have reached to the conclusion that any linear combination of
upper- and lower-sideband signal output of the balanced homodyne
detection {\it never} yields the expectation value
$\langle\hat{b}_{\theta}\rangle$.

%********************************************************************

In these examinations, we discuss more wider class of linear
combinations of two quadratures, which are summarized in
Table~\ref{tab:balanced-homodyne-combination-summary}.
As seen in Table~\ref{tab:balanced-homodyne-combination-summary}, we
found that many types of linear combinations of two quadratures.
However, only two cases which include the situations of the measurement
$\hat{b}_{\theta}$ is not possible.
On the other hand, in these examinations, we dare to assume that the
complex amplitude of the coherent state from the local oscillator is
completely controllable.
One of the aim of this assumption was to find the requirements to the
coherent state from the local oscillator.
In possible cases of
Table~\ref{tab:balanced-homodyne-combination-summary}, we have been
able to obtain these requirements as expected.
First, the complex amplitude of the coherent state from the local
oscillator must have its support at the frequencies
$\omega_{0}\pm\Omega$.
Second, the phase of the complex amplitude at the lower- and
upper-sideband frequency must be chosen according to the possible
cases as shown in Table~\ref{tab:balanced-homodyne-combination-summary}.
The first requirement means that the complex amplitude of the coherent
state from the local oscillator have its broad-band support.
This will not be satisfied by the monochromatic laser sources.
The second requirement for the phase of the complex amplitude is
essential to obtain the desired output signals.

%********************************************************************

Throughout this paper, we assumed the detected observable by
photodetectors is photon number, though there is a long history of the
controversy which is the detected variable by the photodetectors in
the case of the detection of multi-frequency optical
fields~\cite{multi-frequency-filed-detection}.
If this assumption is incorrect, the different arguments will be
necessary.
However, even in this case, we expect that our conclusion will be
correct.
In this paper, we just discussed the calculation procedures of the
expectation values of non-self-adjoint operators from the expectation
values of the self-adjoint operator based on the basic issues of
linear algebra.
Therefore, we expect that our conclusion will not change as far as the
direct observable by photodetectors is self-adjoint.

%********************************************************************

Based on the results in Sec.~\ref{sec:BHD-two-photon}, we
reached to the proposal of the ``{\it double balanced homodyne detection}'' in
Sec.~\ref{sec:Double_balanced_homodyne_detection}.
This double balanced homodyne detection is the combination of two
balanced homodyne detection whose phases of the complex amplitude of
the coherent state from the local oscillator are different with
$\pi/2$ from each other, and its application to the readout scheme
enables us to measure the expectation values
$\langle\hat{b}_{\theta}\rangle$.
Thus, we reached to the correct statement: ``{\it the expectation
  value $\langle\hat{b}_{\theta}\rangle$ an be measured as the linear
  combination of the upper- and lower-sidebands from the output of the
  double balanced homodyne detection}.''
We also rediscovered that the same interferometer setup as the
eight-port homodyne detection in the literature~\cite{N.G.Walker-etal-1986-J.W.Noh-etal-1991-1993-M.G.Raymer-etal-1993}
realizes the double balanced homodyne detection.
We also clarified the requirements for the complex amplitude of the
coherent state from the local oscillator to realize the double
balanced homodyne detection, which are similar to the requirements in Table~\ref{tab:balanced-homodyne-combination-summary}.

%********************************************************************

Even if we do not apply our double balanced homodyne detection, we can
carry out the usual balanced homodyne detection.
In this case, we cannot measure the expectation value
$\langle\hat{b}_{\theta}\rangle$.
Instead, we can measure the expectation value of a linear combination
of $\hat{b}_{1}$ and $\hat{b}_{1}^{\dagger}$, or a linear combination
of $\hat{b}_{2}$ and $\hat{b}_{2}^{\dagger}$, as shown in
Table~\ref{tab:balanced-homodyne-combination-summary}.
Even in these cases, the above requirements to the complex
amplitude from the local oscillator are also crucial.
Therefore, the requirements clarified in this paper are very important
not only for our double balanced homodyne detection but also for the
conventional balanced homodyne detection.
The broad-band support of the complex amplitude will be realized not
by the injection of the monochromatic laser but by the injection of
the optical pulses.
Even if we can fortunately create this broad-band support of the
complex amplitude from the local oscillator, we have to tune the phase
of the upper- and lower-sidebands.
Although the problem of the realization of the optical source from the
local oscillator is beyond the current scope of this paper, the double
balanced homodyne detection will be realized if these requirements are
satisfied.

%********************************************************************

Here, we note that the generation of the homodyne angle in this paper
is different from the that in
Ref.~\cite{H.J.Kimble-Y.Levin-A.B.Matsko-K.S.Thorne-S.P.Vyatchanin-2001}.
In
Ref.~\cite{H.J.Kimble-Y.Levin-A.B.Matsko-K.S.Thorne-S.P.Vyatchanin-2001},
a filter cavity to produce the frequency-dependent homodyne angle is
also discussed.
In their arguments, the signal field associated with the quadrature
$\hat{b}$ is injected to the filter cavity.
However, in this paper, the homodyne angle is generated by the phase
of the complex amplitude of the coherent state from the local
oscillator.
If we want to generate the frequency-dependent homodyne angle along
our double balanced homodyne detection in this paper, we will have to
inject the optical field from the local oscillator to a filter cavity
before the interferometer setup of our double balanced homodyne
detection.

%********************************************************************

As emphasized in Sec.~\ref{sec:introduction}, our double balanced
homodyne detection is not the direct measurement of the
non-self-adjoint operator $\hat{b}_{\theta}$ but just the calculation
procedure which yields the expectation value
$\langle\hat{b}_{\theta}\rangle$ from the four photon number
measurements.
The difference from the direct measurement of the operator
$\hat{b}_{\theta}$ leads the additional fluctuations in the
measurement, which is also evaluated in
Sec.~\ref{sec:DBHD_in_two_photon_formulation}.
Trivially, the imperfection of the interferometer, which is modeled by
the parameter $\eta\neq1/2$ in this paper, gives additional noise.
Even in the ideal case $\eta=1/2$, the additional noise arise in our
measurement.
Furthermore, in Sec.~\ref{sec:Noise_spectrum_GW_detector}, we also
evaluate of the noise spectral density in the case where the main
interferometer is the Fabry-P\'erot gravitational-wave detector in
Ref.~\cite{H.J.Kimble-Y.Levin-A.B.Matsko-K.S.Thorne-S.P.Vyatchanin-2001}
as an example.
This example explicitly shows that the additional noise due to the
readout scheme may break the advantage of the main interferometer.
The total quantum measurement process even for classical gravitational
waves is completed through the inclusion of its readout scheme.
If we want to discuss the signal-noise trade off relation, we have to
discuss the total quantum measurement process including its readout
scheme.
In this sense, researches on the readout scheme is crucial in
gravitational-wave detectors.

%********************************************************************

To develop the discussion on the signal-noise trade off relation in
gravitational-wave detectors through the total quantum measurement
process, we have to discuss the advantages or disadvantages of our
double homodyne detection comparing with other readout scheme, i.e.,
DC readout scheme, and heterodyne detection, and other homodyne
detection.
Furthermore, one of possibilities to improve the signal-noise trade
off relation for our double balanced homodyne detection will be the
injection of the squeezed state from the vacuum ports, from which the
fields associated with the quadratures $\hat{e}_{i}$ and $\hat{f}_{i}$
are injected.
More importantly, we have to examine the other possibilities of the
direct observable in photodetectors than the ``photon number'' in the
frequency domain, since the logic in this paper is entirely based on
the assumption that the observed variable of optical fields by the
photodetector is ``photon number'' in the frequency domain.
As noted in the introduction, there is a long history of the
controversy which variable is the directly detected variable by the
photodetectors in the case of the detection of multi-frequency optical
fields~\cite{multi-frequency-filed-detection}.
This issue will depend on the physical process in photodetectors.
Thus, there are many important issues to be clarified as future work,
though these issues are beyond current scope of this paper.
We hope the ingredients of this paper are useful for the
investigation of these future works.

%********************************************************************

%%%%%%%%%%%%%%%%%%%%%%%%%%%%%%%%%%%%%%%%%%%%%%%%%%%%%%%
%%%%%%%%%%%%%%%%%%%%%%%%%%%%%%%%%%%%%%%%%%%%%%%%%%%%%%%
%%%%%%%%%%%%%%%%%%%%%%%%%%%%%%%%%%%%%%%%%%%%%%%%%%%%%%%
\section*{Acknowledgments}
%%%%%%%%%%%%%%%%%%%%%%%%%%%%%%%%%%%%%%%%%%%%%%%%%%%%%%%
%%%%%%%%%%%%%%%%%%%%%%%%%%%%%%%%%%%%%%%%%%%%%%%%%%%%%%%
%%%%%%%%%%%%%%%%%%%%%%%%%%%%%%%%%%%%%%%%%%%%%%%%%%%%%%%

%********************************************************************

K.N. acknowledges to Dr. Tomotada Akutsu and the other members of the
gravitational-wave project office in NAOJ for their continuous
encouragement to our research.
K.N. appreciate Prof. Akio Hosoya, Prof. Izumi Tsutsui, and
Dr. Hiroyuki Takahashi for their support and continuous encouragement.
We also appreciate Prof. Werner Vogel for valuable information and the
anonymous referee of Progress of Experimental and Theoretical Physics
for fruitful comments.

%********************************************************************

%%%%%%%%%%%%%%%%%%%%%%%%%%%%%%%%%%%%%%%%%%%%%%%%%%%%%%%
%%%%%%%%%%%%%%%%%%%%%%%%%%%%%%%%%%%%%%%%%%%%%%%%%%%%%%%
%%%%%%%%%%%%%%%%%%%%%%%%%%%%%%%%%%%%%%%%%%%%%%%%%%%%%%%
\appendix
%%%%%%%%%%%%%%%%%%%%%%%%%%%%%%%%%%%%%%%%%%%%%%%%%%%%%%%
%%%%%%%%%%%%%%%%%%%%%%%%%%%%%%%%%%%%%%%%%%%%%%%%%%%%%%%
%%%%%%%%%%%%%%%%%%%%%%%%%%%%%%%%%%%%%%%%%%%%%%%%%%%%%%%

%********************************************************************

%%%%%%%%%%%%%%%%%%%%%%%%%%%%%%%%%%%%%%%%%%%%%%%%%%%%%%%
%%%%%%%%%%%%%%%%%%%%%%%%%%%%%%%%%%%%%%%%%%%%%%%%%%%%%%%
\section{Derivations of Eqs.~(\ref{eq:hatt+-def})}
\label{sec:Deriv_of_hatt+_hatt-_operators}
%%%%%%%%%%%%%%%%%%%%%%%%%%%%%%%%%%%%%%%%%%%%%%%%%%%%%%%
%%%%%%%%%%%%%%%%%%%%%%%%%%%%%%%%%%%%%%%%%%%%%%%%%%%%%%%

%*******************************************************************

If we may use a similar way to those in
Sec.~\ref{sec:exp_val_b1_b1dagger_b2_b2dagger}, we can show that the
expectation value of the operator
$\cos\theta\hat{b}_{1}+\sin\theta\hat{b}_{2}$ is also possible as
discussed in Sec.~\ref{sec:exp_val_costhetab1+sinthetab2}.
Here, we show the explicit derivation of the definitions
(\ref{eq:hatt+-def}) of the operators $\hat{t}_{+}$ and $\hat{t}_{-}$,
respectively.
This is the aim of this Appendix.

%*******************************************************************

To reach to this aim, we return to
Eq.~(\ref{eq:alphaexphats++betaexphats-}) and we concentrate only on
the case where the coefficients $\alpha$ and $\beta$ are real.
In this case, the linear combination
(\ref{eq:alphaexphats++betaexphats-}) is written in the form
\begin{eqnarray}
  \alpha \left\langle\hat{s}_{+}\right\rangle
  +
  \beta \left\langle\hat{s}_{-}\right\rangle
  =
  \frac{1}{\sqrt{2}}
  \left\langle
  \kappa
  \hat{b}_{1}
  +
  \lambda
  \hat{b}_{2}
  +
  \kappa^{*}
  \hat{b}_{1}^{\dagger}
  +
  \lambda^{*}
  \hat{b}_{2}^{\dagger}
  \right\rangle
  .
  \label{eq:alphaexphats++betaexphats--kappa-lambda}
\end{eqnarray}
Here, we defined
\begin{eqnarray}
  \kappa := \alpha \gamma_{+}^{*} + \beta \gamma_{-}, \quad
  \lambda := i \left(\alpha \gamma_{+}^{*} - \beta \gamma_{-}\right).
  \label{eq:kappa-lambda-def}
\end{eqnarray}

%*******************************************************************

Here, we denote $\gamma_{\pm}$ as
\begin{eqnarray}
  \label{eq:gammapm-|gammapm|expithetapm}
  \gamma_{\pm} =: |\gamma_{\pm}| e^{i\theta_{\pm}}.
\end{eqnarray}
In terms of $|\gamma_{\pm}|$, $\theta_{\pm}$, $\alpha$, and $\beta$,
from Eqs.~(\ref{eq:kappa-lambda-def}), we obtain the expression of
$\kappa$ and $\lambda$ as
\begin{eqnarray}
  \kappa =: |\kappa| e^{i\varphi_{\kappa}}, \quad \lambda =: |\lambda| e^{i\varphi_{\lambda}},
\end{eqnarray}
where $|\kappa|$, $\varphi_{\kappa}$, $|\lambda|$, and
$\varphi_{\lambda}$ are given by
\begin{eqnarray}
  |\kappa|
  &=&
      \left[
      \alpha^{2} |\gamma_{+}|^{2}
      + \beta^{2} |\gamma_{-}|^{2}
      + 2 \alpha \beta |\gamma_{+}| |\gamma_{-}| \cos(\theta_{+}+\theta_{-})
      \right]^{1/2}
      ,
      \label{eq:|kappa|-expression}
      \\
  \tan\varphi_{\kappa}
  &=&
      \frac{
      -
      \alpha |\gamma_{+}| \sin\theta_{+}
      +
      \beta |\gamma_{-}| \sin\theta_{-}
      }{
      \alpha |\gamma_{+}| \cos\theta_{+}
      +
      \beta |\gamma_{-}| \cos\theta_{-}
      }
      ,
      \label{eq:varphikappa-expression}
      \\
  |\lambda|
  &=&
      \left[
      \alpha^{2} |\gamma_{+}|^{2}
      +
      \beta^{2} |\gamma_{-}|^{2}
      -
      2 \alpha \beta |\gamma_{+}| |\gamma_{-}| \cos(\theta_{+}+\theta_{-})
      \right]^{1/2}
      ,
      \label{eq:|lambda|-expression}
      \\
  \tan\varphi_{\lambda}
  &=&
      \frac{
      \alpha |\gamma_{+}| \cos\theta_{+}
      - \beta |\gamma_{-}| \cos\theta_{-}
      }{
      \alpha |\gamma_{+}| \sin\theta_{+}
      + \beta |\gamma_{-}| \sin\theta_{-}
      }
      .
      \label{eq:varphilambda-expression}
\end{eqnarray}

%*******************************************************************

As seen in Sec.~\ref{sec:exp_val_b1_b1dagger_b2_b2dagger}, we can
expect that the information of
Eq.~(\ref{eq:alphaexphats++betaexphats--kappa-lambda}) with vanishing
phase in some sense and the information of
Eq.~(\ref{eq:alphaexphats++betaexphats--kappa-lambda}) with
$\pi/2$-phase in some sense are necessary.
Further we will be able to derive the expectation value
$\langle\cos\theta\hat{b}_{1}+\sin\theta{b}_{2}\rangle$ from these
information.
As our speculation, we consider the cases
$\varphi_{\kappa}=\varphi_{\lambda}=0$ and
$\varphi_{\kappa}=\varphi_{\lambda}=\pi/2$, respectively.
As the result, we will see that our speculation is justified.

%*******************************************************************

%%%%%%%%%%%%%%%%%%%%%%%%%%%%%%%%%%%%%%%%%%%%%%%%%%%%%%%
%%%%%%%%%%%%%%%%%%%%%%%%%%%%%%%%%%%%%%%%%%%%%%%%%%%%%%%
\subsection{$\varphi_{\kappa}=\varphi_{\lambda}=0$ case}
\label{sec:varphikappa=varphilambda=0-case}
%%%%%%%%%%%%%%%%%%%%%%%%%%%%%%%%%%%%%%%%%%%%%%%%%%%%%%%
%%%%%%%%%%%%%%%%%%%%%%%%%%%%%%%%%%%%%%%%%%%%%%%%%%%%%%%

%*******************************************************************

First, we consider the case where
$\varphi_{\kappa}=\varphi_{\lambda}=0$.
From Eqs.~(\ref{eq:varphikappa-expression}) and
(\ref{eq:varphilambda-expression}), this case is characterized by the
equations
\begin{eqnarray}
  \label{eq:varphikappa=0-condition}
  - \alpha |\gamma_{+}| \sin\theta_{+}
  + \beta |\gamma_{-}| \sin\theta_{-}
  &=&
      0
      ,
  \\
  \label{eq:varphilambda=0-condition}
  \alpha |\gamma_{+}| \cos\theta_{+}
  - \beta |\gamma_{-}| \cos\theta_{-}
  &=&
      0
      .
\end{eqnarray}
Here, we note that $\alpha|\gamma_{+}|\neq 0$ and
$\beta|\gamma_{-}|\neq 0$.
From Eqs.~(\ref{eq:varphikappa=0-condition}) and
(\ref{eq:varphilambda=0-condition}), we obtain
\begin{eqnarray}
  \label{eq:varphikappa=varphilambda=0-condition-1}
  \beta
  =
  \alpha
  \frac{
    |\gamma_{+}| \sin\theta_{+}
  }{
    |\gamma_{-}| \sin\theta_{-}
  }
  =
  \alpha
  \frac{
    |\gamma_{+}| \cos\theta_{+}
  }{
    |\gamma_{-}| \cos\theta_{-}
  }
  .
\end{eqnarray}
The second equality in
Eq.~(\ref{eq:varphikappa=varphilambda=0-condition-1}) gives
\begin{eqnarray}
  \label{eq:tantheta+=tantheta-}
  \tan\theta_{+}
  =
  \tan\theta_{-}
  .
\end{eqnarray}
As a solution to Eq.~(\ref{eq:tantheta+=tantheta-}), we may choose
\begin{eqnarray}
  \label{eq:theta+=theta-}
  \theta_{+}=\theta_{-}
  .
\end{eqnarray}
In this choice, the first equality in
Eq.~(\ref{eq:varphikappa=varphilambda=0-condition-1}) yields
\begin{eqnarray}
  \beta
  =
  \alpha
  \frac{
    |\gamma_{+}|
  }{
   |\gamma_{-}|
  }
  .
  \label{eq:varphikappa=varphilambda=0-condition-2}
\end{eqnarray}

%*******************************************************************

Substituting and Eqs.~(\ref{eq:theta+=theta-}) and
(\ref{eq:varphikappa=varphilambda=0-condition-2}) into
Eqs.~(\ref{eq:|kappa|-expression}) and (\ref{eq:|lambda|-expression}),
we obtain
\begin{eqnarray}
  \label{eq:varphikappa=varphilambda=0-solution-kappa-lambda}
  \kappa
  =
  2 \alpha |\gamma_{+}| |\cos\theta_{+}|
  ,
  \quad
  \lambda
  =
  2 \alpha |\gamma_{+}| |\sin\theta_{+}|
  .
\end{eqnarray}
Furthermore, through Eqs.~(\ref{eq:theta+=theta-}),
(\ref{eq:varphikappa=varphilambda=0-condition-2}), and
(\ref{eq:varphikappa=varphilambda=0-solution-kappa-lambda}), the
linear combination (\ref{eq:alphaexphats++betaexphats--kappa-lambda})
is given by
\begin{eqnarray}
  \frac{1}{\sqrt{2}}
  \left(
  \frac{
  \left\langle\hat{s}_{+}\right\rangle
  }{
  |\gamma_{+}|
  }
  +
  \frac{
  \left\langle\hat{s}_{-}\right\rangle
  }{
  |\gamma_{-}|
  }
  \right)
  =
  \left\langle
  |\cos\theta_{+}|
  \left(
  \hat{b}_{1}
  +
  \hat{b}_{1}^{\dagger}
  \right)
  +
  |\sin\theta_{+}|
  \left(
  \hat{b}_{2}
  +
  \hat{b}_{2}^{\dagger}
  \right)
  \right\rangle
  .
  \label{eq:varphikappa=varphilambda=0-case-final-result}
\end{eqnarray}
This is the expected result and corresponds to
Eq.~(\ref{eq:hatb1-hatb1dagger-case-result-gammapm=gamma-theta0}) in
Sec.~\ref{sec:exp_val_b1_b1dagger_b2_b2dagger}.
Therefore, the half of our speculation was correct.

%*******************************************************************

%%%%%%%%%%%%%%%%%%%%%%%%%%%%%%%%%%%%%%%%%%%%%%%%%%%%%%%
%%%%%%%%%%%%%%%%%%%%%%%%%%%%%%%%%%%%%%%%%%%%%%%%%%%%%%%
\subsection{$\varphi_{\kappa}=\varphi_{\lambda}=\pi/2$ case}
\label{sec:varphikappa=varphilambda=pi/2-case}
%%%%%%%%%%%%%%%%%%%%%%%%%%%%%%%%%%%%%%%%%%%%%%%%%%%%%%%
%%%%%%%%%%%%%%%%%%%%%%%%%%%%%%%%%%%%%%%%%%%%%%%%%%%%%%%

%*******************************************************************

Next, we consider the case where $\varphi_{\kappa}=\varphi_{\lambda}=\pi/2$.
From Eq.~(\ref{eq:varphikappa-expression}) and
(\ref{eq:varphilambda-expression}), this case is characterized by the
equations as
\begin{eqnarray}
  \label{eq:varphikappa=pi/2-condition}
  \alpha |\gamma_{+}| \cos\theta_{+}
  +
  \beta |\gamma_{-}| \cos\theta_{-}
  =
  0
  , \\
  \label{eq:varphilambda=pi/2-condition}
  \alpha |\gamma_{+}| \sin\theta_{+}
  +
  \beta |\gamma_{-}| \sin\theta_{-}
  =
  0.
\end{eqnarray}
Here, we also note that $\alpha|\gamma_{+}|\neq 0$ and
$\beta|\gamma_{-}|\neq 0$.
From Eqs.~(\ref{eq:varphikappa=pi/2-condition}) and
(\ref{eq:varphilambda=pi/2-condition}), we obtain
\begin{eqnarray}
  \label{eq:varphikappa=varphilambda=pi/2-condition-1}
  \beta
  =
  -
  \alpha
  \frac{
    |\gamma_{+}| \cos\theta_{+}
  }{
    |\gamma_{-}| \cos\theta_{-}
  }
  =
  -
  \alpha
  \frac{
    |\gamma_{+}| \sin\theta_{+}
  }{
    |\gamma_{-}| \sin\theta_{-}
  }
  .
\end{eqnarray}
The second equality in
Eq.~(\ref{eq:varphikappa=varphilambda=pi/2-condition-1}) gives
\begin{eqnarray}
  \label{eq:tantheta+=tantheta-2}
  \tan\theta_{+} = \tan\theta_{-}
  .
\end{eqnarray}
As a solution to Eq.~(\ref{eq:tantheta+=tantheta-2}), we may choose
\begin{eqnarray}
  \label{eq:theta+=theta-2}
  \theta_{-}
  =
  \theta_{+}
  ,
\end{eqnarray}
as in the case of $\varphi_{\kappa}=\varphi_{\lambda}=0$ in
Sec.~\ref{sec:varphikappa=varphilambda=0-case}.
In this choice, the first equality in
Eq.~(\ref{eq:varphikappa=varphilambda=pi/2-condition-1}) yields
\begin{eqnarray}
  \label{eq:varphikappa=varphilambda=pi/2-condition-2}
  \beta
  =
  -
  \alpha
  \frac{
    |\gamma_{+}|
  }{
    |\gamma_{-}|
  }
  .
\end{eqnarray}

%*******************************************************************

Substituting Eqs.~(\ref{eq:theta+=theta-2}) and
(\ref{eq:varphikappa=varphilambda=pi/2-condition-2}) into
Eqs.~(\ref{eq:|kappa|-expression}) and (\ref{eq:|lambda|-expression}),
we obtain
\begin{eqnarray}
  \label{eq:varphikappa=varphilambda=pi/2-solution-kappa-lambda}
  \kappa
  =
  2 i \alpha |\gamma_{+}| |\sin\theta_{+}|
  ,
  \quad
  \lambda
  =
  2 i \alpha |\gamma_{+}| |\cos\theta_{+}|
  .
\end{eqnarray}
Furthermore, through Eqs.~(\ref{eq:theta+=theta-2}),
(\ref{eq:varphikappa=varphilambda=pi/2-condition-2}), and
(\ref{eq:varphikappa=varphilambda=pi/2-solution-kappa-lambda}), the
linear combination (\ref{eq:alphaexphats++betaexphats--kappa-lambda})
is given by
\begin{eqnarray}
  \frac{1}{\sqrt{2}i}
  \left(
  \frac{
  \left\langle\hat{s}_{+}\right\rangle
  }{
  |\gamma_{+}|
  }
  -
  \frac{
  \left\langle\hat{s}_{-}\right\rangle
  }{
  |\gamma_{-}|
  }
  \right)
  =
  \left\langle
  |\sin\theta_{+}|
  \left(
  \hat{b}_{1}
  -
  \hat{b}_{1}^{\dagger}
  \right)
  +
  |\cos\theta_{+}|
  \left(
  \hat{b}_{2}
  -
  \hat{b}_{2}^{\dagger}
  \right)
  \right\rangle
  .
  \label{eq:varphikappa=varphilambda=pi/2-case-final-result}
\end{eqnarray}
This is also the expected result and corresponds to
Eq.~(\ref{eq:hatb1-hatb1dagger-case-result-gammapm=gamma-thetapi/2}) in
Sec.~\ref{sec:exp_val_b1_b1dagger_b2_b2dagger}.
Therefore, the remaining part of our speculation was correct.

%*******************************************************************

%%%%%%%%%%%%%%%%%%%%%%%%%%%%%%%%%%%%%%%%%%%%%%%%%%%%%%%
%%%%%%%%%%%%%%%%%%%%%%%%%%%%%%%%%%%%%%%%%%%%%%%%%%%%%%%
\subsection{To calcluate $\cos\theta\hat{b}_{1}+\sin\theta\hat{b}_{2}$}
\label{sec:to_calc_costhetab1+sinthetab2}
%%%%%%%%%%%%%%%%%%%%%%%%%%%%%%%%%%%%%%%%%%%%%%%%%%%%%%%
%%%%%%%%%%%%%%%%%%%%%%%%%%%%%%%%%%%%%%%%%%%%%%%%%%%%%%%

%*******************************************************************

As derived in the previous subsections, we have seen that we can
obtain Eq.~(\ref{eq:varphikappa=varphilambda=0-case-final-result}) or
(\ref{eq:varphikappa=varphilambda=pi/2-case-final-result}) through the
appropriate choice of the complex amplitude $\gamma_{\pm}$ for the
coherent state from the local oscillator.
To obtain Eq.~(\ref{eq:varphikappa=varphilambda=0-case-final-result})
as a signal output, we have to choose the phase of  $\gamma_{\pm}$ as
Eq.~(\ref{eq:theta+=theta-}) and have to take the linear combination
with the coefficients $(\alpha,\beta)$ as
Eq.~(\ref{eq:varphikappa=varphilambda=0-condition-2}).
On the other hand, to obtain
Eq.~(\ref{eq:varphikappa=varphilambda=pi/2-case-final-result}) as a
signal output, we have to choose the phase of $\gamma_{\pm}$ as
Eq.~(\ref{eq:theta+=theta-2}) and have to take the linear combination
with the coefficients $(\alpha,\beta)$ as
Eq.~(\ref{eq:varphikappa=varphilambda=pi/2-condition-2}).
The conditions (\ref{eq:theta+=theta-}) and (\ref{eq:theta+=theta-2})
are same.
Although the choice of the coefficients $(\alpha,\beta)$ are easy in
practice, it is necessary to satisfy $|\gamma_{\pm}|\neq 0$.
Therefore, to accomplish
Eq.~(\ref{eq:varphikappa=varphilambda=0-case-final-result}) or
(\ref{eq:varphikappa=varphilambda=pi/2-case-final-result}) as a signal
output, $|\gamma_{\pm}|\neq 0$ and $\theta_{+}=\theta_{-}$ are
regarded as the requirement for the complex amplitude $\gamma_{\pm}$
for the coherent state from the local oscillator.

%*******************************************************************

As in the cases in Sec.~\ref{sec:exp_val_b1_b1dagger_b2_b2dagger}, we
assume that the expectation values
(\ref{eq:varphikappa=varphilambda=0-case-final-result}) and
(\ref{eq:varphikappa=varphilambda=pi/2-case-final-result}) can be
independently obtained as the signal output at the same time.
This assumption can be realized through the interferometer
configuration depicted in
Fig.~\ref{fig:DoubleBalancedHomodyneDetection-configuration}.

%*******************************************************************

If we choose the phase of the complex amplitude $\gamma_{\pm}$ so that
$\theta_{\pm}=\theta$ for the signal output
(\ref{eq:varphikappa=varphilambda=0-case-final-result}) and choose
$\theta_{\pm}=\theta+\pi/2$ for the signal output
(\ref{eq:varphikappa=varphilambda=pi/2-case-final-result}), these
signal outputs are given by
\begin{eqnarray}
  \left.
  \frac{1}{\sqrt{2}}
  \left(
  \frac{
  \left\langle\hat{s}_{+}\right\rangle
  }{
  |\gamma_{+}|
  }
  +
  \frac{
  \left\langle\hat{s}_{-}\right\rangle
  }{
  |\gamma_{-}|
  }
  \right)
  \right|_{\theta_{\pm}=\theta}
  &=&
      \left\langle
      |\cos\theta|
      \left(
      \hat{b}_{1}
      +
      \hat{b}_{1}^{\dagger}
      \right)
      +
      |\sin\theta|
      \left(
      \hat{b}_{2}
      +
      \hat{b}_{2}^{\dagger}
      \right)
      \right\rangle
      ,
      \label{eq:varphikappa=varphilambda=0-case-theteapm=theta}
  \\
  \left.
  \frac{1}{\sqrt{2}i}
  \left(
  \frac{
  \left\langle\hat{s}_{+}\right\rangle
  }{
  |\gamma_{+}|
  }
  -
  \frac{
  \left\langle\hat{s}_{-}\right\rangle
  }{
  |\gamma_{-}|
  }
  \right)
  \right|_{\theta_{\pm}=\pi/2+\theta}
  &=&
      \left\langle
      |\cos\theta|
      \left(
      \hat{b}_{1}
      -
      \hat{b}_{1}^{\dagger}
      \right)
      +
      |\sin\theta|
      \left(
      \hat{b}_{2}
      -
      \hat{b}_{2}^{\dagger}
      \right)
      \right\rangle
      .
      \label{eq:varphikappa=varphilambda=pi/2-case-theteapm=pi/2+theta}
\end{eqnarray}
From Eqs.~(\ref{eq:varphikappa=varphilambda=0-case-theteapm=theta})
and (\ref{eq:varphikappa=varphilambda=pi/2-case-theteapm=pi/2+theta}), we
obtain
\begin{eqnarray}
  &&
     \!\!\!\!\!\!\!\!\!\!\!
     \frac{1}{2}
     \left\{
     \left.
     \frac{1}{\sqrt{2}}
     \left(
     \frac{
     \left\langle\hat{s}_{+}\right\rangle
     }{
     |\gamma_{+}|
     }
     +
     \frac{
     \left\langle\hat{s}_{-}\right\rangle
     }{
     |\gamma_{-}|
     }
     \right)
     \right|_{\theta_{\pm}=\theta}
     +
     \left.
     \frac{1}{\sqrt{2}i}
     \left(
     \frac{
     \left\langle\hat{s}_{+}\right\rangle
     }{
     |\gamma_{+}|
     }
     -
     \frac{
     \left\langle\hat{s}_{-}\right\rangle
     }{
     |\gamma_{-}|
     }
     \right)
     \right|_{\theta_{\pm}=\pi/2+\theta}
     \right\}
     \nonumber\\
  &=&
  \left\langle
    |\cos\theta| \hat{b}_{1} + |\sin\theta| \hat{b}_{2}
  \right\rangle
  .
  \label{eq:DoubleBalancedHomodyneDetection-pre-result}
\end{eqnarray}
The factors $|\cos\theta|$ and $|\sin\theta|$ in
Eq.~(\ref{eq:DoubleBalancedHomodyneDetection-pre-result}) is naturally
replaced by $\cos\theta$ and $\sin\theta$, respectively, in
Sec.~\ref{sec:exp_val_costhetab1+sinthetab2}.
Thus, we can see that if the expectation values
(\ref{eq:varphikappa=varphilambda=0-case-final-result}) and (\ref{eq:varphikappa=varphilambda=pi/2-case-final-result}) can be
independently obtained as the signal output at the same time.
We can measure the expectation value
$\langle\cos\theta\hat{b}_{1}+\sin\theta\hat{b}_{2}\rangle$.

%*******************************************************************

Before closing this appendix, we emphasize that the operators
\begin{eqnarray}
  \frac{1}{\sqrt{2}}\left(
  \frac{\hat{s}_{+}}{|\gamma_{+}|}
  +
  \frac{\hat{s}_{-}}{|\gamma_{-}|}
  \right)
  ,
  \quad
  \frac{1}{\sqrt{2}i}\left(
  \frac{\hat{s}_{+}}{|\gamma_{+}|}
  -
  \frac{\hat{s}_{-}}{|\gamma_{-}|}
  \right)
\end{eqnarray}
play the most important role for the measurement of
$\langle\cos\theta\hat{b}_{1}+\sin\theta\hat{b}_{2}\rangle$.
Therefore, in the main text, we begin our discussion from the explicit
form of these operators.

%*******************************************************************

%%%%%%%%%%%%%%%%%%%%%%%%%%%%%%%%%%%%%%%%%%%%%%%%%%%%%%%
%%%%%%%%%%%%%%%%%%%%%%%%%%%%%%%%%%%%%%%%%%%%%%%%%%%%%%%
%%%%%%%%%%%%%%%%%%%%%%%%%%%%%%%%%%%%%%%%%%%%%%%%%%%%%%%

\end{document}